%% file: main.tex
\documentclass[longauth]{aaEC}

\usepackage{euclid}
\usepackage{natbib}
\usepackage{graphicx}
\usepackage{tabularx}
\usepackage{subcaption}
\usepackage{txfonts,textcomp}
\usepackage{hyperref}
\hypersetup{
    colorlinks=true,
    linkcolor=blue,
    filecolor=magenta,      
    urlcolor=blue,
    citecolor=blue
}

\usepackage{orcidlink} 
\newcommand{\orcid}[1]{\orcidlink{#1}}

\makeatletter
\renewcommand*\aa@pageof{, page \thepage{} of \pageref*{LastPage}}
\makeatother

\graphicspath{ {./images/} }

\newcommand*{\DUNE}{DUNE\xspace}
\newcommand*{\SPACE}{SPACE\xspace}
\newcommand*{\Gaia}{\textit{Gaia}\xspace}

\usepackage[switch, modulo]{lineno}

\usepackage{color}

\begin{document} 

\title{\textit{Euclid}}
\subtitle{II. The VIS instrument}    
\titlerunning{The \Euclid VIS Instrument} 
\authorrunning{Euclid Collaboration: M. Cropper et al.}

\input{authors}

\date{\today}

\abstract{This paper presents the specification, design, and development of the Visible Camera (VIS) on the European Space Agency's \Euclid mission. VIS is a large optical-band imager with a field of view of 0.54\,deg$^2$ sampled at \ang{;;0.1} with an array of 609\,Megapixels and a spatial resolution of \ang{;;0.18}. It will be used to survey approximately 14\,000\,deg$^2$ of extragalactic sky to measure the distortion of galaxies in the redshift range $z=0.1$--1.5 resulting from weak gravitational lensing, one of the two principal cosmology probes leveraged by \Euclid. With photometric redshifts, the distribution of dark matter can be mapped in three dimensions, and the extent to which this has changed with look-back time can be used to constrain the nature of dark energy and theories of gravity. The entire VIS focal plane will be transmitted to provide the largest images of the Universe from space to date, specified to reach $m_{\rm AB}\geq24.5$ with a signal-to-noise ratio ${\rm S/N}\geq10$ in a single broad $I_{\scriptscriptstyle\rm E} \simeq (r+i+z)$ band over a six-year survey. The particularly challenging aspects of the instrument are the control and calibration of observational biases, which lead to stringent performance requirements and calibration regimes. With its combination of spatial resolution, calibration knowledge, depth, and area covering most of the extra-Galactic sky, VIS will also provide a legacy data set for many other fields. This paper discusses the rationale behind the conception of VIS  and describes the instrument design and development, before reporting the prelaunch performance derived from ground calibrations and brief results from the in-orbit commissioning. VIS should reach fainter than $m_{\rm AB}=25$ with ${\rm S/N}\geq 10$ for galaxies with a full width at half maximum of \ang{;;0.3} in a \ang{;;1.3} diameter aperture over the Wide Survey, and $m_{\rm AB}\geq26.4$ for a Deep Survey that will cover more than 50\,deg$^2$. The paper also describes how the instrument works with the \Euclid telescope and survey, and with the science data processing, to extract the cosmological information.} 


\maketitle

\section{Introduction}
\label{sec:intro}

The Visible Camera (VIS) is a large optical-band imager built for the European Space Agency (ESA) \Euclid cosmology mission, the principle aim of which is to investigate the nature of dark energy and dark matter \citep{Refregier:cospar, Laureijs:11, Mellier:16, Racca:16, EuclidSkyOverview}. By measuring the correlated distortions of galaxies at different spatial scales and redshifts caused by weak gravitational lensing, the growth of structure in the Universe and its expansion history can be deduced \citep[for an early review see][]{Mellier:99}. From this, parameters for cosmological models can be determined, including the cosmological parameters for the $\Lambda$CDM concordance model and those for the modified version that includes a linear description of the equation of state of dark energy. 

Weak lensing is one of the most powerful probes of dark matter and large-scale structure, and can be used to constrain the nature of dark energy if the requisite level of control of observational biases can be achieved \citep{Albrecht:06,Peacock:06,Frieman:08}. A large number of galaxies must be imaged to measure the correlated distortions with sufficient precision, and so a characteristic of VIS is its exceptionally large field of view sampled at high spatial resolution. \Euclid also measures the growth of structure through galaxy clustering measurements using infrared spectroscopy and infrared imaging provided by the other instrument on the payload, the Near Infrared Spectrometer Photometer \citep[NISP;][]{EuclidSkyNISP}. The infrared measurements provide constraints on the cosmological parameters complementary to those derived from measurements of weak lensing. With these two instruments, \Euclid is expected to provide the most advanced observational data on the large-scale constituents of the Universe, with power that extends beyond testing standard cosmological models to constraining modified gravity alternatives and the neutrino mass hierarchy \citep{Amendola:18}. In addition, the combination of high-spatial-resolution visible-band imaging from VIS with infrared spectroscopy and deep imaging over the majority of the extragalactic sky will facilitate a wide range of science beyond cosmology alone \citep{Refregier:08,Refregier:10,Laureijs:11,EuclidSkyOverview}. \Euclid will also be a resource for other facilities, especially those with narrower fields, such as the {\it James Webb} Space Telescope, and surveys, such as those carried out by the\textit{ Vera Rubin} Observatory and the {\it Nancy Roman} Telescope, which will match the VIS spatial resolution in the infrared.

\Euclid grew from the merger of two proposals from the scientific community, the Dark Universe Explorer (\DUNE) and the Spectroscopic All-sky Cosmic Explorer (\SPACE), submitted to ESA in October 2007  for the second Medium Mission call in their Cosmic Vision Programme. \SPACE was to carry out spectroscopic galaxy clustering measurements, and \DUNE weak lensing observations (with a VIS precursor responsible for the shape measurements). Through the gravitational bending of light by the density inhomogeneity in the Universe, the shapes of distant galaxies are distorted by the matter distribution along the line of sight. Creating a three-dimensional matter map -- that is dominated by dark matter -- requires successively distant foregrounds to be removed in a tomographic process in order to measure the more distant matter distributions; hence the redshifts of the galaxies are also required. With careful calibration, photometrically derived redshifts are sufficient for this purpose, and so \DUNE also included an infrared imaging channel, with the optical bands secured through ground-based imaging. In the merged concept that became \Euclid, this infrared channel was incorporated into the previous SPACE infrared instrument to create NISP \citep{EuclidSkyNISP}, while the visible channel, VIS, was largely unchanged. After phase-A and -B1 studies, \Euclid was selected for flight in mid-2011, and the satellite was launched on 2023 July 1. 

 With the launch of \Euclid, VIS will be the second largest focal plane in space, after that in the ESA \Gaia satellite \citep{Gaia}. Whereas in the case of \Gaia, only small regions associated with each object are transmitted to Earth, full images are available from the \Euclid VIS focal plane. These will therefore be the largest images acquired from space of the external Universe. A  major challenge for the  \Euclid\ team is the required level of control of observational
biases, which requires an exceptional level of knowledge of the VIS instrument.

In terms of the  organisation of the ESA mission, the spacecraft, including the telescope and the mission operations, are the responsibility of ESA, while the instruments and the Science Ground Segment that will process the data are provided by consortia of scientific institutes funded by their national agencies. In the case of VIS, the agencies are those of the United Kingdom, France, Italy, and Switzerland. The activities associated with the two instruments and the Science Ground Segment, as well as a number of Science Working Groups for both cosmology and the wider astronomical fields, are coordinated by the Euclid Consortium,\footnote{\url{https://www.euclid-ec.org}} a grouping of some 2000 scientists and engineers working with ESA to deliver the mission. 

Biennial reports made throughout the development of VIS can be found in \cite{Cropper:18} and references therein. The present paper begins with the rationale behind the design  of VIS (Sect.~\ref{sec:rationale}), and continues with the instrument design (Sect.~\ref{sec:design}). Shorter sections on the instrument assembly (Sect.~\ref{sec:AIT}) and operation (Sect.~\ref{sec:operation}) are followed by the VIS prelaunch and immediate post-launch performance (Sect.~\ref{sec:performance}). Sect.\ref{sec:open_points} reflects on the open points in meeting the weak-lensing goals of the \Euclid\ mission, and where recent advances have identified margins. We end with a final summary of the main challenges faced in developing VIS and comment on some of the advances that may be anticipated.
(Sect.~\ref{sec:summary}).

\vspace*{-4mm}
\section{Rationale behind the conception of VIS }
\label{sec:rationale}

In order to provide a context for the design of VIS, we start by providing a general overview of the considerations for the implementation of an advanced weak-lensing survey. This includes the telescope and instrument, as well as operations and the data analysis. 

The weak-lensing requirements for \Euclid originate in the Phase-0 Study by the French Space Agency CNES for the \DUNE precursor \citep[][summarised in their Table 1]{Refregier:06}. This identified that in order to reach the necessary statistical precision mediated by the number of galaxies available for measurement, a survey covering approximately 20\,000\,deg$^2$ would be required, and that to measure galaxy shapes out to a median redshift of $z\,{\sim}\,1$, typical galaxy sizes would require an image quality of \ang{;;0.23}, spatially Nyquist sampled. This set a requirement of $2\times10^{13}$ on the total survey pixel number. Given a mission duration of 3\,years, considered feasible at the time, an exposure of 1300 seconds imposes a focal plane of $2.7\times10^8$ pixels, or an array of 4\texttimes4 detectors of 4k\texttimes4k pixels. Such quick calculations should be adjusted by the efficiency of the survey and the required exposure times, as well as the survey duration, but this established the feasibility for an instrument of achievable scale in an ESA Medium Mission. 

The next step was to consider the signal-to-noise ratio (S/N) in order to size the telescope and define the passband as well as the quantum efficiency of the detectors. \cite{Refregier:06} adopted a single wide passband from $566\,{\rm nm}$ to 1$\,\micron$ to maximise the S/N for shape measurements, with red-sensitive CCDs, and determined that a telescope aperture of $1.2\,{\rm m}$, would provide a S/N $\geq 10$ for the faintest sources, just adequate for weak-lensing measurements. 
We step back to consider these in more detail.

\vspace*{-4mm}\subsection{Mission and spacecraft}
\label{subsec:mission}

The complex task of setting the scientific requirements for the \Euclid mission, and hence the weak-lensing science performance, is carried out by the Science Working Groups within the Euclid Consortium. This is itself an advanced scientific activity which has progressed in understanding with the development of the mission. 

The weak-lensing science considerations established during the \Euclid study phases were consolidated in The Euclid Imaging Consortium Science Book \citep{Refregier:10} with those at the higher level in \cite{Amendola:18}. Briefly, the logic was (and remains) as follows.
\begin{enumerate}
    \item The primary function of \Euclid is to distinguish between the $\Lambda$CDM cosmology (i.e., with a cosmological constant) and alternative cosmologies. It was considered (\citealt[][Sect. 2.1.2]{Laureijs:11}; \citealt[][Sect. 1.5.1]{Amendola:18}; \citealt{Taylor:07}) that there should be $\leq 0.01$ chance that, from the \Euclid weak-lensing data, $\Lambda$CDM might be incorrectly established as an acceptable cosmology, corresponding to a Bayes factor of 1:100 that a dark energy equation of state $w=-1$ at all redshifts, compared to evidence of any deviation. This requires measurements that permit a figure of merit (FoM) greater than or equal to 375. The FoM, as forecast using a Fisher matrix analysis, is proportional to the inverse of the area of the error ellipse in the $(w_0,w_a)$ parameter space of the first-order Taylor expansion $w(a)=w_{0}+w_{a}(1-a)$, where $a = 1/(1+z)$ is a scale factor of the Universe. To achieve this, the error ellipse constrained by \Euclid observations should confine the uncertainty of $w_0$ to be less than 0.01 and  that of $w_a$ to be less than 0.1. These constraints assume the combined analysis of the weak-lensing and galaxy clustering data -- on their own each probe results in larger uncertainties. Requirements for weak lensing were therefore derived from cosmological forecasts assuming a prior corresponding to the expected galaxy clustering constraints (\citealt{Laureijs:11}).
    \item \cite{Amara:08} considered the number of galaxies required to shrink the error ellipse to within a range of permitted maximum width. This is effectively driven by the shot noise. They plotted this as a function of survey area (their fig. 11) assuming 35 suitable galaxies/arcmin$^2$. A width of 1\% (corresponding to ${\rm FoM}=375$) was reached with a survey of 20\,000\,deg$^2$. This sets the scale of the survey and the S/N to be reached to permit the shape of the galaxies to be measurable. This creates the first three \Euclid weak-lensing top-level science requirements that are captured in the Science Requirements Document \cite{SciRD}. 
    \item Accounting for only the shot noise, this number of galaxies provides the maximum precision achievable. However, any incorrect calibrations of the data will leave residual biases which will displace the location of the error ellipse, so an upper value limiting this displacement sets the requirement for additive biases $\sigma_{\rm{SYS}}<10^{-7}$ and multiplicative biases $m_0<10^{-3}$ \citep[][Eqs. 21 and 22]{Amara:08}. This creates the fourth top-level science requirement.
    \item \Euclid weak lensing requires the distance to the galaxies to be known, and this is achieved through photometric redshifts with the contribution of external data and infrared photometry provided by NISP. The final three top-level requirements control the performance of external data biases, for example the photometric redshifts 
\end{enumerate}

\begin{table*}[htbp!]
\begin{tabularx}{2\columnwidth}{lX}
\includegraphics[width=1.5\columnwidth]{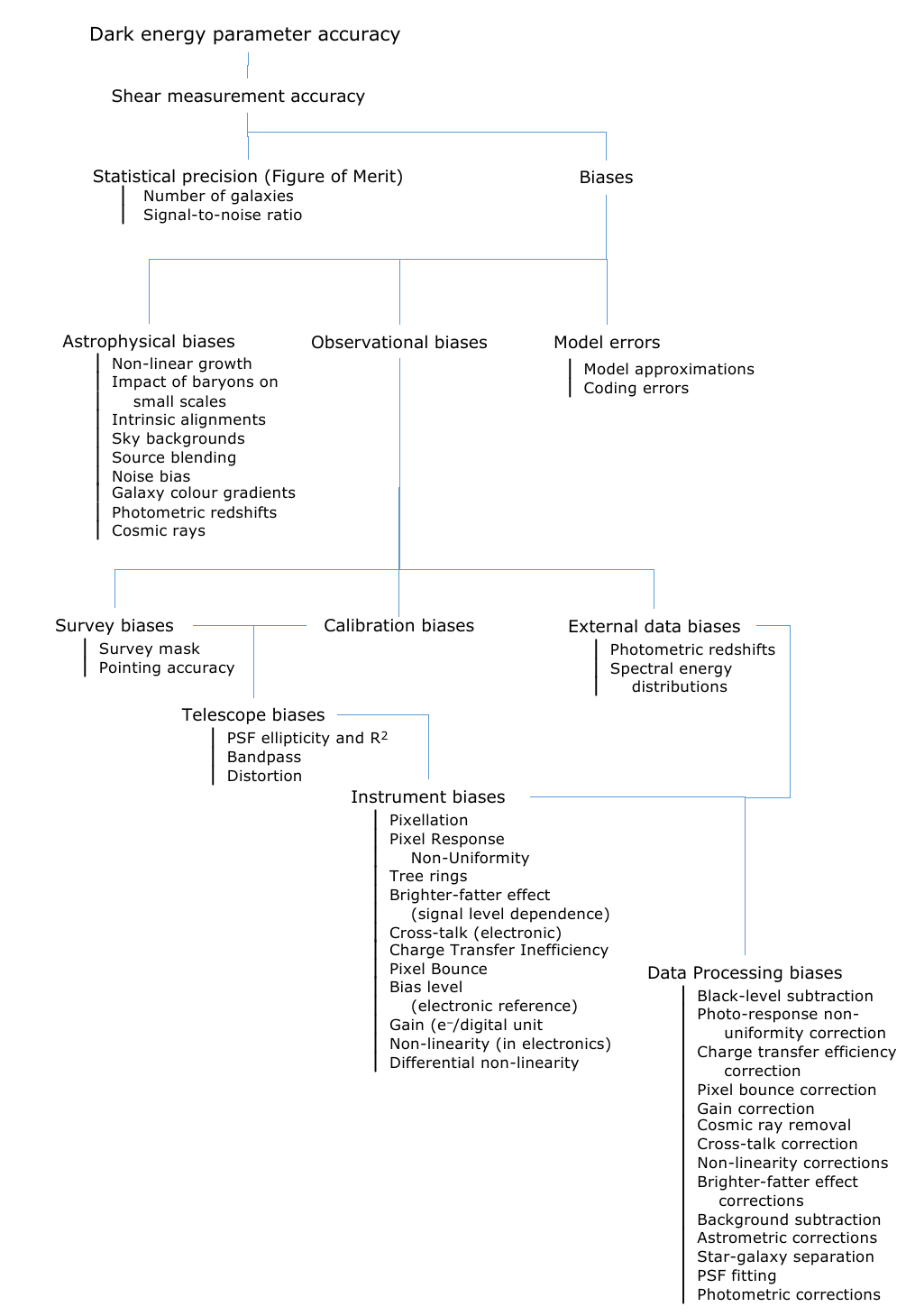}
    & \vspace*{-18cm} \captionof{figure}{An (incomplete) overview of the factors relevant to reaching sufficient constraining power in a weak-lensing survey and their relationship to the top-level weak-lensing requirements in the Science Requirements Document \citep{SciRD}. Once sufficient statistical precision is reached, attention has to be paid to astrophysical biases arising from the nature and disposition of the contents of the Universe, those arising from the observational strategy and the observations themselves, and those from the cosmological modelling. The final accuracy of the experiment is set by the residuals after treatment of all of these effects.} 
\label{fig:flowdown}
\end{tabularx}
\vspace*{-5mm}
\end{table*}

A schematic overview of what is required is shown in Fig.~\ref{fig:flowdown}. Achieving the requisite survey area and S/N requires significant analyses and trade-offs, but what really sets \Euclid apart is the requirement in item~4 above in respect of the control of biases. Ultimately, there will be residuals after calibration in all of these contributions, and it is these residuals which will be judged against the requirements. 

The overall mission summary is shown in Table~\ref{tab:mission_summary} as presented in the \Euclid Definition Study Report \citep[the `Red Book',][]{Laureijs:11}. With the successful selection of \Euclid in mid-2011, the science requirements were more strictly developed -- for weak lensing see \cite{Cropper:13}, formalised in \cite{SciRD}.  The requirements flowdown for the mission to achieve the science was expanded significantly. Starting from the Science Requirements Document, the requirements flowdown continued with a Mission Requirements Document \citep{MRD:13}; these in turn lead to hierarchies of requirements documents for the overall system, the instruments, the data processing, mission operations, and science calibrations. Allocations were made for many of the biases in Fig.~\ref{fig:flowdown}, with the \cite{Massey:13} formulation for translating image quality to weak-lensing bias. The requirements for the mission, satellite, telescope, and instruments were carefully separated and allocations were apportioned and assigned throughout the \Euclid system in a substantial flowdown, eventually reaching, for example, individual subsystems in the telescope and instruments, processing functions in the science ground segment, and repetitive observational sequences and calibrations in the survey. These and other effects were combined in an inverted-tree structure, the top level of which was required to meet the allocations in point 3 above. The document directly setting the VIS requirements is the Payload Elements Requirements Document \citep{PERD}, but other documents also influence the VIS design, testing, and operation. For example, calibration requirements are in the Calibration Concept Document Part-B \citep{CalCD-B} and in respect of operation during the survey they are in the Mission Operations Concept Document \citep{MOCD-A}. In addition to performance requirements, a number of technical documents apply, including the Experiment Interface Document \citep{EID-A} and the suite of ESA European Cooperation for Space Standards (ECSS)\footnote{\url{https://ecss.nl}} which also specify formal procedures, such as for testing. From these, lower level requirements were derived. Within the VIS instrument itself, the number of requirements in the flowdown eventually exceeded 4000 \citep{PERD,Requirements}.

\begin{table*}[htbp!]
\caption{Summary of the characteristics of the \Euclid mission as set out in the \Euclid `Red Book'\tablefootmark{a}.}
\begin{center}
\begin{tabular}{c}
\includegraphics[width=1.5\columnwidth] {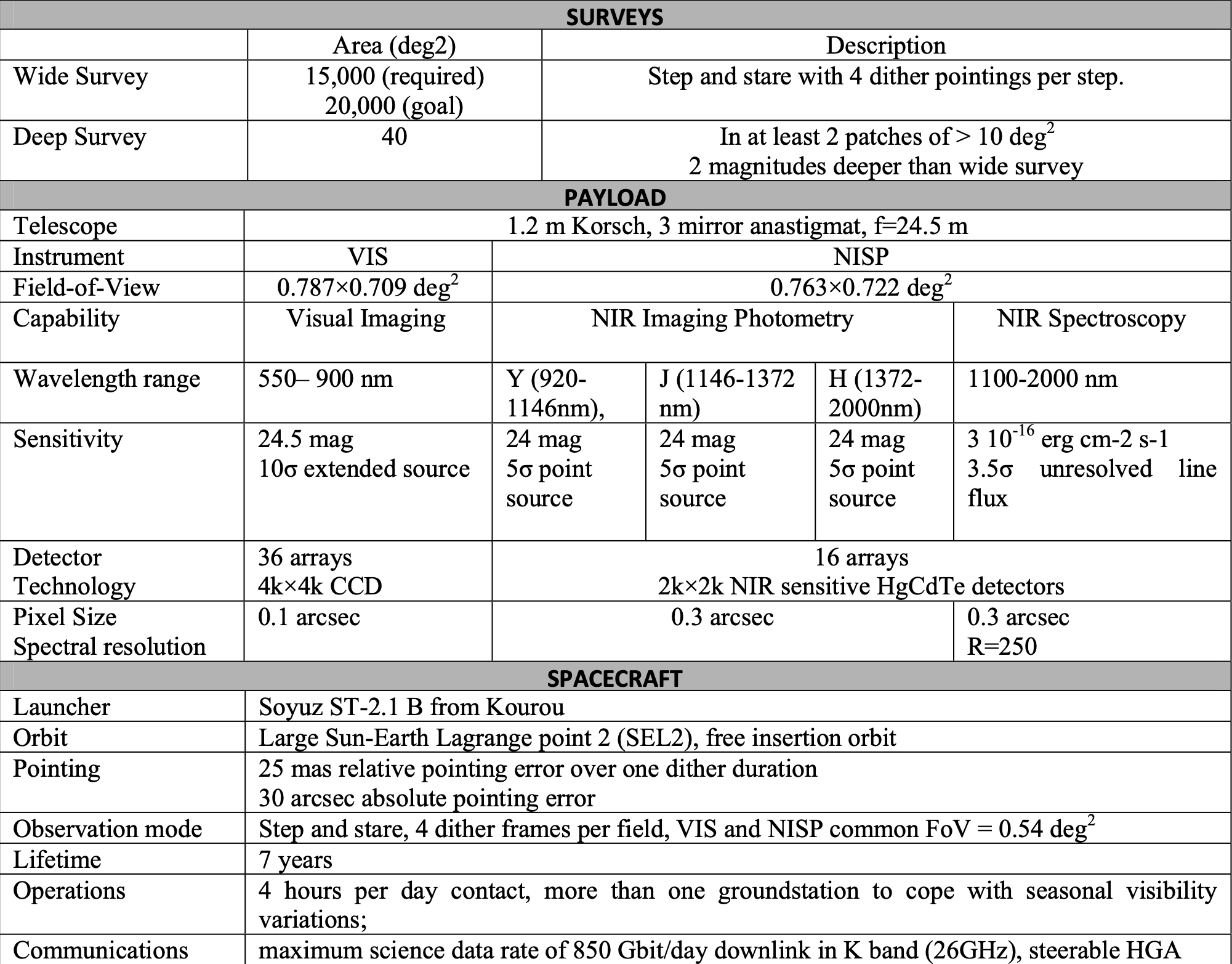}
\end{tabular}
\tablefoot{\tablefoottext{a}{\citep{Laureijs:11}. At the time of publication, a Soyuz launch was anticipated.}}
\end{center}
\label{tab:mission_summary}
\end{table*}

It should be noted that because the \Euclid mission does not produce cosmological parameters, only the means by which they can be computed  by the wider community, there are no higher-level requirements above those in the Science Requirements Document. It has however been demonstrated through a higher level of analysis \citep[for example][]{SPV2} that the Primary Science Objectives in Table 1 of the Science Requirements Document will be met, which includes the necessity to reach a ${\rm FoM} \geq 375$ as discussed in Sect.~\ref{subsec:mission}.

In the intervening period a greater understanding has been achieved of how the different biases interact. It was recognised that the hierarchical combination of requirements is too restrictive, and a different formalism was developed to address this \citep{EuclidCollaboration:19c} and used for System Performance Verification at the \Euclid Mission Critical Design Review \citep{SPV2}. More recent explorations of how these requirements may be formulated and combined are in, for example, \cite{Kitching:19} which examined the spatial effect of biases across the survey. Margins have been found, while on the other hand, some of the more recently identified effects in Fig.~\ref{fig:flowdown} remain without allocations. \Euclid project reviews still compare outcomes to the formal flowdown, as it has proved too complex and in most cases, for example in commercial contracts, unfeasible to reformulate iteratively the morphology of the entire structure. Fortunately, at the top level, the outcomes remain consistent or conservative.

\subsubsection{Statistical precision}
\label{subsubsec:precision}

Measuring the distortions (or `shear') caused by weak gravitational lensing of galaxies is challenging for a number of reasons. The projected shapes of galaxies are non-circular in general. Hence many galaxies are required in order to average out this intrinsic `shape noise' and to establish an underlying distortion in any particular direction on the sky. The distortions are small, approximately 1\%, and typical galaxy diameters in the redshift range $z\sim 0.1$--1.5 (where the accelerated expansion of the Universe from dark energy is most prevalent) are sub-arcsecond, so that their images will be coarsely sampled. This combination of effects can be overcome only by measuring a very large number of galaxies. Assuming the measurement process itself is perfect, the ultimate precision of any derived parameters will be set by the scale of the survey and the S/N that can be achieved. The imperative of this statistical precision places it at the top level of Fig.~\ref{fig:flowdown}.

\subsubsection{Systematic effects}
\label{subsubsec:systematics}

Having achieved sufficient statistical precision, the main challenge for the \Euclid weak lensing is to maintain the requisite control of the systematic effects to meet the scientific requirements. There are a number of sources of bias in the measurements (Fig.~\ref{fig:flowdown}), arising in the telescope and instrument (for example nonlinearity of response); in the interaction of the instrument with the Universe (for example selection effects arising from the instrument passband); and in the Universe itself (for example the intrinsic alignments of galaxies around mass concentrations). These biases require calibration and/or modelling for correction, and, for the levels of accuracy demanded in \Euclid, the corrections can all be considered at least marginally incorrect. The ultimate accuracy of the experiment will therefore not only depend on the initial precision from the survey, but also the level of success in making these systematic corrections. It is the residual biases after the corrections which will set the accuracy of the cosmological parameters (Fig.~\ref{fig:flowdown}).

It was understood from the outset that it would be essential to achieve a high level of calibration knowledge in all respects. A key principle in achieving this would be simplicity of operation. Wherever possible these calibrations should be derived at the same time as and from  within the science image itself, for example by monitoring the telescope performance using stellar images. This should be backed up by a demanding level of opto-thermo-mechanical and electrical stability to be achieved through careful hardware and survey design. In order to maintain both opto-mechanical and electrical stability both in VIS and in the external optical path, the permitted range of variation in temperature must be limited, placing strong constraints on the changes in thermal dissipation within the instrument, and on the spacecraft and survey. With respect to the operation of the instrument, if dissipation cannot be entirely constant, then at least the operations should consist of the same repeated sequence throughout the survey, resulting in a predictable thermal state of the instrument.

\subsection{The current \Euclid configuration}
\label{subsection:configuration}

\Euclid's fundamental parameters derived in the \DUNE Phase-0 Study \citep{Refregier:06} have evolved. Instead of 1300\,s, the total exposure is approximately twice that, resulting in a higher S/N. The survey now takes 6\,years instead of the envisaged 3 to observe 15\,000\,deg$^2$ with a reduced survey efficiency caused in part by calibration and pointing constraints. In the optical, images from a telescope with a 1.2-m diameter primary mirror providing a 0.787\texttimes0.709\,deg$^2$ field of view are focused on a much larger array of 6\texttimes6 detectors with 4k\texttimes4k pixels, each pixel subtending \ang{;;0.1}. 
 With the telescope and detector sensitivity and a passband from 550--$920\,{\rm nm}$, $m_{\rm AB}\geq24.5$ at S/N $\geq 10$ was required for the fields in the main survey. The total mission lifetime is to be 6.5 years. Other characteristics of the survey and orbit are in Table~\ref{tab:mission_summary}.

The characterisation of the morphology of the optical point spread function (PSF) provided by the telescope is one of the most important calibrations, because the measured shape of a galaxy will be significantly rounder if the PSF is wider. The required knowledge of the shape of the PSF to correct this effect is extreme. It drives the temporal and spatial stability of the optical system across the field of view, and the shaping of the passband. In order to minimise the complications arising from colour-dependence and the tendency for ghost images in transmissive optics, the optical system should be fully reflective. Diffraction effects, such as those introduced by secondary mirror supports, should be minimised. In early configurations for \Euclid the focal planes of the two instruments would sample adjacent fields, with the overlap between the two instruments achieved by stepping though the survey in such a way that one instrument would subsequently record the field from the other. However, besides constraining the survey, this required the field of view provided by the telescope to be impractically large to accommodate the already large fields of view of both instruments. The change was therefore made to separate the optical feed by wavelength, using a dichroic. As this is constructed from a large number of interfering layers, different wavelengths are reflected from different layers, each with their own optical properties. In addition there is the propensity to create optical ghosts from the rays passing through the multilayer dielectric stack and reflecting from the back of the optical element. This element, and some of the others in the optical system that use multilayer dielectric stacks to shape the VIS and NISP passbands, cannot therefore be considered simple reflectors.

\subsection{VIS instrument concept}
\label{subsec:instrument}

To minimise biases, simplicity of design and operation would be vital for VIS. We elaborate here the considerations that shaped the VIS instrument concept. For example, no filter wheel could be contemplated because the variability induced by changes in image location and shape resulting from a moving mechanism could not be sufficiently characterised for the level of accuracy required for weak lensing.\footnote{The NISP instrument incorporates a filter wheel, but NISP is optimised for galaxy clustering and does not require the same level of PSF knowledge as that required for weak lensing.} 

\subsubsection{Detectors and focal plane}
\label{subsubsec:detectors}

The accurate knowledge required of the effects within the detectors and associated detection chain was a driver on the choice of detector technology. With their high sensitivity from infrared to red optical wavelengths, and their relative immunity to radiation damage by ions from cosmic rays and Solar protons, HgCdTe detectors were one option for \Euclid VIS. However, the high level of knowledge of, and stability of, charge-coupled devices (CCDs) were ultimately more critical characteristics, despite their susceptibility to radiation damage by ions, even for a mission duration of 6.5 years. A further consideration was the ambition and cost of populating a focal plane of the size of VIS by infrared detectors within the envelope of an ESA M-class mission. 

Photons arriving at a CCD can generate electron-hole pairs in a depleted region in the pixel \citep{Janesick:01}. The electrons can be collected at the end of an exposure by passing them along the rows and columns of the CCD to a readout node. The design of the CCD would be critical to the performance of VIS. It should be of the highest possible sensitivity (Detective Quantum Efficiency) especially at red wavelengths to exploit the smoother shapes of red galaxies for shape measurements; hence it should be back-illuminated so that the pixel electrode structure is not in the light path, and also sufficiently thick that it is not transparent at longer wavelengths. On the other hand, diffusion effects within the pixels modify the measured PSF as a function of wavelength and of measured flux, and as the detector thickness is increased these effects become problematic.

Detector pixel sizes should be commensurate with adequate sampling of the PSF, the size of which is set by the optical system and the satellite pointing performance provided by the attitude, orientation, and control system (AOCS). Pixels should be large enough to store sufficient charge to maximise the dynamic range of signals the instrument can record without saturating, but not too large to drive unnecessarily the physical size of the focal plane and therefore the mass of the instrument. In line with the drive for calibratability, the internal structure of pixels should be as simple as possible. 

While, if care is taken with the operation of the CCD, the integrity of the charge packet as it is transferred to the readout node is remarkably preserved, intrinsic or radiation-induced damage sites in the Si lattice can temporarily trap electrons, releasing them into the charge packet of subsequent pixels and hence changing the recorded shapes of images. It is important therefore that the overall number of transfers be limited by the provision of sufficient readout nodes on the CCD. This must be balanced by the consideration that each readout node requires a dedicated set of electronics to digitise the signals, with the associated system resources of power and spatial accommodation.

As the radiation damage increases during the mission, the charge trapping during transfer becomes a driving factor, as it directly affects the shape measurement by eroding the leading edge of a galaxy image and adding a trail of released electrons as it is transferred through the CCD. Any design decisions to the CCD itself to mitigate the effect should be considered. An extensive campaign to understand the scientific effects would be necessary, and the means to quantify and calibrate these evaluated and incorporated in operations and in the data processing. Given the uncertainties in this calibration, a conservative approach must be adopted in determining the end-of-survey radiation damage.

CCD pixels have slightly different sensitivities as a result of their fabrication process. This photo-response non-uniformity (PRNU) can be calibrated by flooding the CCD by a uniform or at least smoothly varying illumination. A calibrating unit to produce these flat illuminations at several wavelengths is required. These flats are also useful for calibrating other effects within the detector, such as the degree of independence of one pixel from another. 

There will be gaps between individual detectors in a large focal plane. In order to maximise the spatial uniformity of exposure depth in any survey, these gaps should be as small as possible. However, this constrains the mechanical packaging of the detectors, and their means of connection to their associated electronics. The stringent requirements on the thermal and mechanical stability drive the design of the detector support structure, and the positioning of the detectors relative to each other, especially in the focus direction.  

\subsubsection{Shutter}
\label{subsubsec:shutter}

The \DUNE Phase-0 Study baselined CCDs derived from ESA's \Gaia mission operating in Time Delay Integration (TDI) mode. In this mode, the satellite is made to scan the sky at the same rate as that which the CCDs are being read out row-by-row, resulting in an ever-extending ribbon of image with the width of the CCD. This technique allows stable operation of the detector and its associated electronics, unchanging for long periods until interrupted for other reasons. It also eliminates the need for a shutter: images falling initially at the top of the CCD accumulate as they are shifted row-by-row in synchronisation with the satellite scanning down to the readout register, where they are read out and digitised. There are disadvantages too, in that the data rate will be high unless the scanning rate is slow, which then imposes very stringent satellite pointing requirements. Lower exposure levels are disproportionately affected by radiation damage because they have yet to accumulate much charge while being subject to radiation damage impacts as all other areas, so that weaker signals at the top of the CCD are more strongly affected. This is not the case for standard expose-and-repoint operation where all pixels are transferred only at the end of the exposure with consequently higher signal levels. Ultimately, however, as it is not yet possible to operate HgCdTe detectors in TDI mode, and no adequate design could be found for a de-scan mechanism for the infrared focal planes in \DUNE or \SPACE\!\!, for \Euclid this major trade-off between TDI and standard expose-and-repoint operation was decided in favour of the latter.

However, in expose-and-repoint operation, the image collected during the exposure should be shielded from further accumulation during its transfer to the readout node in order to prevent trailed images of the scene being recorded. Although some CCDs can be configured to shift rapidly the recorded images under a light shield (frame-transfer CCDs), these shields reduce (typically by half) the light-sensitive region of the CCD, creating areas of dead space on the focal plane. While these can be filled using an appropriate survey pattern, for the same active area the optical field of view and the focal plane dimensions are significantly larger -- which may be unfeasible when these are already challenging. Hence, and without the option of TDI operation, a shutter would be necessary in the instrument. Given the criticality of the knowledge of the PSF shape noted above, operating the shutter must perturb the satellite pointing only minutely, imposing a tight requirement for internal momentum compensation, both linear and angular. In addition, as a single point of failure, the highest level of reliability would be needed and this essential mechanism should be the only one permitted in the whole instrument. 

\subsubsection{Detector electronics}
\label{subsubsec:detectorchain}

The transfer of charge through the CCD when it is being read out is achieved by toggling (`clocking') the voltages of electrodes associated with each pixel sequentially so that the charge is moved to the adjoining pixel. All pixels in a row are moved to the next row simultaneously. The charge is prevented from moving laterally by electric fields created by doping the Si as part of the manufacturing process: these barriers define columns, down which the charge is transferred. When a row reaches the end of the CCD photo-sensitive area, it is transferred to a readout register. To move the charge from each pixel to the output node, the electrodes in this register are clocked separately, and faster than the pixel electrodes, as the process must complete before the next row is transferred. External electronics are used to generate the requisite waveforms to activate the electrodes. The timing and shapes of these clocking waveforms play a critical role in the performance of the detector and require careful optimisation. 

The output node of the CCD provides a packet of charge which is directly, though not necessarily linearly, related to the number of photons incident on that pixel. A small amount of noise, the readout noise (RON), is inherent in this process. This charge must then be measured and digitised by electronics external to the detector at the time it becomes available on the node, as determined by the clocking. Careful design is necessary to minimise the addition of further noise, and the maintenance of the stability of the supplied and internally created reference voltages is critical. The availability of electronics components with suitable performance and radiation tolerance to both ions and electrons is limited; in particular, the digitising element, the analogue-to-digital converter (ADC), should have sufficiently fine digital resolution compared to the noise in the system. The isolation of one channel from another is also important. This drives the layout of the circuitry and requires pixels to be transferred, measured and digitised synchronously. 

These external electronics are required to operate in different modes, for example to allow rapid flushing of pixels, so it must be possible to configure them and also to adjust the most critical operating parameters (voltages, timings) if necessary. Detailed knowledge of their internal state is also required.

\subsubsection{Data handling and instrument control}
\label{subsubsec:datahandling}

Once a digitised number is created for a pixel, the information must then be prepared for transfer elsewhere in the system. Pixel data will be arriving synchronously from each CCD readout node through its electronics and these must be arranged in the correct sequence to maintain their relationship so as to rebuild the image recorded on the focal plane. This is a challenging real-time operation given that there are 144 readout nodes and a large number of pixels. To minimise the telemetry bandwidth required for the images to be transmitted to Earth, the data must then be losslessly compressed\footnote{Lossy compression, while generally more efficient, modifies the image in ways that cannot be known or recovered at the levels of accuracy required by a weak-lensing survey.} and prepared in the appropriate form to be transferred to the satellite mass memory; from there it can be retrieved for transmission during ground contact. 

The survey observations require a sequence of instrument preparations, shutter operations, exposures, and calibrations and science data transfers. VIS must therefore be able to receive and interpret commands from the spacecraft. At the same time, knowledge of the internal instrument status and parameters (temperatures, voltages and currents) will be required to ensure its correct routine operation and to deal with non-standard events. This information must also be passed to the spacecraft for transmission to the ground.

\subsection{Operations}
\label{subsec:ops}

The operation of the satellite should be planned to maximise the stability of the telescope and instruments so that calibration errors, and hence biases, are minimised. Simple repeatable observation sequences minimise the parameter space over which the calibrations must be made, and enhance the understanding of the trajectories of the calibration parameters on short and long timescales as the survey progresses. This is more important than is often realised, as there are many complicated effects, such as the different thermal relaxation timescales for different components in the optical path, hysteresis and the changing background that are impossible to predict with sufficient accuracy.

Consequently, the impact of interruptions to the regular survey pattern should be carefully evaluated and their frequency minimised. Some, such as orbit maintenance, are impossible to avoid; others, such as to make specific calibrations, may inadvertently introduce perturbations within the satellite which are complex to analyse and ultimately counterproductive to the quality of the calibrations overall. 

Equipment safe modes and failures also cause interruptions. The on-board mitigation measures should minimise the impact of these, particularly if the trigger to a safe state is a minor one. For example, recovering the normal levels of thermal equilibrium after switching off a full instrument, rather than only the element that has indicated an unsafe condition, can potentially take days, forcing changes to the survey or creating holes in the spatial coverage which impact significantly on the science. 

Finally, a sufficient downlink data rate is particularly important for survey instruments with large focal planes. The operations should take into account the impact of repointing of the high gain antenna on the satellite pointing stability, as well as that from the dissipation by the transmitter on the thermal stability during downlink periods. The advent of K-band telemetry at the time of the \DUNE/\SPACE merger was a critical development for the feasibility of the \Euclid concept. 

\subsection{Data processing}
\label{subsec:data-processing}

Given the imperative to minimise biases, the data processing on the ground is a critical element in a weak-lensing survey. This is where the biases are identified and corrected through generating and applying calibrations (Fig.~\ref{fig:flowdown}). The scale and breadth of the task is considerable. This component of the mission must be configurable to respond to the evolution of the spacecraft behaviour, and also, importantly, in the later data releases to encapsulate the understanding that has grown from working with the data. This is where the ultimate performance of the mission is reached.

\section{VIS design}
\label{sec:design}

\begin{figure*}[htbp!]
\center{\includegraphics[width=1.58\columnwidth] {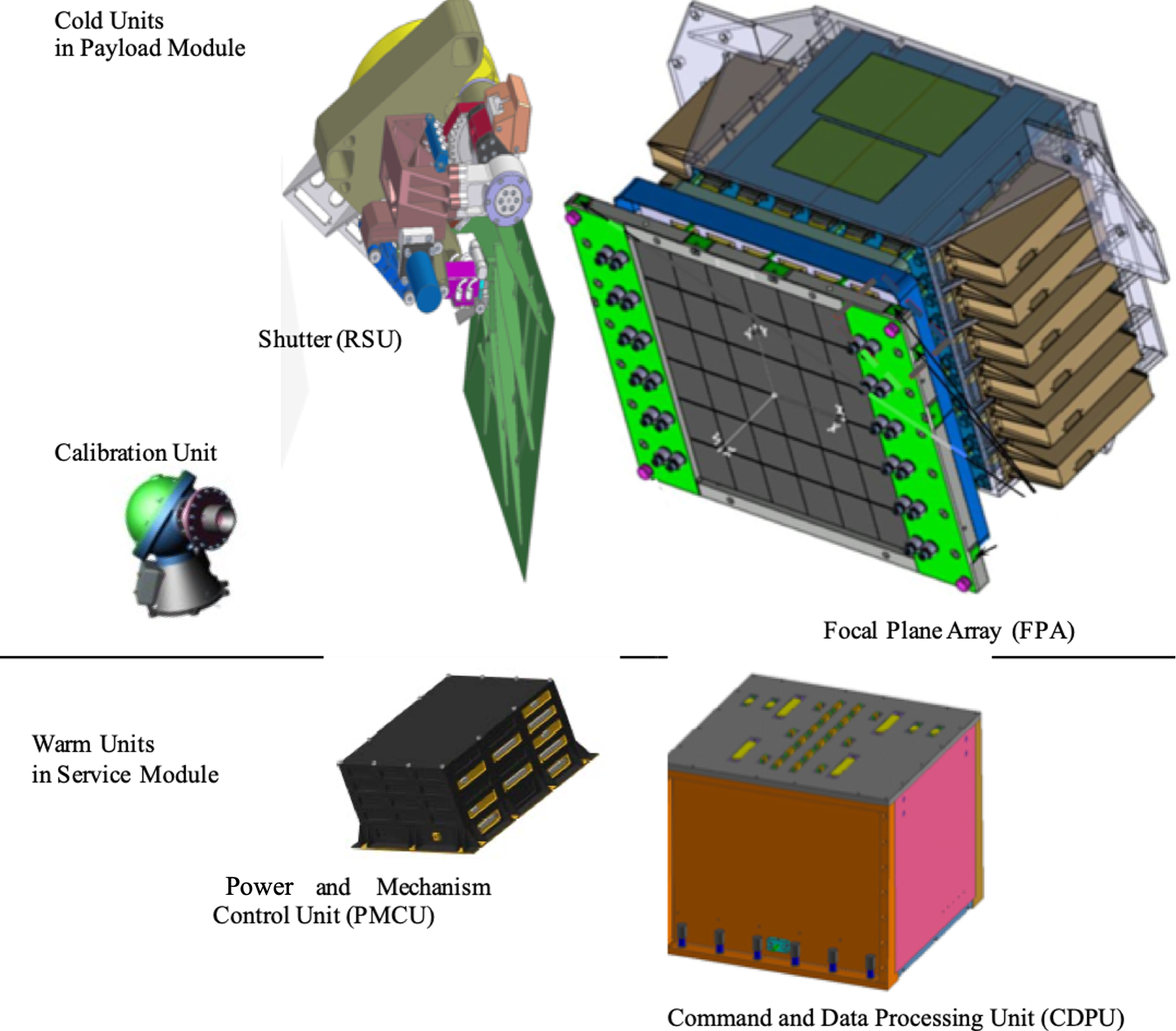}}
\caption{Computer-aided design view of the VIS instrument. VIS consists of five units, three in the Payload Module and two in the Service Module. The Focal Plane Array dimensions are approximately $600\times500\times400$\,mm; the Shutter $475\,{\rm mm}\times375\,{\rm mm}\times250$\,mm; the Calibration Unit $170\,{\rm mm}\times160\,{\rm mm}\times130$\,mm; the Control and Data Processing Unit $285\,{\rm mm}\times285\,{\rm mm}\times230$\,mm; and the Power and Mechanism Control Unit $335\,{\rm mm}\times285\,{\rm mm}\times130$\,mm. The harness connecting these units together, and to the Service Module, are not shown, nor is the radiator to space. This is attached to the rear of the Focal Plane Array to radiate the power dissipated by the Readout Electronics in the Focal Plane Array. The harness and radiator can be seen in Fig.~\ref{fig:PLM-CAD}.}
\label{fig:VIS_components}

\center{\includegraphics[width=1.4\columnwidth] {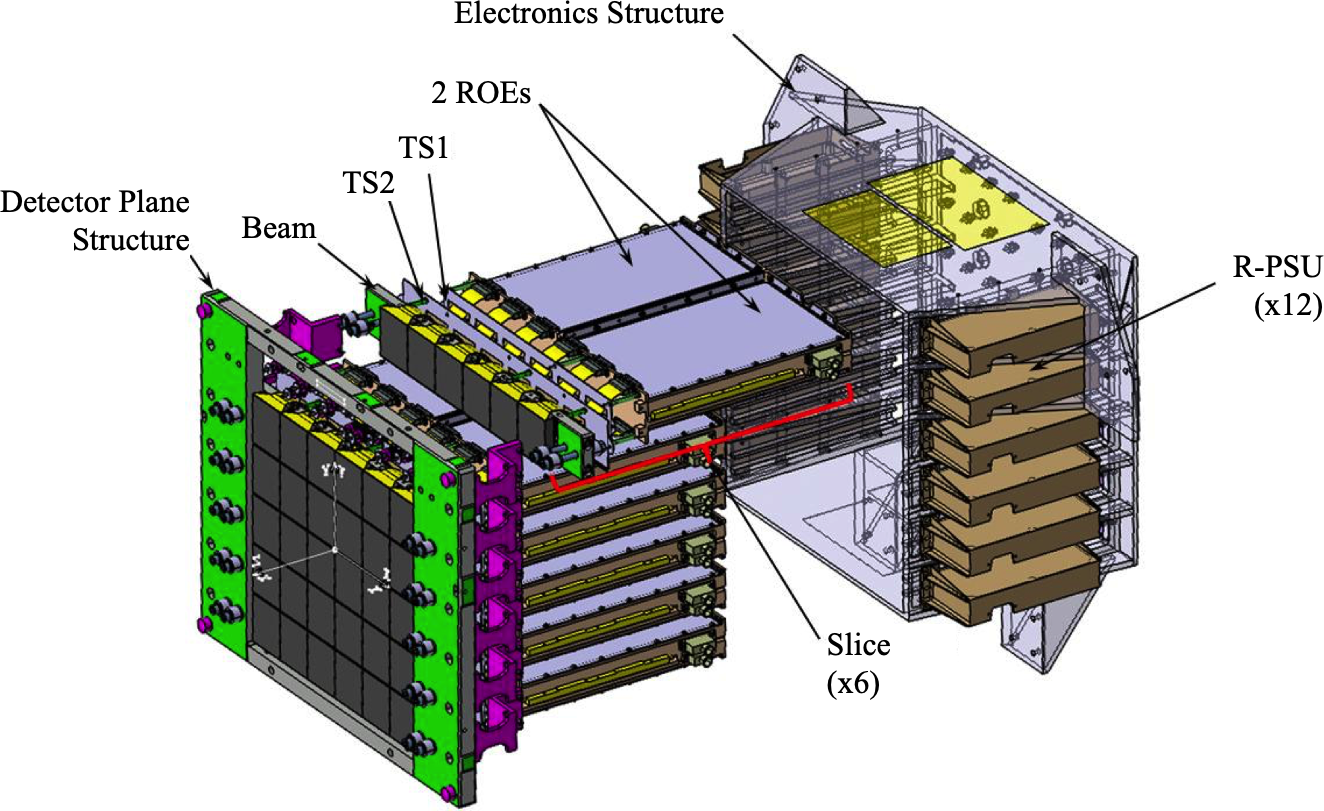}}
\caption{Computer-aided design expanded view of the Focal Plane Array showing the arrangement of the 12 detector blocks, each consisting of three CCDs, a Readout Electronics (ROE) and its Power Supply (R-PSU). Pairs of these are arranged in a slice, and so there are six slices. Also visible in this diagram are the two thermal shrouds (TS1, 2) and the Detector Plane Structure and Electronics Structure.}
\label{fig:FPA_expanded}
\end{figure*}

The VIS instrument design evolved within the context of the considerations enumerated in Sect.~\ref{sec:rationale}. By \Euclid mission selection in 2011 it had attained its current format and layout \citep[see][]{Laureijs:11}. This was largely as a result of the merger in mid-2010 of the previous \DUNE and \SPACE infrared focal planes, leaving VIS a stand-alone optical imager. Also, by this time, TDI operation had been discarded so that a shutter was incorporated, a square array of 6\texttimes6 CCDs had replaced earlier rectangular layouts for the Focal Plane Array, a flat-field calibration unit was included, and the common digital unit in previous designs was replaced by two digital units specific to VIS -- the Control and Data Processing Unit (CDPU) and the Power and Mechanisms Control Unit (PMCU). These components are shown diagrammatically in Fig.~\ref{fig:VIS_components}. The detectors and associated electronics -- the detector chains -- are packaged within the Focal Plane Array as shown in Fig.~\ref{fig:FPA_expanded}. 

\begin{figure*}
\center{\includegraphics[width=1.7\columnwidth] {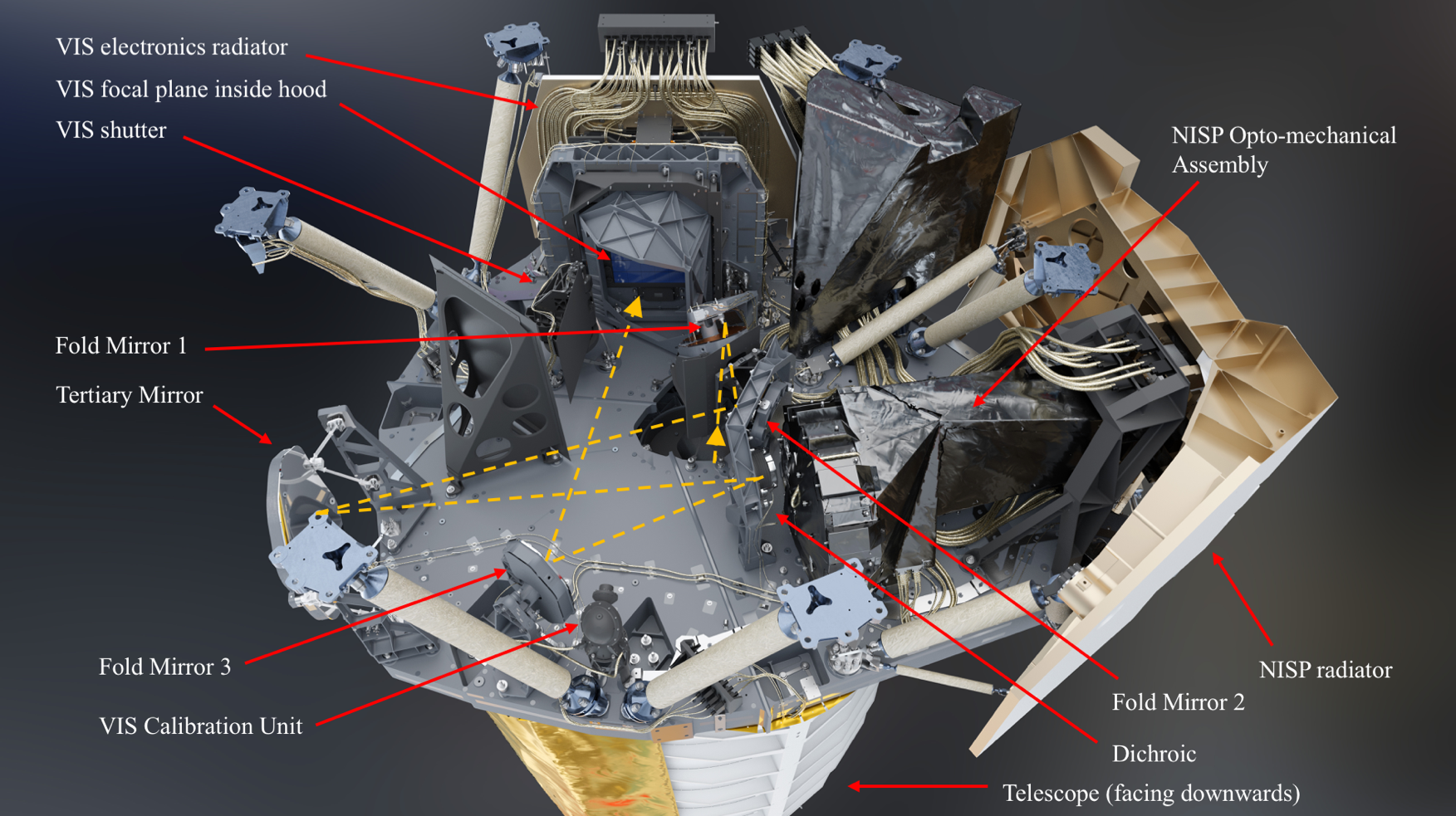}}
\caption{Computer-aided design view of the Payload Module. The telescope is at the bottom of the image, looking downwards, and the telescope beam (in orange) enters upwards through a hole in the centre of the Payload Module Baseplate, which supports both telescope and instruments, to the Korsch tertiary. From there, the beam is either transmitted to NISP or reflected to VIS. The placement of the three VIS units is shown, with the Focal Plane Array shrouded within a hood, which limits scattered light and reduces radiation damage to the CCDs. Image courtesy of Airbus.}
\label{fig:PLM-CAD}

\center{\includegraphics[width=1.7\columnwidth] {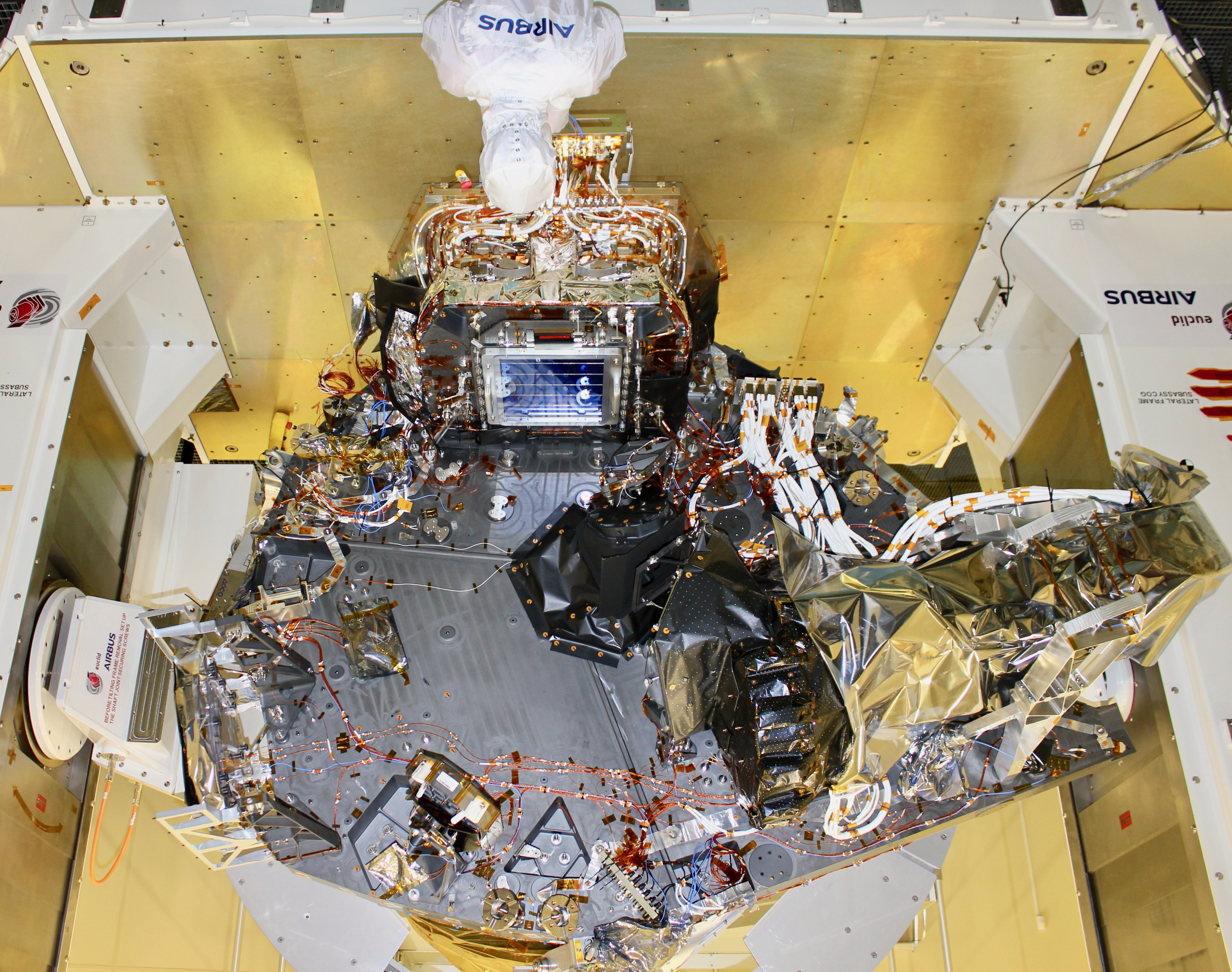}}
\caption{Disposition of the VIS Focal Plane Array on the Payload Module baseplate. In this image, which is in the same orientation as Fig.~\ref{fig:PLM-CAD}, the detector array is visible under a protective transparent cover, as the hood is not yet in place. The Shutter and the Calibration Unit are not yet integrated. Image courtesy of Airbus.}
\label{fig:PLM}
\end{figure*}

\begin{figure}
\center{\includegraphics[width=\columnwidth] {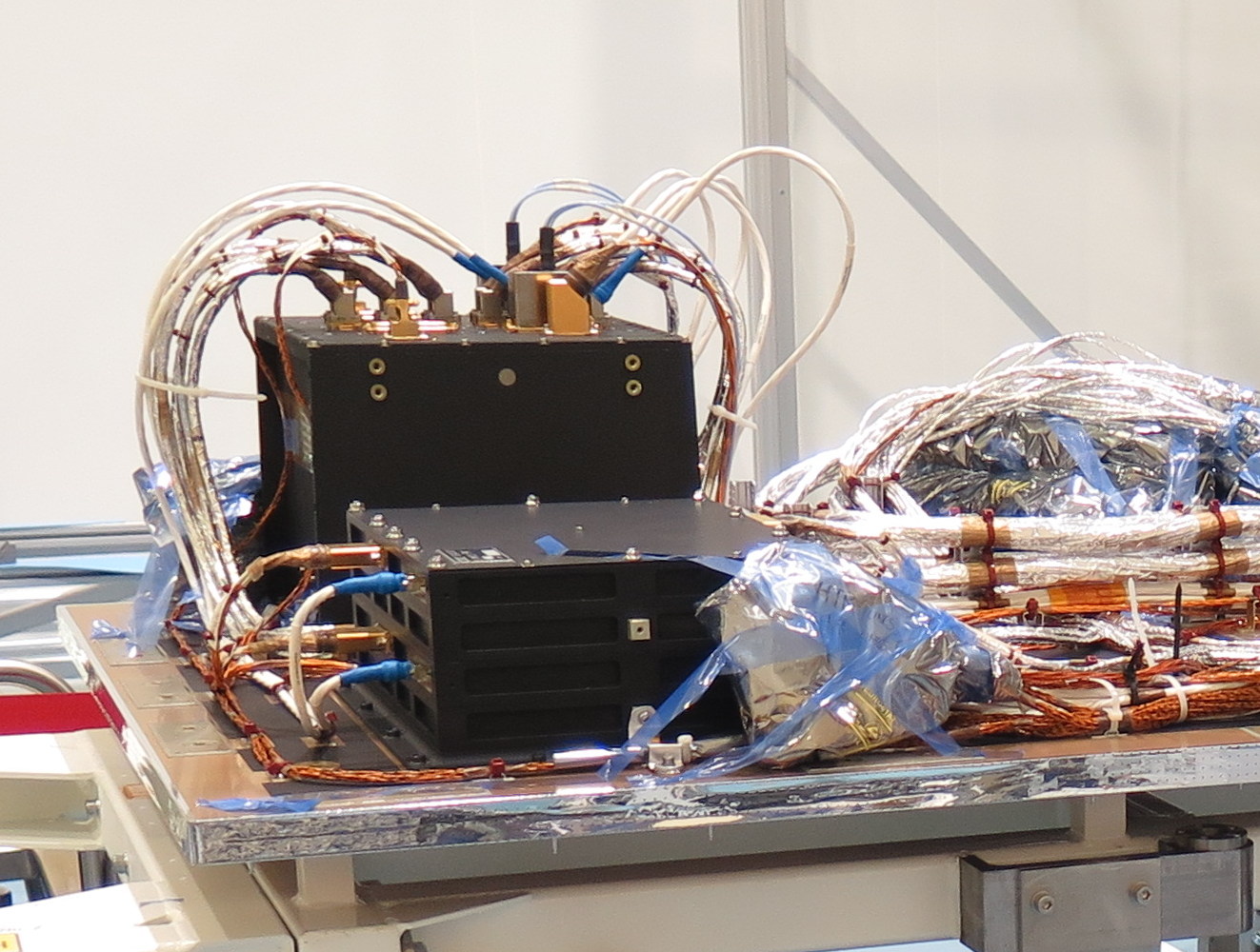}}
\caption{Disposition of the VIS warm units on a Service Module side panel. The Power and Mechanism Control Unit is the dark unit in the foreground, and the Control and Data Processing Unit is behind it. Stacked harnesses for attachment to the Focal Plane Array are visible to their right.}
\label{fig:SVM}
\end{figure}

Three of the five VIS units, the focal plane, the shutter, and the calibration unit, are located within the \Euclid Payload Module \citep{Racca:16}. The 1.2-m Korsch telescope primary mirror with its truss supporting the secondary is located on one side of a Silicon Carbide (SiC) baseplate. This has a central aperture through which the optical beam passes via two flat fold mirrors to the Korsch tertiary mirror on the other side of the baseplate. This corrects the intermediate image produced by the first two telescope mirrors, providing an f/20 beam for the instruments. A pupil image is formed part-way along this converging beam, and at this point a dichroic separates the beam by wavelength, with wavelengths $\lambda<0.95\,\micron$ reflected via a third flat fold mirror to the VIS focal plane. The transmitted wavelengths $\lambda>0.95\,\micron$ pass through the NISP optical system to the NISP focal plane. The layout of the optical elements in the Payload Module is shown in Fig.~\ref{fig:PLM-CAD}, with an indication of the optical path to VIS. The disposition of the VIS units during integration in the Flight Payload Module is shown in Fig.~\ref{fig:PLM}. The other two VIS units, the Control and Data Processing Unit and the Power and Mechanisms Control Unit, in the \Euclid Service Module, are located on one of its side panels, alongside similar units from NISP. They are shown before integration onto the Service Module itself in Fig.~\ref{fig:SVM}.

While the optics are not part of the VIS instrument, it is instructive to understand briefly their role and impact on the shaping of the VIS passband. Details of the optical design are in \cite{Venancio:14}. The Korsch telescope provides excellent image quality over the large \Euclid VIS field of view, with distortions and image ellipticities within specification. Thermal stability is provided by the SiC structures and SiC optical elements. Very detailed modelling has been carried out of the thermo-elastic effects on the optical performance \citep{Anselmi:18}. Even so, for the PSF shape, a further level of knowledge using stars detected on the VIS focal plane is required as a function of many satellite parameters to achieve the required weak-lensing accuracy (Miller, L. et al., in prep.).

\begin{figure}
\includegraphics[width=\columnwidth] {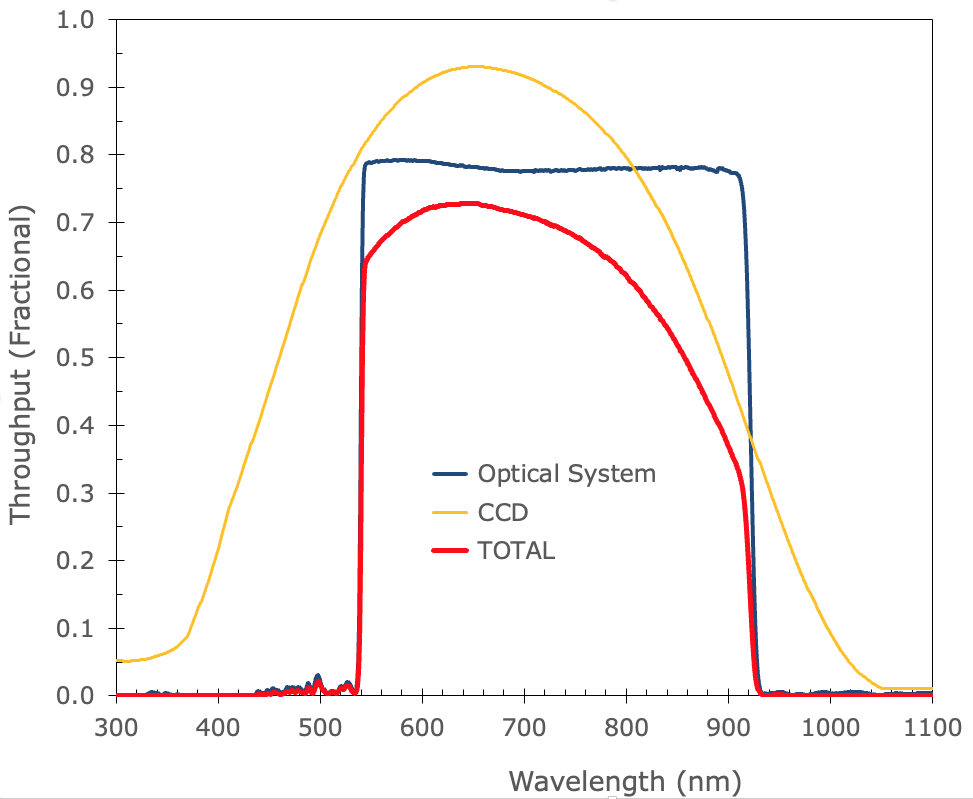}

{\hspace*{2mm}\includegraphics[width=0.95\columnwidth] {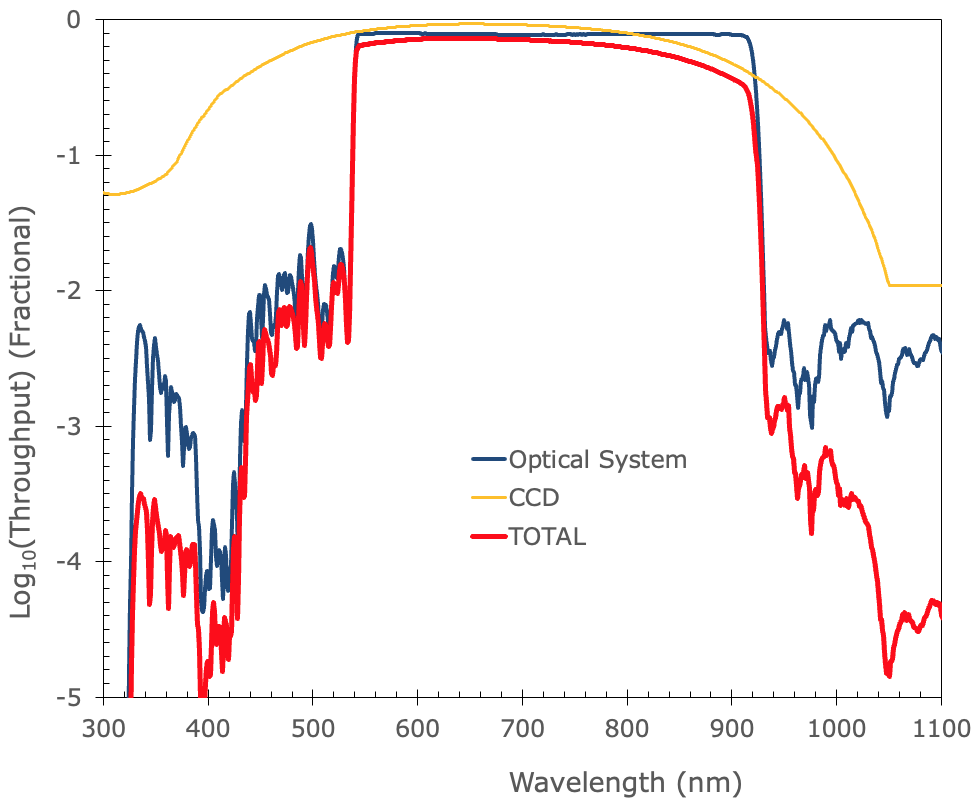}}
\caption{The \Euclid throughput as a function of wavelength of the optical elements to VIS (blue) and the Detective Quantum Efficiency of the CCDs (orange), giving the total throughput for this channel (red). {\it Top}: Throughput on a linear scale to show the VIS passband. {\it Bottom}: As above, but on a log scale to show the rejection of wavelengths outside of the VIS passband. }
\label{fig:throughput}
\end{figure}

The throughput of the optical elements in the VIS channel is shown in Fig.~\ref{fig:throughput}. The reflective coating on the three telescope mirrors and Fold Mirror~3 is protected silver, which provides a high reflectivity in the VIS passband. Fold Mirrors 1 and 2 and the dichroic have complex multilayer dielectric coatings which act to shape the passbands of both instruments. These multilayers are required to provide steep passband boundaries and high levels of rejection outside the passband over a wide range of incidence angles, given the large field of view. The layers are ordered in such a way as to minimise the complexity of the aberrations introduced into the PSF by the non-uniformity of the layers. It should be appreciated that in this system, each layer, and hence wavelength, introduces a slightly different end-to-end wavefront error owing to spatial non-uniformities across the layer. The two dielectrically coated fold mirrors have, in addition, different beam footprint sizes and positions depending on the field angles. These factors complicate the PSF modelling significantly, and are discussed in Miller, L. et al., (in prep.). 

The dichroic dielectric stacks are state of the art, and present on both sides of the element substrate, but a small fraction of the incident beam -- including wavelengths outside of the required VIS passband -- will be reflected by the rear surface of the dichroic and be transmitted back through the front surface to the VIS detectors where it forms an optical ghost. Because the dichroic is angled with respect to the incoming beam from the telescope, the ghost is displaced with respect to the direct image; however, in order to limit aberrations to NISP the angle is insufficient to throw the ghosts off the VIS focal plane entirely.

Returning to the layout of the units in the Payload Module related to VIS (Fig.~\ref{fig:PLM-CAD}), there is a stray light- and radiation-shielding hood surrounding the Focal Plane Array detector plane. When closed, the VIS shutter is in close proximity to this hood, but does not make contact with it. The Focal Plane Array is attached to the telescope baseplate by a substantial SiC bracket surrounding the hood; this is evident in Fig.~\ref{fig:PLM-CAD}. At the far side of the Focal Plane Array is the VIS electronics radiator which maintains the temperature of the large block of detector chain electronics within the Focal Plane Array at approximately 255\,K by radiating passively to space. The telescope temperature is maintained at about 130\,K where the SiC coefficient of expansion is extremely low, and the VIS detector plane is held at 153\,K where the VIS CCD operation in the presence of irradiating ions was determined to be optimal. 

Also shown in Fig.~\ref{fig:PLM-CAD} is the substantial harness of connections from VIS to the other two units, the Control and Data Processing Unit and the Power and Mechanism Control Unit, in the \Euclid Service Module. As is standard, the Service Module, and hence also these two units, are maintained at approximately ambient temperature.

\subsection{Focal Plane Array}
\label{subsec:FPA}

\begin{figure}
\center{\includegraphics[width=\columnwidth] {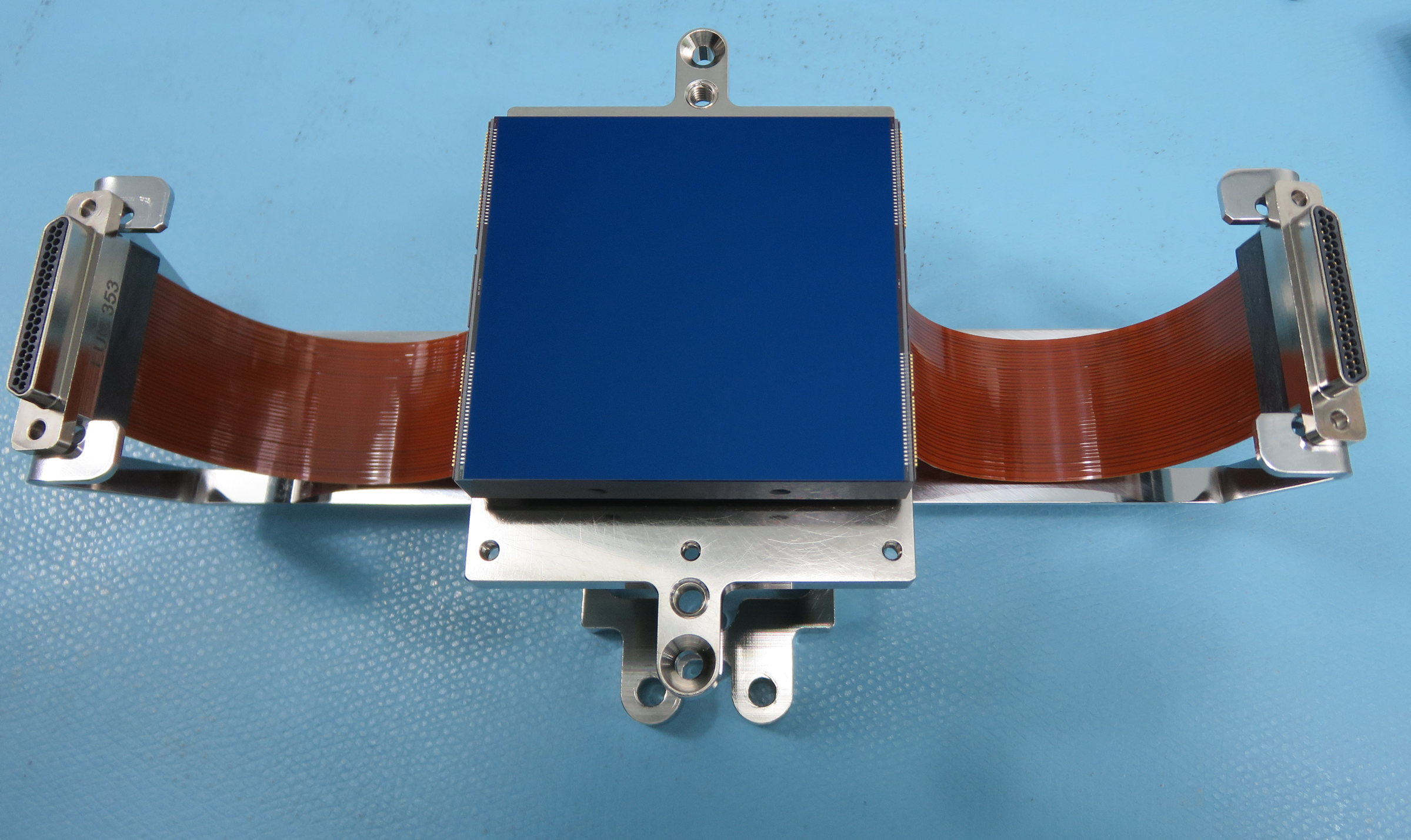}}
\vspace*{1mm}
\caption{ CCD273-84 developed specifically for VIS by e2v Technologies.}
\label{fig:CCD}
\end{figure}

\begin{figure}[htbp!]
\center{\includegraphics[width=\columnwidth] {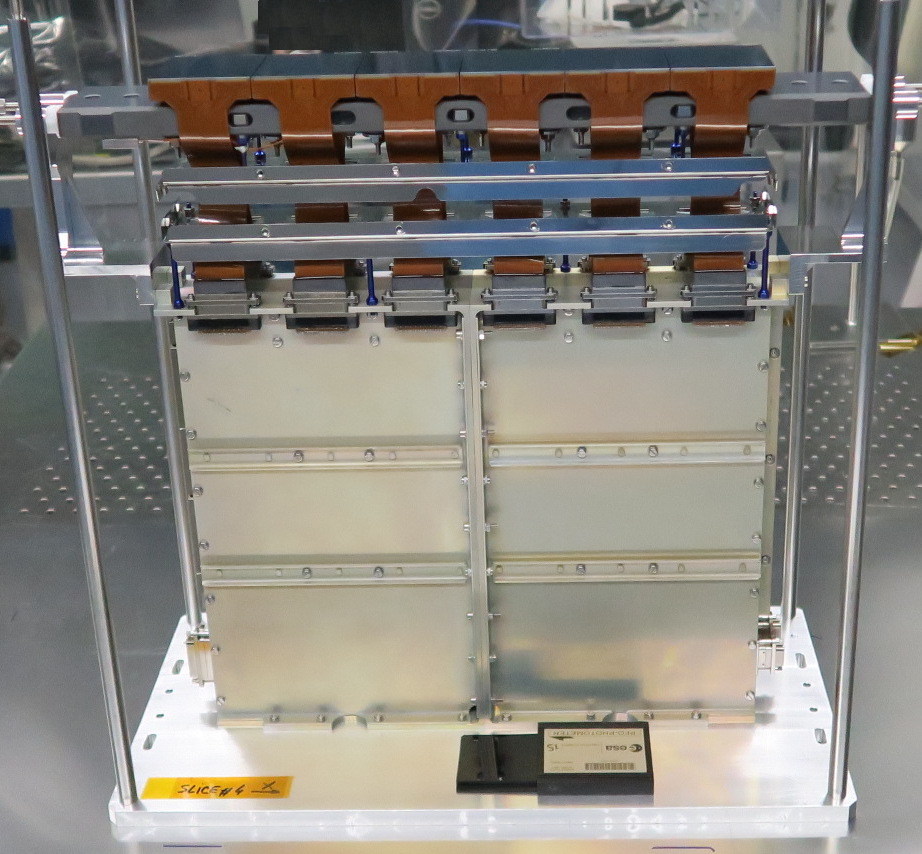}}
\vspace*{2mm}
\caption{Integrated `slice' of six CCDs connected to two Readout Electronics. The grey SiC beam supporting the CCDs near the top is not yet connected to the detector plane structure frame. Here it can be seen how the flexible connections from the CCDs pass through the two thermal shields. The item in the foreground is used for cleanliness monitoring. The base with corner columns is a frame for maintaining cleanliness and is not part of the slice.\vspace*{3mm} }
\label{fig:slice}
\end{figure}

The Focal Plane Array is the largest VIS unit, and contains the VIS detectors and associated electronics. The detectors are Charge Coupled Devices (CCDs) custom made for VIS (CCD273-84;
see Fig.~\ref{fig:CCD}). Because the detectors must be located with tight constraints relative to the telescope beam under all flight conditions, they are held in a detector plane structure (Fig.~\ref{fig:FPA_expanded}) consisting of a SiC frame across which the CCDs (Figs.~\ref{fig:CCD} and \ref{fig:slice}) are held on six SiC beams in six rows. The detector plane structure is supported directly from the substantial SiC baseplate bracket seen surrounding the VIS Focal Plane Array in Fig.~\ref{fig:PLM-CAD}. Each CCD has two flexible connections which pass behind to their associated electronics through two levels of thermal isolation to minimise the parasitic heating of the detectors by their electronics, which operate at a much warmer temperature. These electronics, the Readout Electronics, service three CCDs in one row, and two of these are connected side-by-side to service a row of six CCDs to produce a `slice'. All of the Readout Electronics are mechanically identical rather than being in mirror image pairs, so one of the two is upside down, and this is reflected in its associated three CCDs also orientated upside down. This arrangement is shown in Fig.~\ref{fig:slice}.

The 12 Readout Electronics are held within a substantial Aluminium structure, the Electronics Structure, also supported on the SiC baseplate bracket, but separately from the detector plane structure. This arrangement maximally isolates the Detector Plane from any mechanical displacements arising from varying power dissipation in the Readout Electronics during different operating modes. The requirements for the Electronic Structure location are set only by the constraints for the 72 flexible connections and are hence more relaxed. The 12 Readout Electronics Power Supply Units are located on the sides of the Electronics Structure immediately adjacent to their Readout Electronics so as to minimise the susceptibility of their power lines to radiated electromagnetic fields. 

\begin{figure*}[t!]
\center{\includegraphics[width=1.5\columnwidth] {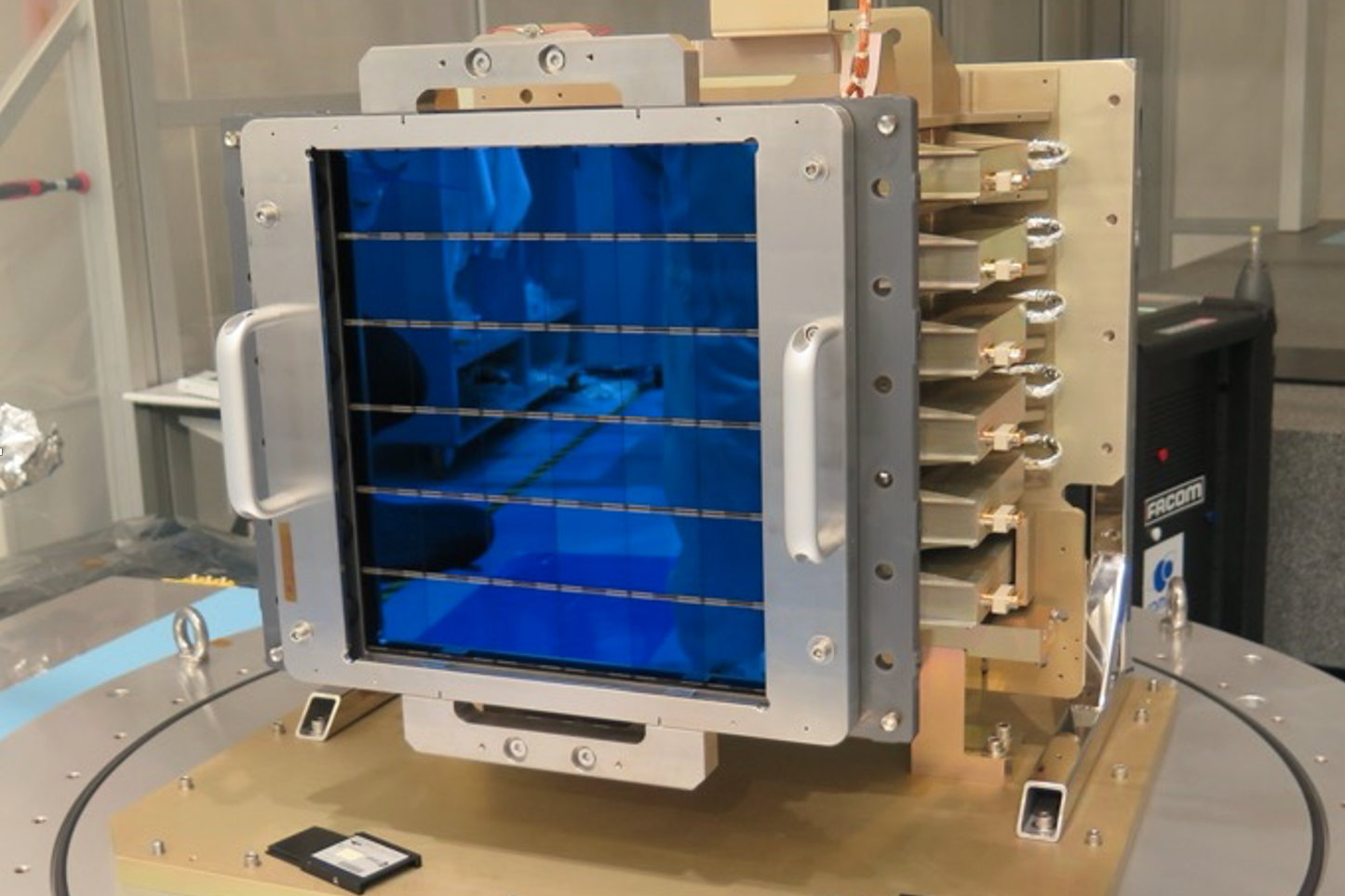}}
\center{\includegraphics[width=1.5\columnwidth] {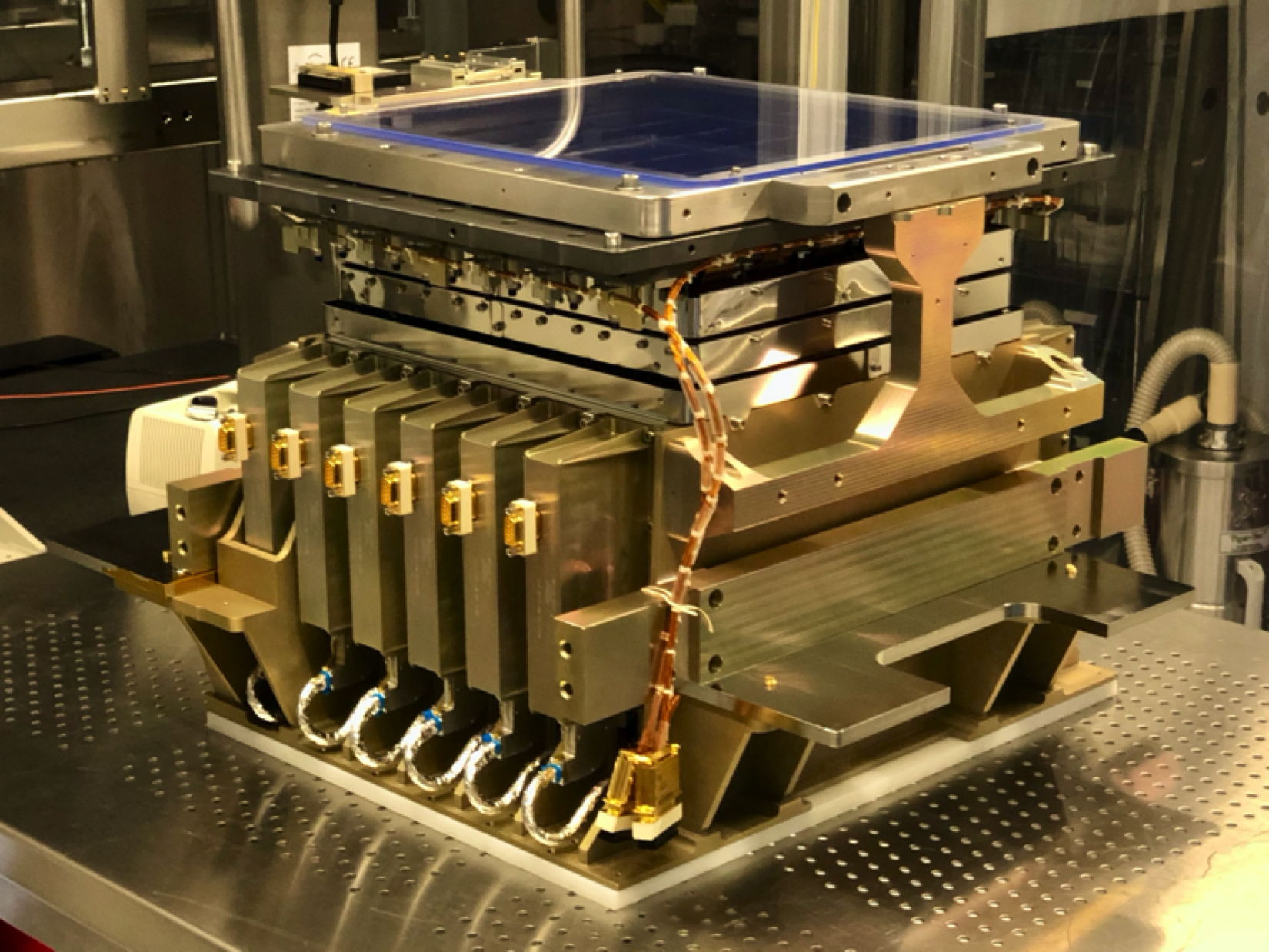}}
\caption{Two views of the fully integrated Focal Plane Array. The 36 blue CCDs supported on the grey SiC detector plane structure are visible under a protective cover. Below them are two levels of thermal isolation to minimise the parasitic heating of the detectors by the 12 Readout Electronics, which are located below the CCDs (see Figs.~\ref{fig:FPA_expanded} and \ref{fig:slice}) in the body of the Focal Plane Array Electronics Structure. Six Power Supplies for the Readout Electronics are located on two sides of this structure. Spacecraft power is delivered to the connections visible on the Power Supplies, while a SpaceWire data connection to the Control and Data Processing Unit is made to each Readout Electronics in the space below each Power Supply. The harness in the foreground in the lower image is the connection from the Power and Mechanism Control Unit to temperature sensors on the detector plane structure.}
\label{fig:FPA}
\end{figure*}

Figure~\ref{fig:FPA} shows the fully integrated Flight Model Focal Plane Array. The 36 CCDs are located on the grey SiC detector plane structure under a protective cover. When integrated on the Payload Module, the protective cover is removed, as is the structure supporting it and the SiC detector plane structure, because the two parts of the Focal Plane Array are maintained in position by the SiC baseplate bracket as described above. 

The VIS radiator (see Fig.~\ref{fig:PLM-CAD}), a passive element supplied as part of the Payload Module, is interfaced to the base of the Electronic Structure in Fig.~\ref{fig:FPA}. It is sized to radiate to space the total  136W dissipated in the 12 Readout Electronics and their Power Supply Units with sufficient control margin to maintain their temperature at ambient on the other side of the thermal isolations. The CCD array radiates and conducts its low levels of internal dissipation and parasitically derived heat into the Payload Module.

\begin{figure}
\center{\includegraphics[width=1.0\columnwidth] {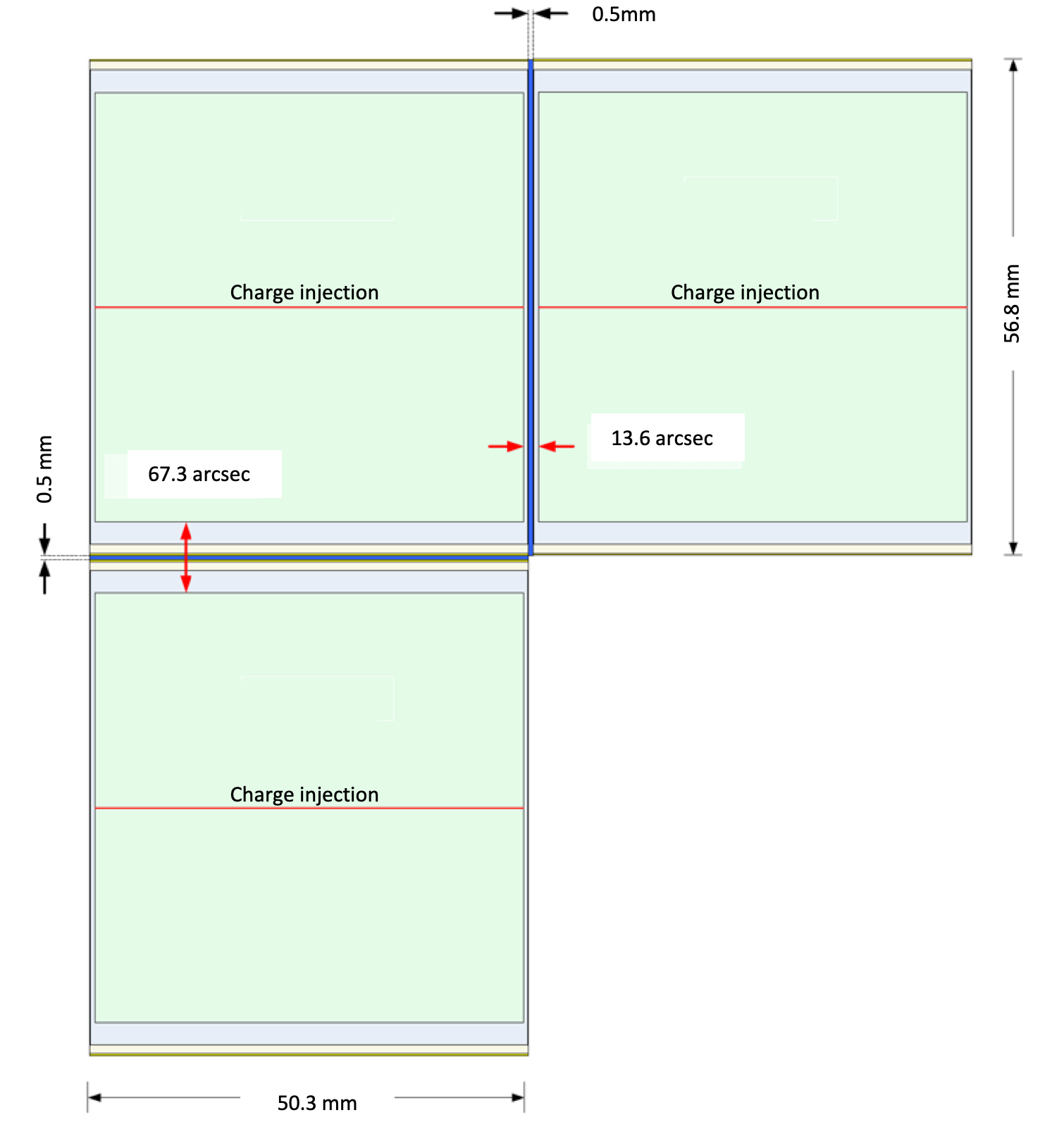}}
\center{\includegraphics[width=0.95\columnwidth] {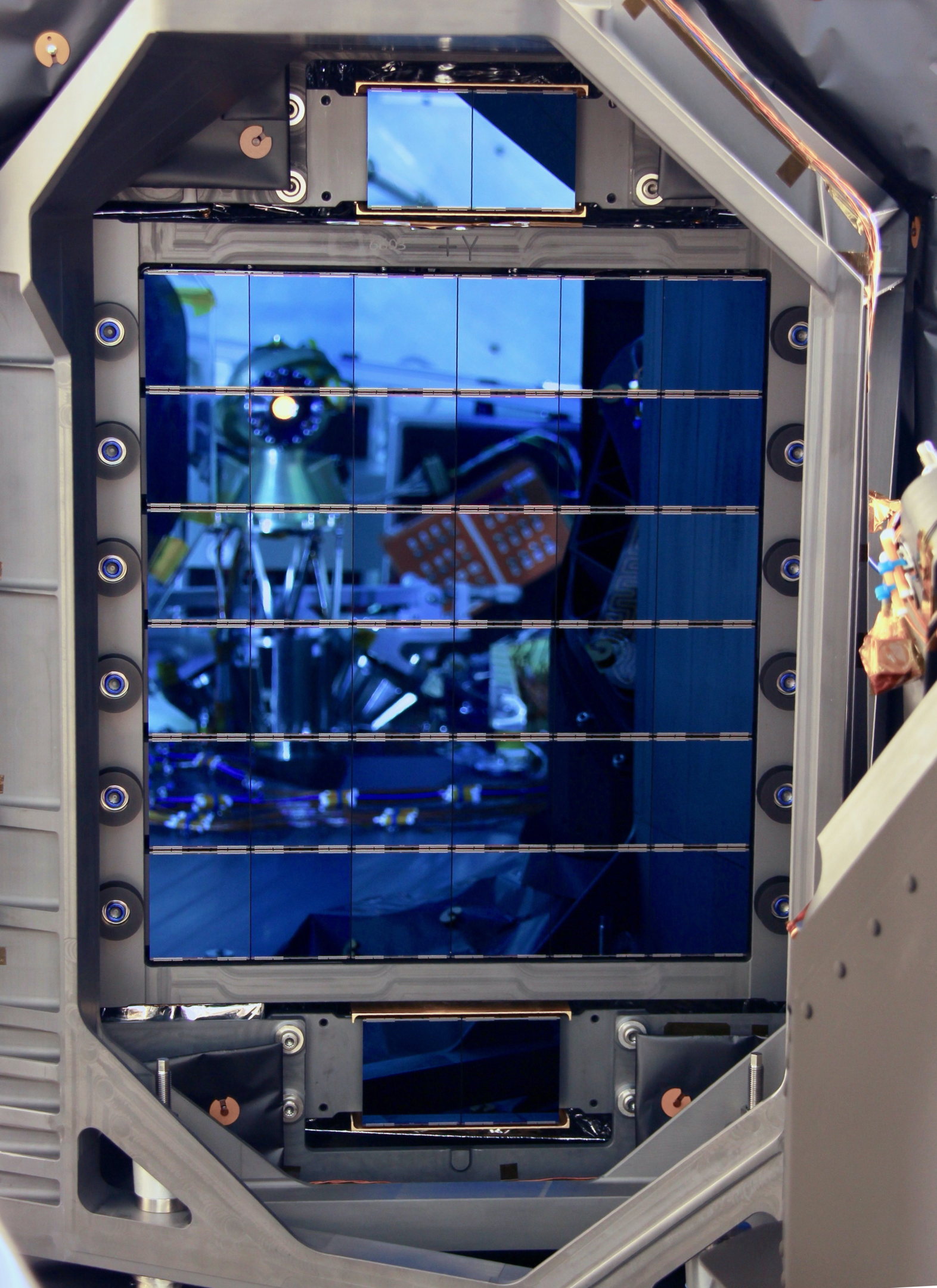}}
\caption{The arrangement of the CCDs in the focal plane. 
{\it Top}: Schematic of the butting of the CCDs. Green indicates the active area of Si. {\it Bottom}: The detector array integrated into the VIS bracket on the telescope baseplate, showing the spacing of the CCD and the wire bonds from the active Si to the flexible connections at the top and bottom of each CCD, as well as the two pairs of Fine Guidance Sensor CCDs above and below the VIS Focal Plane Array.}
\label{fig:FPA_DP}
\end{figure}

A full description of the Focal Plane Array is in \cite{Martignac:14}. The 36 CCDs on the detector plane structure are close-packed to maximise the filling factor of active Si. Some dead space is required at the top and bottom of each CCD for the readout registers and the connections by fine wire bonds to the flexible connections leading to the Readout Electronics.  A detailed schematic of the focal plane is shown in Fig.~\ref{fig:FPA_DP}. The fractional filling factor achieved is 0.86. In order to eliminate the possibility of scattered light from sources imaged on the detector plane structure, the CCD active surface is above all of the other elements on the Structure except the unavoidable 12 attachments for the CCD-supporting SiC beams to the frame. These therefore carry top-hat baffles.

The salient metrics for the VIS detector plane are given in Table~\ref{tab:FPA}. Each of the VIS detectors was ranked for a number of different characteristics as measured during their pre-delivery testing, including detective quantum efficiency, the number of cosmetic defects, readout noise, etc. and assigned to a position on the Focal Plane Array with the aim of providing an even spread of characteristics. This layout and information for each CCD is available in \cite{FPA_Matrix}.

\begin{table}[htbp!]
\caption{Interesting parameters for the VIS CCDs and Focal Plane Array.}
\begin{flushleft}
\begin{tabular}{lrrr}
\hline\hline
\noalign{\vskip 2pt}
{\bf Optical parameters} & & & \\
Focal length & \multicolumn{3}{r}{24.87 m} \\
Image scale &  \multicolumn{3}{r}{ 1 arcsec = 120.57\,$\micron$}  \\
{\bf CCD} & & & \\
Number of rows (height) & \multicolumn{3}{r}{4132}  \\
Number of columns (width) & \multicolumn{3}{r}{4096}  \\
Number of pixels & \multicolumn{3}{r}{16\,924\,672}  \\
{\bf Array} & & & \\
Number of rows & \multicolumn{3}{r}{6}  \\
Number of columns & \multicolumn{3}{r}{6}  \\
Number of CCDs & \multicolumn{3}{r}{36}  \\
Number of pixels & \multicolumn{3}{r}{609\,288\,192} \\
CCD spacing & \multicolumn{3}{r}{0.5 mm}  \\
Fill factor & \multicolumn{3}{r}{0.861}  \\
\noalign{\vskip 4pt}
\hline
\noalign{\vskip 4mm}
                               & Height & Width & Area  \\
\hline
\noalign{\vskip 4pt}
{\bf CCD} & & & \\
Pixel dimensions      \hfill [$\micron$] & 12.0  & 12.0  &        \\
                          \hfill [arcsec] & 0.100 & 0.100 &        \\
Active Area dimensions    \hfill [mm] & 49.6  & 49.2  &        \\
            \hfill [arcmin or arcmin$^2$] & 6.85  & 6.79  & 46.6  \\
Physical Dimension        \hfill [mm] & 56.8  & 50.3  &      \\
{\bf Array} & & & \\ 
Active Area dimensions  \hfill [mm] & 336.6 & 302.7 &       \\
                  \hfill [deg or deg$^2$] & 0.775 & 0.697 & 0.540  \\
Physical dimensions     \hfill [mm] & 343.3 & 304.3 &       \\
\noalign{\vskip 2pt}
\hline
\end{tabular}
\end{flushleft}
\tablefoot{The active area dimensions refer to the enclosing rectangle about the active area of the array of CCDs,  i.e., including the CCD spacing.}
\label{tab:FPA}
\end{table}

\subsubsection{CCDs}
\label{subsubsec:CCDs}

As the entities responsible for photon detecting, the CCDs are central to the performance of VIS. Their essential characteristics in the VIS context are enumerated in Sect.~\ref{subsubsec:detectors}. The earlier \DUNE studies had considered variants on the e2v CCD91-72 used in \Gaia \citep{Gaia} with square $13.5\,\micron$ pixels, but finer sampling would result in the benefit of a physically smaller focal plane. A pixel size of $12\,\micron$ would retain sufficient Full Well Capacity, and devices in a 4k\texttimes4k format were available in an existing CCD, the e2v CCD203-82, which also had been used in a space context. This was therefore baselined in the \DUNE proposal to ESA \citep{Refregier:08}. Subsequently, a series of discussions with the manufacturer led to modifications to be incorporated in a custom design optimised for \Euclid\!\!, designated the CCD273-84, shown in Fig.~\ref{fig:CCD}. This was a careful upgrade with already proven elements and revisions -- particularly improving the  tolerance of the device to radiation damage by ions -- which were technically feasible and cost-effective. 

The specification for the VIS CCD is contained in \cite{CCD_Spec} and the device characteristics are provided in \cite{Endicott:12} and \cite{Short2014}. The format is slightly larger than that of the CCD203-82 at 4096\texttimes4132 $12\,\micron$ square pixels in four quadrants, each with a readout node located at the corners of the CCD, and therefore each CCD requires four channels of associated electronics. The readout nodes are improved dual MOSFET stages to reduce readout noise. To ease calibratability, the pixel structure is simple, containing neither antiblooming drains nor supplementary buried channels. A thinner gate dielectric on 1500 $\Omega$\,cm resistivity Si increased the full well capacity and hence the dynamic range above that usually achieved with 12-$\micron$ pixels, while also reducing the susceptibility to ionising radiation, such as high energy electrons. The pixels are laid out in `stitch blocks' of 512\texttimes256 pixels.

The devices themselves are back-illuminated and moderately thinned to $40\,\micron$, as in the red-sensitive variant of the \Gaia CCDs, and coated to reach a Detective Quantum Efficiency greater than 83\% from 550--$750\,{\rm nm}$, and greater than 49\% at $900\,{\rm nm}$. The average of the Detective Quantum Efficiencies achieved for the flight CCDs peaks at 94\% at $650\,{\rm nm}$, as shown in Fig.~\ref{fig:throughput}. The reduction in sensitivity towards longer wavelengths results from the increasing transparency of the Si. It was considered whether greater red sensitivity should be achieved by using a thicker CCD, but these lacked flight heritage, and have subsequently also found to be more subject to intensity-dependent charge diffusion between pixels -- the `brighter-fatter effect', so on balance this decision was probably correct.

It was clear from the extensive programme undertaken to characterise the performance degradation caused by Si lattice damage from non-ionising radiation (such as high-energy protons) in the \Gaia programme that this would also require significant attention for \Euclid VIS. Most CCDs are n-channel (p-doped) devices where the incoming photons generate electrons which are eventually transferred to the readout node. P-channel variants (where the photons generate holes) were considered in the early stages of specifying the VIS CCDs, and a careful trade-off was undertaken as to whether these would be more robust to radiation damage. Given the limited information at the time, p-channel variants of the CCD204-42 (itself a cut-down CCD203-82) were fabricated, irradiated and tested \citep{Gow:16}, but long before this, n-channel devices had been selected for VIS on the grounds of the lack of p-channel flight heritage and the difficulty of sourcing the highest quality silicon. 

However, two measures were taken to improve the tolerance to non-ionising radiation in the CCD273-84. Because in \Euclid VIS, pixels are not summed together on the device, the charge handling capacity of the serial readout register was reduced to a value just exceeding that of each pixel, specified as greater than 175\,000 e$^{-}$: the smaller volume of Si reduces the number of radiation damage sites and hence improves the charge transfer in the serial register. Then, a charge-injection structure was introduced between the top and bottom pairs of quadrants. This allowed user-defined packets of charge to be introduced into the pixels at the top of each CCD column. Multiple injections during readout result in a block of rows with the specified charge level. This is a primary facility for calibrating the extent of the radiation damage, as it allows the erosion of the front of the image and the extent of the trails behind it to be quantified. The charge injection capability of the CCD273-84 was enhanced compared to that of previous e2v devices, both in the improved uniformity of injection across the CCD and, through a novel notch design, in the provision of low levels of charge injection.

The charge is moved from pixel to pixel down a column using four electrodes (four `phases') of size 4, 2, 4, and $2\,\micron$, while there are three equally sized $4\,\micron$ electrodes to move the charge in the serial register. As the boundaries of a pixel during an exposure are set by an appropriate voltage on one of these phases to create an electrostatic barrier, charge is stored under the other three phases, with a confinement which depends on whether a 2- or 4-$\micron$ phase is used. It should also be noted that because the CCD273-84 uses much of the photolithographic mask set of the CCD203-82, the sequence of phases is not mirror-imaged about the charge injection structure between the upper and lower quadrants. The consequence is that the behaviour of the charge injection into the first pixel, and the termination into the readout register, is slightly different between the upper and lower pairs of quadrants, so that different operational parameter values are required for optimal performance.

A great deal of attention was paid to the CCD operation. The initial considerations were the optimal operating temperature and rates of charge transfer down columns and then, slightly more than 2048 times faster, through the readout register. These are constrained by the time constants of the release times of electrons from Si lattice radiation damage trap species. The temperature modifies the time constants. To minimise the effect of the trapping, and hence distortion of the galaxy shape measurement, the duration of each charge transfer step (for both rates) should not be commensurate with the trap time constants. At the same time, slow transfer rates result in more dead time between exposures while fast transfer rates limit the accuracy with which the charge packet can be measured, increasing the readout noise, and are at some point also limited by the technology available to digitise it sufficiently (Sect.~\ref{subsubsec:ROE}). The operating temperature selected is $153\,{\rm K}$, which avoids any commensurate trap time constants for a pixel readout duration of $14.3\,\mu$s (a rate of 70 kHz) and a row transfer duration of $4.02\,$ms. With a prescan of 51 pixels, a postscan of 29 pixels in the row and 20 in the column direction for each quadrant, the CCD takes $72\,$s to read out through four nodes \citep{UMan}.  At $153\,{\rm K}$, the dark noise within the pixel is negligible even after long exposures.

Besides these top level questions, many detailed analyses were performed for the operation of the CCD. Much of this work was carried out under the auspices of a group which contained members of the VIS Team, the e2v manufacturer, ESA, and the Science Ground Segment. This informal forum concentrated on achieving a detailed knowledge of the CCD273-84 in all of its aspects, and especially the optimisation of its radiation tolerance, which included significant testing programmes. More information is in \cite{Gow:12,Clarke2012,Prod'homme2014,Skottfelt:15, israel2015, Skottfelt2016,Skottfelt:21}. A particular innovation was the use of tri-level clocking, in which the four pixel phases were activated in sequence not between only high and low values, but also at an intermediate value, with the effect of further encouraging de-trapped electrons to rejoin their charge packet rather than falling back into the following pixel during transfer. A second calibration of the radiation damage effects called trap pumping was introduced to VIS. Here, instead of moving the charge sequentially to the readout register or readout node, it is clocked backwards and forwards. Pairs of bright and dark pixels identify the location of lattice damage traps. This information can be used in the Science Ground Segment data reduction algorithms which correct for the radiation damage effects. New techniques were developed to create the appropriate trap-pumping sequences for the four-phase pixel structure. Trap pumping, especially that in the readout register, is novel in a space instrument. 

Further, radiation damage testing of the CCDs was carried out at operational (cryogenic) temperatures, with keep-cold capability, to reproduce the conditions in orbit as closely as possible. This identified that under these conditions the time constants of the traps created were no longer discrete values, but a continuum \citep[see][]{Skottfelt:24} for this behaviour in early VIS inflight measurements. This is a consequence of the greater rigidity of the Si lattice at lower temperatures, and the inability of the traps subsequently to relax to a common set of characteristics for each trap type. Relaxation takes place in ground testing if the device experiences ambient conditions. Meeting for more than a decade, this forum ensured that the VIS CCD parameters are optimised to the maximum extent possible, and informed the modelling of the radiation damage effects in the Science Ground Segment processing functions.

After a development programme, 50 Flight Model CCD273-84s (Fig.~\ref{fig:CCD}) were produced by e2v, all of which, except with occasional minor infringements, met or exceeded specifications. This program included significant characterisation and qualification at CCD level, and many of the calibrations, such as the Detective Quantum Efficiency in Fig.~\ref{fig:throughput}, were incorporated directly into the \Euclid calibration database.

\subsubsection{Readout Electronics and Power Supply}
\label{subsubsec:ROE}

The Readout Electronics convert the signals at the output node of the CCDs into digital values, and also control their operation by providing the correct operating conditions and clocking signals to read out the images. Each has an associated Power Supply Unit to condition the raw spacecraft power and provide the different voltage levels the Readout Electronics require. As noted above they are all located in the Electronics Structure within the Focal Plane Array. These units are described in detail in \cite{ROE_Design} and \cite{RPSU_Design}.

\begin{figure*}[htbp!]
\center{\includegraphics[width=1.5\columnwidth] {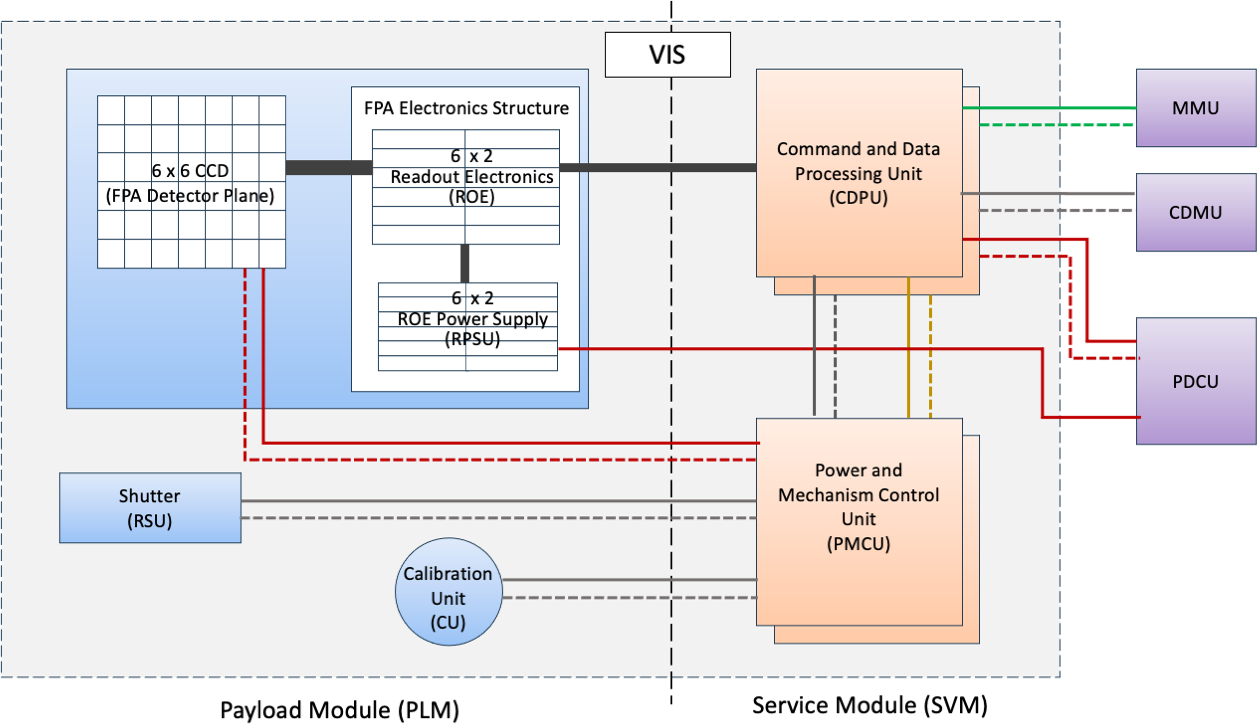}}
\caption{The VIS electronics interfaces. Units in the Payload Module are shown in blue and those in the Service Module in orange (these are dual-redundant). Modules in purple are those belonging to the spacecraft: the Mass Memory Unit (MMU) mostly filled with science images, the Command and Data Management Unit (CDMU), and the Power Control and Distribution Unit (PCDU).}
\label{fig:electronics}
\end{figure*}

With 36 CCDs each with a readout node per quadrant, 144 independent channels are required for the readout nodes. As noted in Sect.~\ref{subsec:FPA} above, an early decision was that groups of three CCDs would be serviced by a single Readout Electronics and Power Supply to minimise the system resource usage while nevertheless providing sufficient redundancy. This was compliant with the requirement that the failure of a single readout electronics/power supply should result in the loss of a maximum of 10\% of the active focal plane. There would therefore be 12 sets of readout electronics and power supplies, each connected to three CCDs and providing 12 channels of image data. It was also feasible to prescribe 12 sets of harnesses from the 12 Readout Electronics to the Control and Data Processing Unit, and to command and receive data to/from 12 units. This architecture, shown in Fig.~\ref{fig:electronics}, was therefore resilient and parsimonious in terms of system resources, requiring a total of less than 142\,W (136\,W average) in operation, despite stringent performance requirements. It lent itself to the cycle of multiple unit fabrication, ground calibration and then also the assembly of the Focal Plane Array, starting with pairs of Readout Electronics as seen in Fig.~\ref{fig:slice}.

Many of the CCD voltage and clocking parameters can be set up per CCD, while some, such as the charge injection parameters, must be set up per half of the CCD. This level of controllability has implications for the digital-analogue converters which must set the parameters, and for which parameters should be monitored. Mass, power, and spatial accommodation constraints require these resources to be assigned judiciously, while nevertheless ensuring that the most critical parameters for the CCD operation could be optimised in orbit, if necessary. 

Figure~\ref{fig:electronics} shows the VIS-level diagram of the electronics layout. The connection between the CCDs is through 36 pairs of flexi-circuits evident in Fig.~\ref{fig:CCD}. The 12 digital connections between the Readout Electronics and the Control and Data Processing Unit are made using the SpaceWire\footnote{\href{https://ecss.nl/standard/ecss-e-70-41a-ground-systems-and-operations-telemetry-and-telecommand-packet-utilization/}{ECSS-E-ST-50-12A}} protocol, running on LVDS\footnote{ANSI/TIA/EIA-644-A} (low-voltage differential signalling) hardware layers; these carry both science data and telecommand/telemetry. A 40 MHz master clock and a synchronisation signal are also distributed through the LVDS connection to ensure all Readout Electronics clocks and reading operations of the CCDs are executed synchronously. This allows the simultaneous reception on the Control and Data Processing Unit of the data from the 12 Readout Electronics. The 12 Power Supplies are fed directly from the spacecraft via latching current limiters, bypassing the Control and Data Processing Unit, and hence are directly under spacecraft control.

\begin{figure}
\center{\includegraphics[width=0.95\columnwidth] {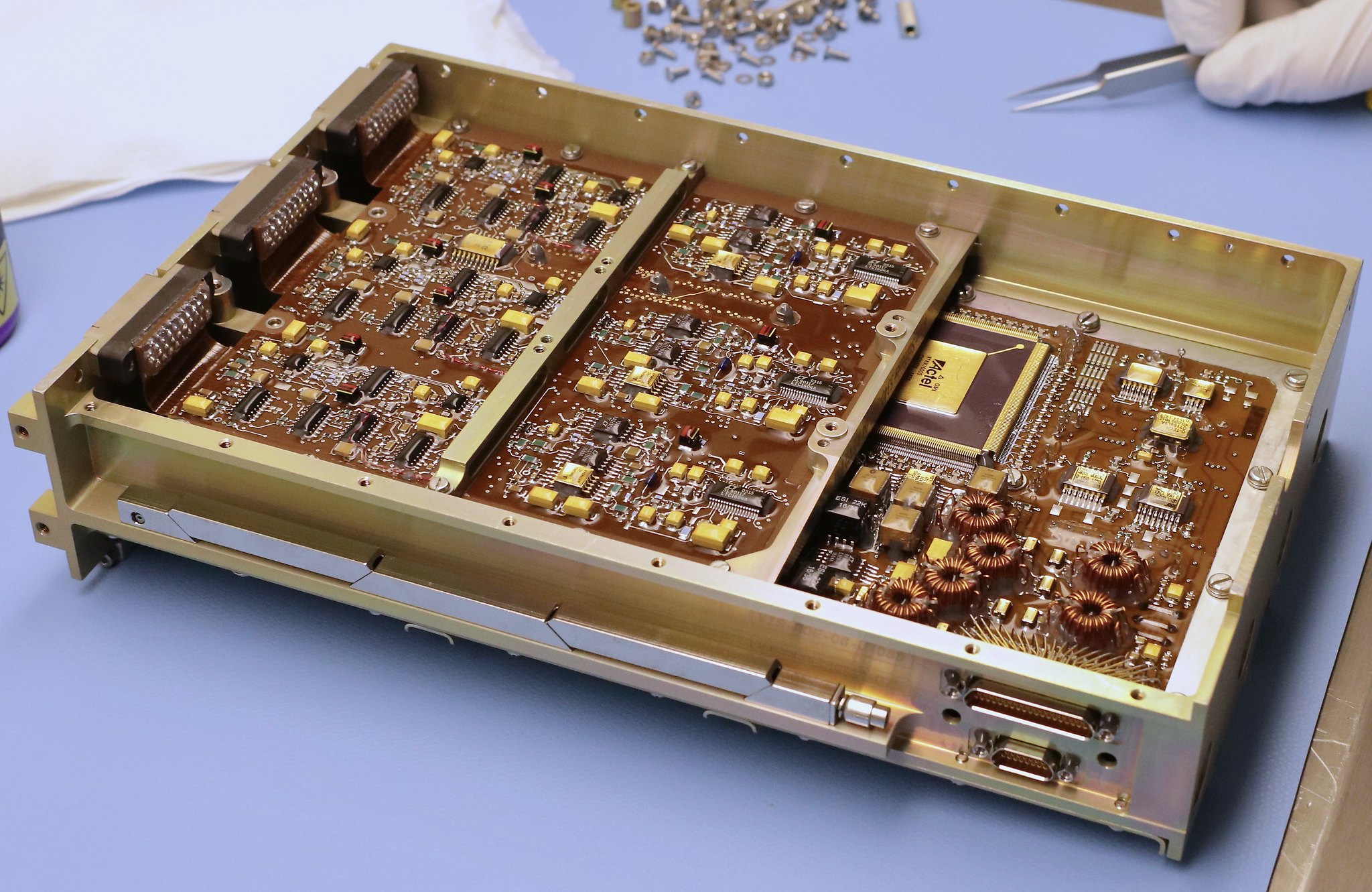}}
\center{\includegraphics[width=0.45\columnwidth] {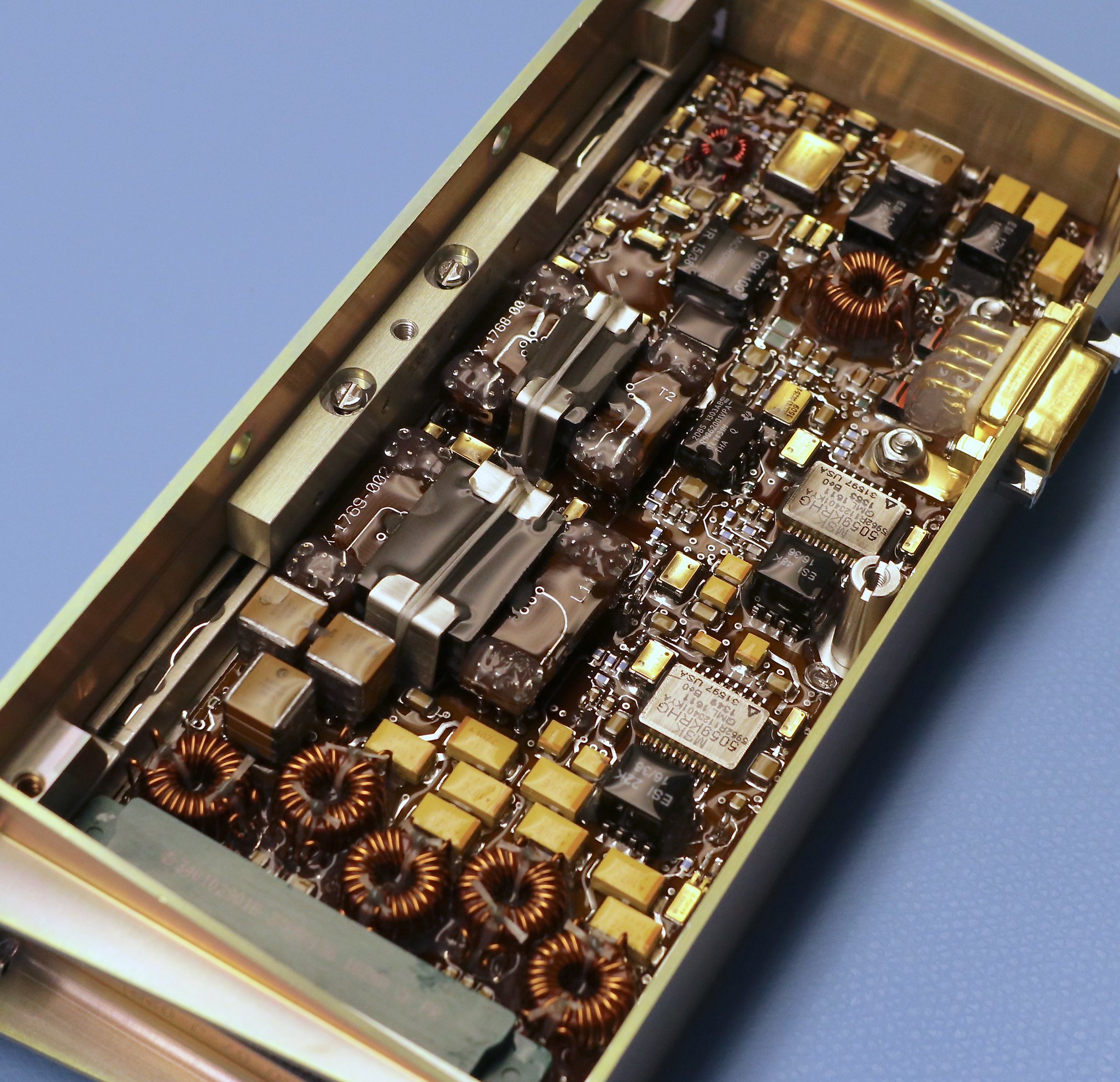}}
\caption{VIS Readout Electronics {\it (top)} and Power Supply Unit {\it (bottom)}. The Readout Electronics consists of two double-sided circuit boards containing the 12 analogue channels -- three repeating analogue circuits can be seen in the topmost board -- with the digital board containing the Field Programmable Gate Array between them and to the rear of the unit. The connectors to the upper flexible connections from the CCDs are at the front of the unit. The Power Supply Unit contains also a tightly populated double-sided circuit board. The planar transformers can be seen on the centre-left of the unit.}
\label{fig:ROE+RPSU}
\end{figure}

The measurement of the pixel charges presented at the CCD output node is performed by the Readout Electronics (Fig.~\ref{fig:ROE+RPSU}) using analogue correlated-double-sampling circuitry \citep{ROE_Design,RPSU_Design} with the associated common mode rejection and low-pass filtering to minimise noise. Digital sampling was considered, but rejected on the grounds of its greater power consumption, and because the achievable electronics noise of the analogue circuitry was already at the level where it was less than that from the CCD readout node, and within the overall requirement. The CCD output is sampled at two stages: first at the reset level, and then when the charge is available, and the difference is applied to a 16-bit analogue-to-digital converter to provide the digital output. As the circuitry should not add significantly to the noise generated by the CCD readout node, requiring noise levels less than 1 part in $2^{16}$ ($1.5\times 10^{-5}$), great care is taken in the timing of the sampling instants, on preventing feedthrough of unwanted signals and on the careful shaping by the filters to match the signal characteristics. Within those that were qualified for the space environment, judicious choices of components were made, both active and passive. Radiation-hard and latch-up resistant 16-bit analogue-to-digital converters became available at the time of the early \Euclid Readout Electronics development, enabling for the first time an intrinsically robust 16-bit solution. After the conversion, the digital values pass to a Field Programmable Gate Array (FPGA) which handles the SpaceWire communication to the Control and Data Processing Unit.

The Field Programmable Gate Array is a device which can be configured to provide extensive high-speed digital processing. Each Readout Electronics contains a single one of these to provide its digital functionality. This includes the interpretation and execution of commands from the Control and Data Processing Unit and the collection of housekeeping information such as temperatures, currents and voltages. Importantly, it also generates the sequencing of the clocking signals to the 12 CCDs. Producing the firmware for the Readout Electronics Field Programmable Gate Array was a significant task, and drove the schedule for the instrument for a time.

The Readout Electronics is required to perform several different operations, including setting the correct parameters for an exposure, flushing the CCD beforehand to clear it, inverting some CCD voltages to remove persistence effects, controlling the charge injection lines when required, and controlling the trap pumping. In order to minimise the readout noise levels, clocking and reading of the CCDs is performed synchronously for all 12 Readout Electronics, aligned to the master clock and synchronisation lines from the Control and Data Processing Unit. However an internal oscillator is available, and this also allows housekeeping and status to be transmitted to the Control and Data Processing Unit in case of a failure condition on the master clock line, which would otherwise cause the Readout Electronics units to halt, though it is possible to operate only one Readout Electronics unit at a time in this condition. It also facilitated stand-alone testing. Switching from external to internal oscillator and back required particular attention in the firmware to avoid failure states.

The Readout Electronics Power Supply Unit (Fig.~\ref{fig:ROE+RPSU}) produces six voltage levels in a tight space envelope. The noise levels on these multiple secondary outputs, particularly those which service the Readout Electronics analogue circuitry, are challenging given the available power and space allocation. A single switch forward topology was used \citep{ROE_Design,RPSU_Design}, with planar transformers to provide better coupling and hence efficiency. The converter switching frequency is synchronised to the CCD readout clocking, to ensure that the conditions provided by the Unit are always in the same state when the readout node is sampled. Each Power Supply Unit is controlled and monitored by its Readout Electronics. Particular attention was paid to the common mode noise isolation and to the grounding of the two units with each other and the external environment. Digital and analogue grounds are kept separated. The layout of both units was optimised over several evolution cycles to minimise internal cross-talk between channels and in order to meet the specified high level of immunity from conducted and radiated interference.

\subsubsection{Block-level integration and testing}
\label{subsubsec:block}

Once the final design of the Readout Electronics and Power Supply Unit was qualified through a Qualification Model, the 12 Flight Model units and two Flight Spares were fabricated and tested in a complex sequence. Automated procedures enabled extensive testing. After conformal coating of the circuit boards and integration in their enclosures, final tests followed in ISO5 clean facilities.

\begin{figure}[htbp!]
\center{\includegraphics[width=0.95\columnwidth] {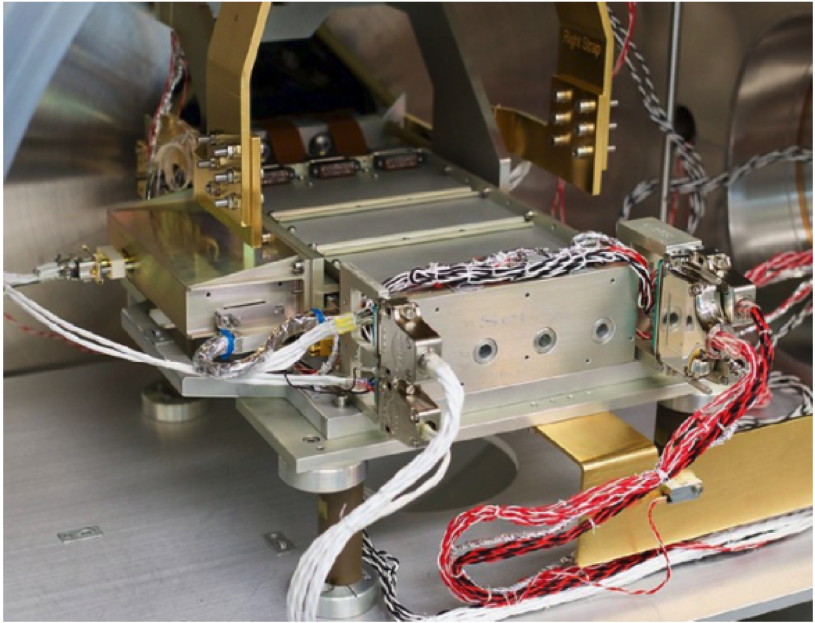}}
\caption{First of the Flight Model blocks configured for testing in the calibration chamber. The upper flexible connections from the three CCDs are visible at the furthest part of the unit, and the Power Supply is connected to the left of it. Gold-coated cold fingers above provide a temperature of $153\,{\rm K}$ at the CCDs while ambient conditions are maintained for the electronics via the structure below them. The cleanliness of the environment during testing is maintained using cold scavenger plates. The light from the optical sources enters through the dark window ahead of the CCDs. To produce the point sources, a lens array is inserted beyond the window to focus five images on each quadrant of each CCD.}
\label{fig:calibration}
\end{figure}

The last stage before delivery for integration into the Focal Plane Array structure was the calibration of each of the fourteen blocks of three CCDs, Readout Electronics, and Power Supply to provide the reference for the understanding of the performance of the instrument before launch; \cite{Azzollini:21} provides a full description. Except for some measurements, such as the noise induced by one set of Readout Electronics in another, this could be achieved at block level, which was simpler, and screened out any deficient units before delivery to the main structure. A comprehensive programme was carried out both using the internal calibrations such as charge injection, and external optical sources -- for example point source measurements and flat-field illuminations -- at several wavelengths. Again, automated sequences and processing of the approximately 12\,TB data set was employed, itself a substantial development. An earlier release of the \cite{Azzollini:21} final report was provided for instrument delivery.
Figure~\ref{fig:calibration} shows the first of the blocks configured for testing prior to insertion into the chamber.

Reduced versions of the Focal Plane Array, with Engineering Model and Qualification Model units including respectively two and three complete detector chains, were tested for electromagnetic compatibility (EMC) at Airbus in Portsmouth \citep{Candini:17a} and the ESA facilities at ESTEC \citep{Candini:17b,Candini:19}. These included a measurement of the interaction and induced noise between the units and the units from the Fine Guidance Sensor. 

\subsection{Shutter}
\label{subsec:Shutter}

\begin{figure}[htbp!]
\center{\includegraphics[width=0.95\columnwidth] {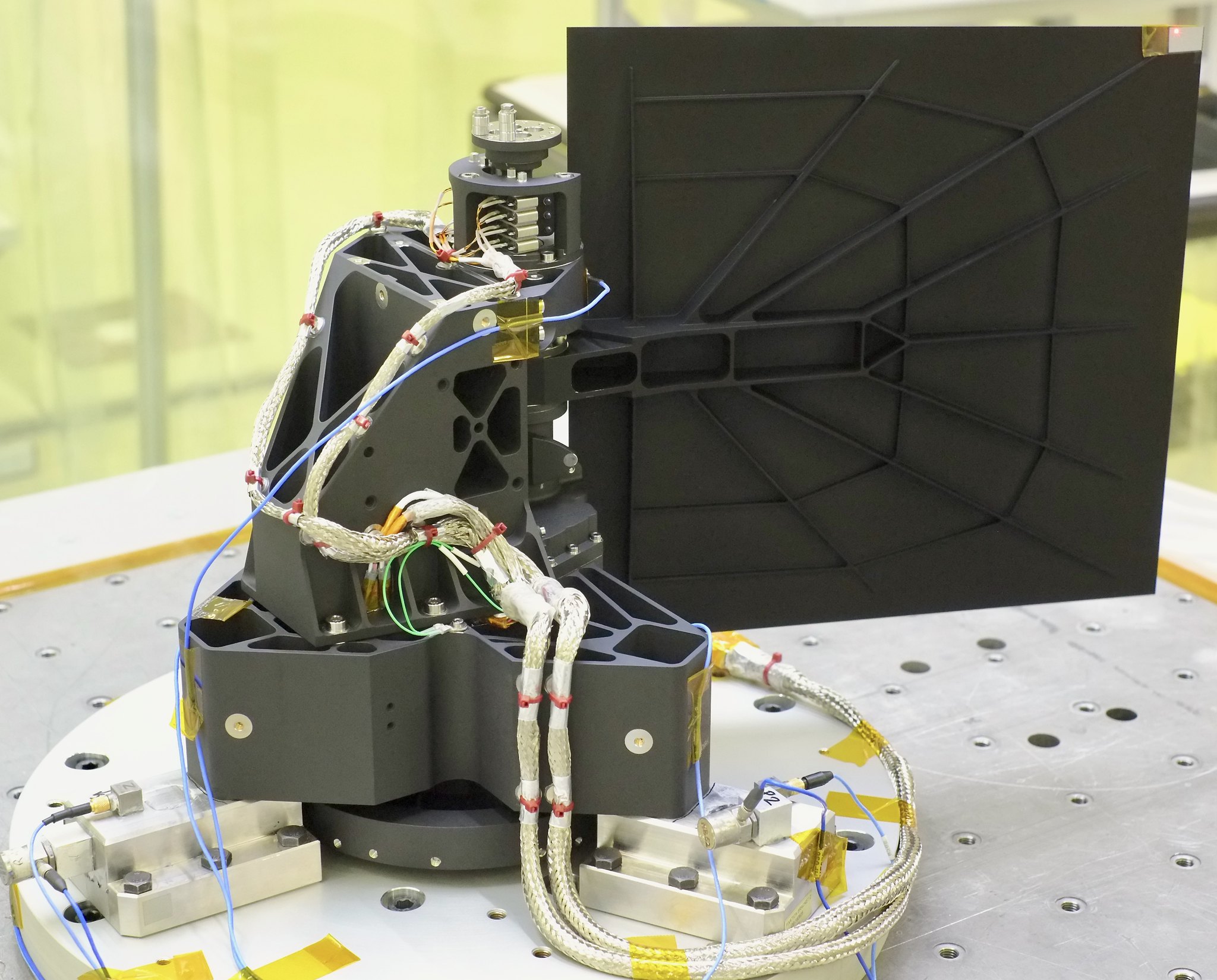}}
\caption{Flight Model Shutter as prepared for its acceptance vibration test. The shutter leaf is the large square structure on the right of the image. The momentum-compensation wheel is at the bottom of the unit in this test orientation, with the drive motor within the structure above it. To minimise scattered light, the Shutter surface treatment is highly absorptive.}
\label{fig:shutter}
\end{figure}

As noted in Sect.~\ref{subsubsec:shutter} the VIS shutter is required so that during the readout of the CCDs, light from the telescope or the Calibration Unit is blocked from the Focal Plane Array. This prevents charge trails being recorded behind sources accumulated during the exposure. It is located in front of the Focal Plane Array, as shown in Fig.~\ref{fig:PLM-CAD}, and is therefore required to operate at the cryogenic temperatures of the Payload Module, approximately 145\,K. The Shutter is described in \cite{Genolet2016} and \cite{shutter} and is shown in Fig.~\ref{fig:shutter}.

Because images are being recorded as soon as the shutter starts to open, and until it is fully closed, it is imperative -- as noted in Sect.~\ref{subsubsec:shutter} -- that its motion disturbs the satellite pointing to the minimum possible extent. To minimise the angular momentum change, designs with two closing leaves were initially considered, but required two motor drives and more complex mounting, and, moreover, doubled the chance of at least one leaf failing, so the VIS shutter consists of a single motor-driven leaf, with an angular momentum compensating wheel driven from the motor shaft. 

The Shutter was required to open and close within 10 seconds, so that the difference in exposure duration from one side of the focal plane to the other was limited, and a 2.74-s duration was ultimately adopted. During development it was found that microvibration levels were in excess of the acceptable levels, and anti-backlash was incorporated into the gear drive. However, the acceptable disturbance levels were subsequently relaxed, so for the Flight Model the gears reverted to the original design which had been found to be superior in lifetime tests. The Shutter can be operated either in closed-loop where the traverse end points are recorded by the end-switches, or in open loop which terminates the traverse just before the end switches are contacted. This increases the repeatability of the exposure duration and reduces the wear of the end-switches.

While the stepper motor has redundant electrical windings, each connected to one half of the Power and Mechanism Control Unit, mechanically the mechanism is a single failure point. Initially therefore a single-shot release mechanism was incorporated to allow the shutter to spring open should the motor drive fail. A failure mode analysis concluded however that its electrical feeds introduced non-repeatability in the delicate balancing of the mechanism, and increased the required actuation torque, while not increasing the overall reliability significantly, so the mechanism was eliminated. A second mechanism to hold the shutter in a launch position was also eliminated, after modelling showed that, given the fine balancing and momentum compensation, launch disturbances would not exceed the motor detent torque which was therefore sufficient to hold the shutter in position. These simplifications significantly improved the design.

\subsection{Calibration Unit}
\label{subsec:CU}

CCDs are not perfectly uniform from pixel to pixel in their sensitivity. Colour independent effects are caused by percent-level variations in the area of each pixel as outlined by the lithographic mask set used in fabrication, while colour dependent effects arise because of small-scale structure caused by the thinning process by ion beam etching and by the intrinsic variations in the Si itself. Calibration of these non-uniformities requires a uniform wavelength-selectable illumination to be projected onto the CCD. The pixel variations can then be recorded by taking an image, and the science images corrected by this flat field map.

\begin{figure}
\center{\includegraphics[width=0.8\columnwidth] {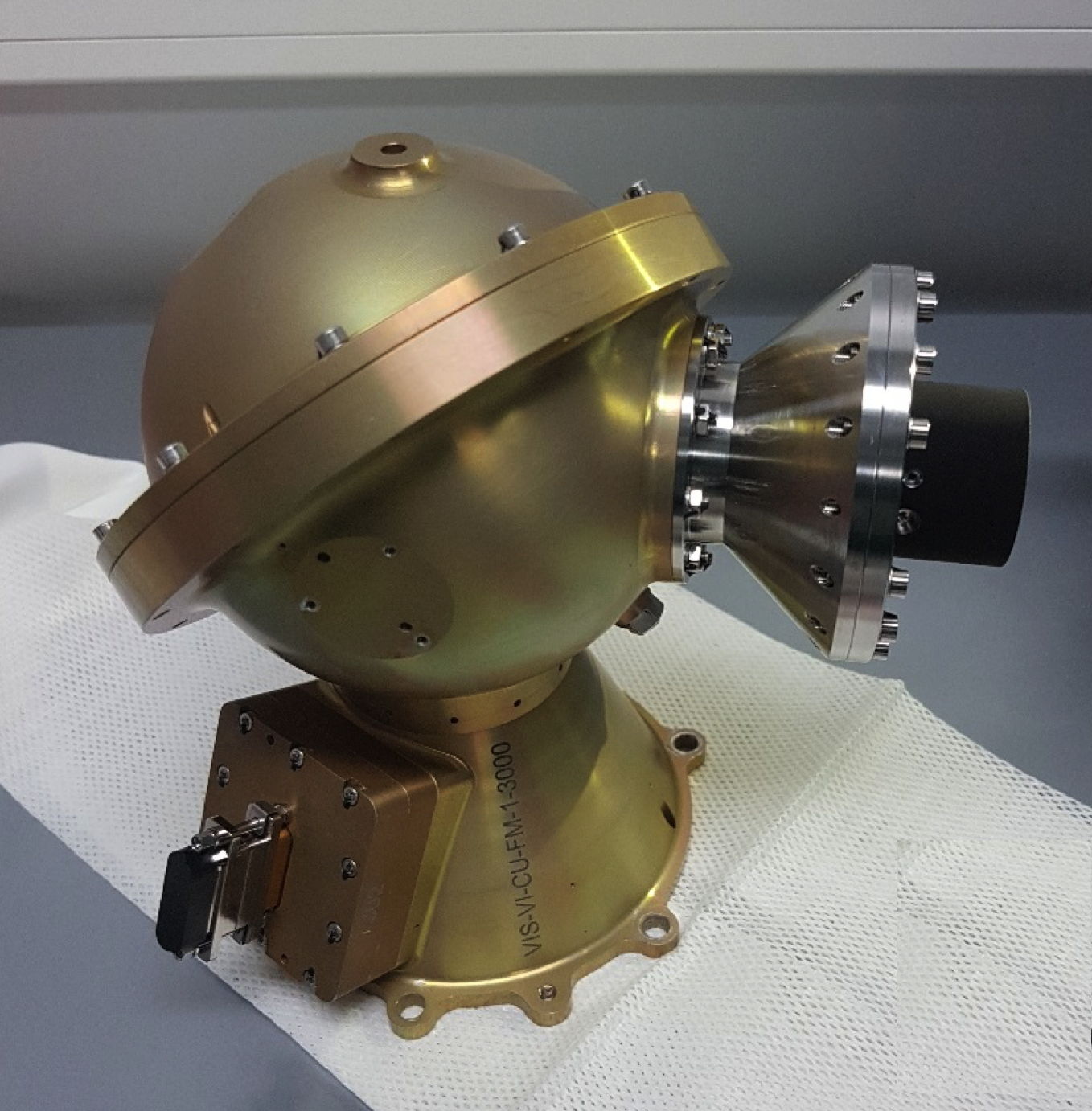}}
\caption{The Calibration Unit. The projection optics on the integrating sphere face to the right in this image.}
\label{fig:CU}
\end{figure}

In VIS the illumination is provided by the Calibration Unit. This is located across the Payload Module, facing VIS (see Fig.~\ref{fig:PLM}). The unit is shown in Fig.~\ref{fig:CU} and described in detail in \cite{CUDesign}. In order to record the illumination from the Calibration Unit, the Shutter must be open and hence the images from the telescope will also be recorded, and these must be eliminated in the data processing if necessary.

The Calibration Unit consists of an integrating sphere containing two redundant sets of six LED illumination sources covering the VIS passband (Fig.~\ref{fig:throughput} and Table~\ref{tab:cal}), and a single aspheric Fused Si lens to project the flux at the output port of the integrating sphere onto the VIS Focal Plane Array. The projected area is controlled by a field stop at the sphere output port and a baffle around the lens in order not to generate scattered light within the Payload Module or around the VIS focal plane itself. As the uniformity on larger spatial scales will be measured in orbit using photometry of stellar sources, the uniformity requirement is set for small and medium spatial scales, and the large-scale uniformity is controlled only to the extent that the exposure levels seen by each CCD are to be similar. Similarly, although in practice the emitted flux is expected to be relatively stable on short and medium timescales, the requirement on the temporal stability of the Calibration Unit is modest because the unit is not designed to provide a reference flux level. 

Each LED-type has different characteristics. As the LEDs age or are damaged by radiation, the flux levels can be adjusted by command.

\begin{table}[htbp!]
\caption{The central wavelengths and full width half-maximum (FWHM) for the six LEDs in the Calibration Unit.}
\begin{center}
\begin{tabular}{ccc}
\hline\hline
\noalign{\vskip 2pt}
& & Full width at \\
No. & Wavelength [nm] & half maximum [nm] \\
\noalign{\vskip 1pt}
\hline
\noalign{\vskip 2pt}
1 & 573 & 7.5 \\
2 & 592 & 8.1 \\
3 & 638 & 7.6 \\
4 & 697 & 17.2 \\
5 & 840 & 25.0 \\
6 & 855 & 34.4 \\
\noalign{\vskip 1pt}
\hline
\end{tabular}
\end{center}
\tablefoot{At the operational temperature of $146\,{\rm K}$ \citep{CUDesign}.}
\label{tab:cal}
\end{table}

Although primarily for measuring the pixel non-uniformity, the Calibration Unit will also be used for other calibrations, in particular the gain (the number of electrons per least significant bit of the digitised signal) for each channel, the dependence of the charge diffusion across pixel boundaries on the integrated flux level (the brighter-fatter effect), and the nonlinearity. More generally it is useful for generating the illumination required to measure the extent to which a pixel acts independently from its neighbours. To perform these calibrations, exposures are taken for different durations, or the emitted flux is controlled by the Payload and Mechanism Control Unit at different levels. The Unit initially included filter networks to inhibit currents in its wiring harness induced by the expected levels of electromagnetic fields in the satellite, but analysis established that these were negligible and the filters were removed.

\subsection{Control and Data Processing Unit}
\label{subsec:CDPU}

The Control and Data Processing Unit \citep{DiGiorgio2010,DiGiorgio:12,CDPUDesign} is located on the inside of one of the Service Module external wall panels, alongside the Power Mechanism and Control Unit (Sect.~\ref{subsec:PMCU}), as seen in Fig.~\ref{fig:SVM}. It has two tasks: firstly to control and monitor VIS and secondly to accept the science data from the 12 Readout Electronics in the VIS focal plane, reorder them to reconstruct the image, and compress and transfer it to the spacecraft. The Unit has redundant halves except for the connections to the 12 Readout Electronics which are accessed by both halves through multiplexers. The active half is selected by the spacecraft power, and this then powers the corresponding Power and Mechanism and Control Unit, Shutter motor windings, and Calibration Unit LEDs -- to avoid single points of failure there is no cross-strapping of these units.

The Control and Data Processing Unit is the primary VIS interface to the satellite, with the only exception being the raw power distributed directly to the 12 Readout Electronics Power Supplies (Fig.~\ref{fig:electronics}). A SpaceWire protocol is used for transferring science data to the Mass Memory Unit, while a Mil-Std-1553 bus provides the VIS interface for telemetry and telecommands to the spacecraft central computer; there are two such interfaces for each redundant half. Within the instrument, the Unit routes power to the Power and Mechanism Control Unit and interfaces with the Readout Electronics through SpaceWire and LVDS links as described in Sect.~\ref{subsubsec:ROE}. The internal architecture is arranged to service these interfaces, around a Maxwell SCS750 triple-redundant Central Processing Unit. The ECSS Packet Utilisation Standard\footnote{\href{https://https://ecss.nl/standard/ecss-e-70-41a-ground-systems-and-operations-telemetry-and-telecommand-packet-utilization/}{ECSS-E-ST-70-41A}} is the protocol used for data, telemetry, and telecommands on the digital links. Telemetry consists of status conditions, temperatures, voltages, and currents, and is taken at 10-, 5-, and 2-s intervals, depending on the cadence required for the ground monitoring.

The application software to handle the challenging data from such a large focal plane  is described in \cite{Galli:14,CDPU-SW}. Pixel data from each CCD quadrant are available as whole rows at the same time at their corresponding Readout Electronics, so the Control and Data Processing Unit accesses each Readout Electronics in sequence and reconstructs each CCD quadrant image in its memory. As they become available, the rows are compressed using a CCSDS-121 lossless compression scheme\footnote{\url{https://public.ccsds.org/Pubs/121x0b2ec1s.pdf}} with an additional option to reorder the incoming bit stream \citep{Giusi14:}. Image packets are then formatted and transferred to the Mass Memory Unit on the spacecraft as requested. These operations are time-constrained by the readout sequence and the requirement to free memory in time for the following exposure. Hence, during the readout period cyclic telemetry from the Readout Electronics is embedded in the image data packets. 

The boot software \citep{CDPU-HDSW} checks the integrity of the memory of the Control and Data Processing Unit and loads the application software. From here, the VIS state can transition to `Science' through intermediate states \citep{IOCD} which are tailored to provide checking at each stage, and fallback positions during operation, should these be necessary. The monitoring of some 2600 parameters representing the internal state of VIS is carried out by the Control and Data Processing Unit, and it takes action as appropriate. The control of the power to the Readout Electronics Power Supply Units is by request to the spacecraft central computer. The failure modes and criticalities have been analysed to ensure communication can be maintained with the spacecraft if at all possible, and to ensure the minimum thermal disturbance to the instrument and Payload Module consistent with the seriousness of the failure, which in many cases may be minor and recoverable. 

\subsection{Power and Mechanism Control Unit}
\label{subsec:PMCU}

The Power and Mechanism Control Unit \citep{PMCUDesign,PMCU} interfaces to the Shutter and the Calibration Unit to operate and monitor them, and is also connected to the Detector Plane Structure of the Focal Plane Array to monitor its temperature. It is a dual redundant unit, with each half interfacing without cross-strapping via SpaceWire and secondary power links to the associated half of the Control and Data Processing Unit (Fig.~\ref{fig:electronics}).  The Unit is accommodated alongside the Control and Data Processing Unit in the Service Module (Fig.~\ref{fig:SVM}).

Originally the Unit was designed to provide power to the Readout Electronics Power Supply Unit, but reliability considerations in the operation of latching current limiters resulted in their connections made directly to the Spacecraft (Sect.~\ref{subsubsec:ROE}), so this functionality was removed. Similarly, the simplification of the Shutter design to remove the launch Hold-Down Mechanism and the Fail-Safe Mechanism (Sect.~\ref{subsec:Shutter}) allowed the removal of circuitry for their actuation.

Twelve platinum thermistors allow temperature monitoring of the Detector Plane Structure of the Focal Plane Array with a resolution of $<0.1\,$K. The temperature is set by Payload Module heaters under spacecraft command.

The LEDs in the Calibration Unit are driven by current sources with 12-bit current and millisecond timing resolution. Care was taken to ensure that when commanded off, no residual currents will flow which might produce a low level of illumination.  Multiple LEDs can be driven at the same time. 

The Shutter stepper motor is also driven by current source power amplifiers which are controlled from trajectory tables stored in the Unit. These construct the acceleration, coast, and deceleration profiles for the Shutter opening and closing. The shutter can be driven either in closed loop until end-switches are actuated, or open loop in which the trajectory stops just short of the switches. The shutter orientation is assumed known from the step-count, and there is no direct monitoring of the Shutter leaf position -- this will anyway be evident from the CCD outputs on the Focal Plane Array. The orientation of the  leaf can be reset by a calibration operation which drives to both sets of end-switches.

\section{VIS assembly, integration, and testing}
\label{sec:AIT}

The VIS instrument development programme followed the standard ESA model as described in \cite{EID-A} with a set of prototypes, followed after a Preliminary Design Review by mechanically and thermally representative structure Thermal Models and functionally representative Engineering Models. Flight-representative Qualification Models were also produced for the Detector Chains, the Power and Mechanism Control Unit, and the Calibration Unit. All of these  required environmental testing consisting of vibration and shock tests, thermal vacuum tests during which the thermal behaviour of the unit was measured, and electromagnetic susceptibility and emissivity tests. The test results were compared to the predictions from the mathematical models to check for compliance. The model units were assembled into a partial VIS with some unit simulators and then used by ESA and industry for their proving of the equivalent stages of the programme for the satellite. 

In order to pursue this programme -- and this applied also to the unit-level assembly and test -- a suite of ground support equipment and associated software was developed, and facilities created or modified for \Euclid needs. In its entirety, this was in itself a major development programme. Two identical subsystem checkout equipments providing representative spacecraft interfaces were supplied by ESA for VIS so that there would be no late-stage incompatibilities with the spacecraft subsystems, and the instrument workstations were integrated with this system.

After the Critical Design Review ending on 2018 February 6, approval was given to commence with the Flight Model programme, which also included some Flight Spare units or parts. On the completion of the production and testing of the five VIS units, two of which reside in the Service Module and three in the Payload Module, VIS was required to be assembled as an entire instrument, with its substantial Flight Model harnesses, and then tested and verified to show that it meets its requirements, as described briefly in Sect.~\ref{sec:rationale}. The distributed nature of VIS allowed this process to be planned in stages to be schedule efficient. The Focal Plane Array assembly consisted of a two-stage integration, firstly of the detector chains on the one hand and the structure on the other, and then the integration of both of these to provide the final Focal Plane Array. Hence this was pursued in one integration flow, with the support of a flight-representative engineering model Control and Data Processing Unit. The other four units were integrated in a second integration flow. Firstly the Power and Mechanism Control Unit was tested with the Calibration Unit, then the Control and Data Processing Unit was added and finally the Shutter, with engineering or qualification models to substitute while the Flight Model units were not available. 

Besides being schedule efficient, these two flows allowed the early identification and addressing of any issues before the final integration to complete the instrument. Indeed, the principle of testing at the lowest applicable level in order to catch problems early was followed throughout. In particular there were no thermal vacuum tests or vibration tests at VIS level as its constituent units had already been tested, and the only additions were the harnesses. Also, while there was a case for electromagnetic susceptibility and emissivity tests at VIS level, the individual units had already been tested and there was little additional information to be had until they were assembled within the spacecraft with the appropriate grounding arrangements. Budgeting of the test cycles and testing levels prevented over-testing from the accumulation of tests at each stage. This verification programme also took into account that the permitted number of actuations of the Shutter at ambient pressure was limited by its MoS$_{2}$ lubrication.

At VIS level, the testing was limited to functional and long-duration testing at Airbus in Toulouse \citep{Candini:20a,Candini:20b}. This period lasted two months during which difficulties were encountered both within VIS and the spacecraft interface equipment -- which was not fully representative -- as well as the testing environment (such as the electrical earthing). Lessons learned included longer prior coupling tests between individual units to identify infrequent communication problems, a higher level of capability in the VIS Instrument Workstation, and a larger integration team. However, by 2020 February 28 the communications with the Spacecraft Interface were fully secured, the testing programme completed and the instrument handed over to ESA and Airbus.

\begin{figure*}[htbp!]
\center{\includegraphics[width=0.985\columnwidth] {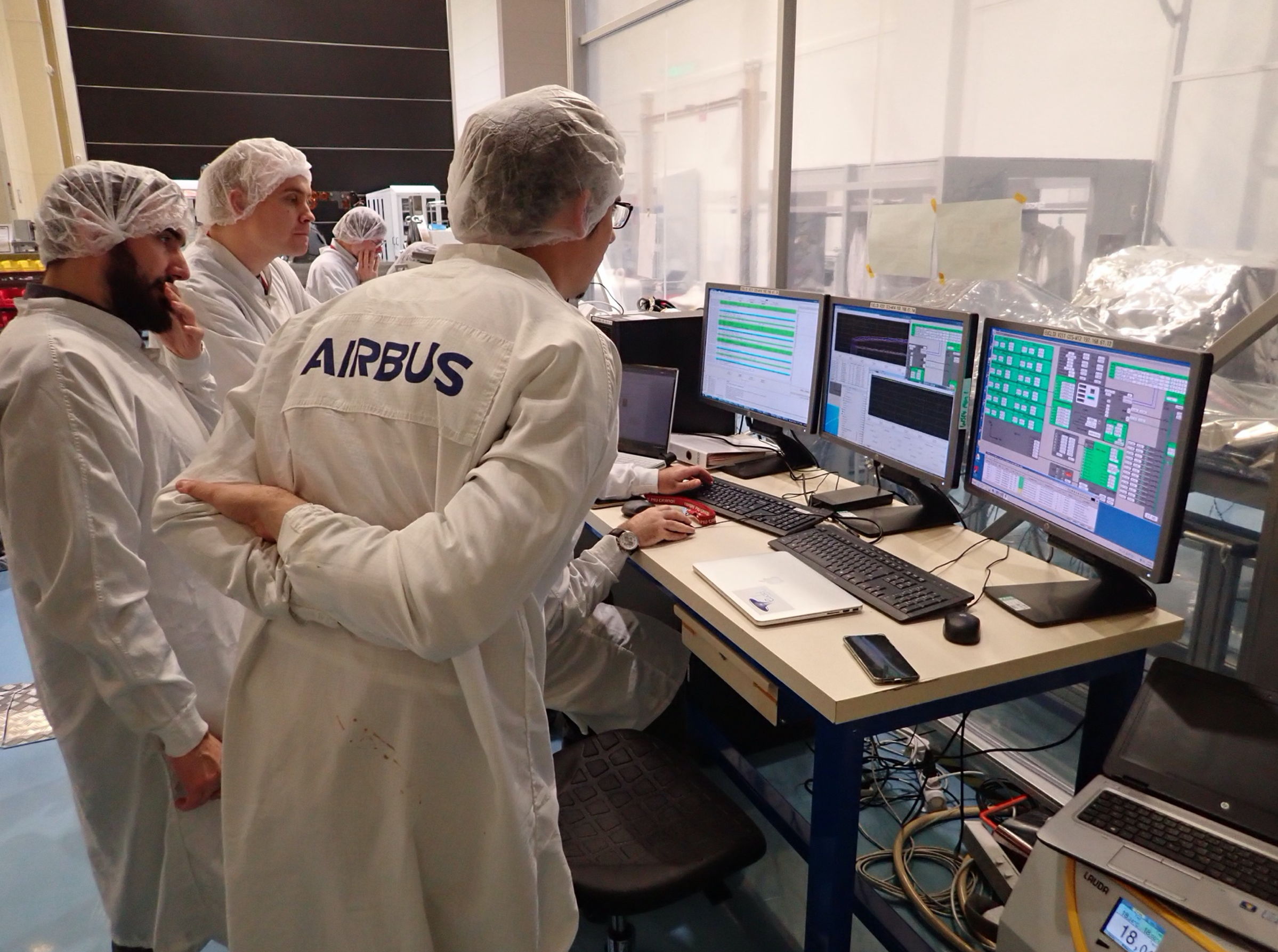}\hspace*{2mm}\includegraphics[width=0.98\columnwidth] {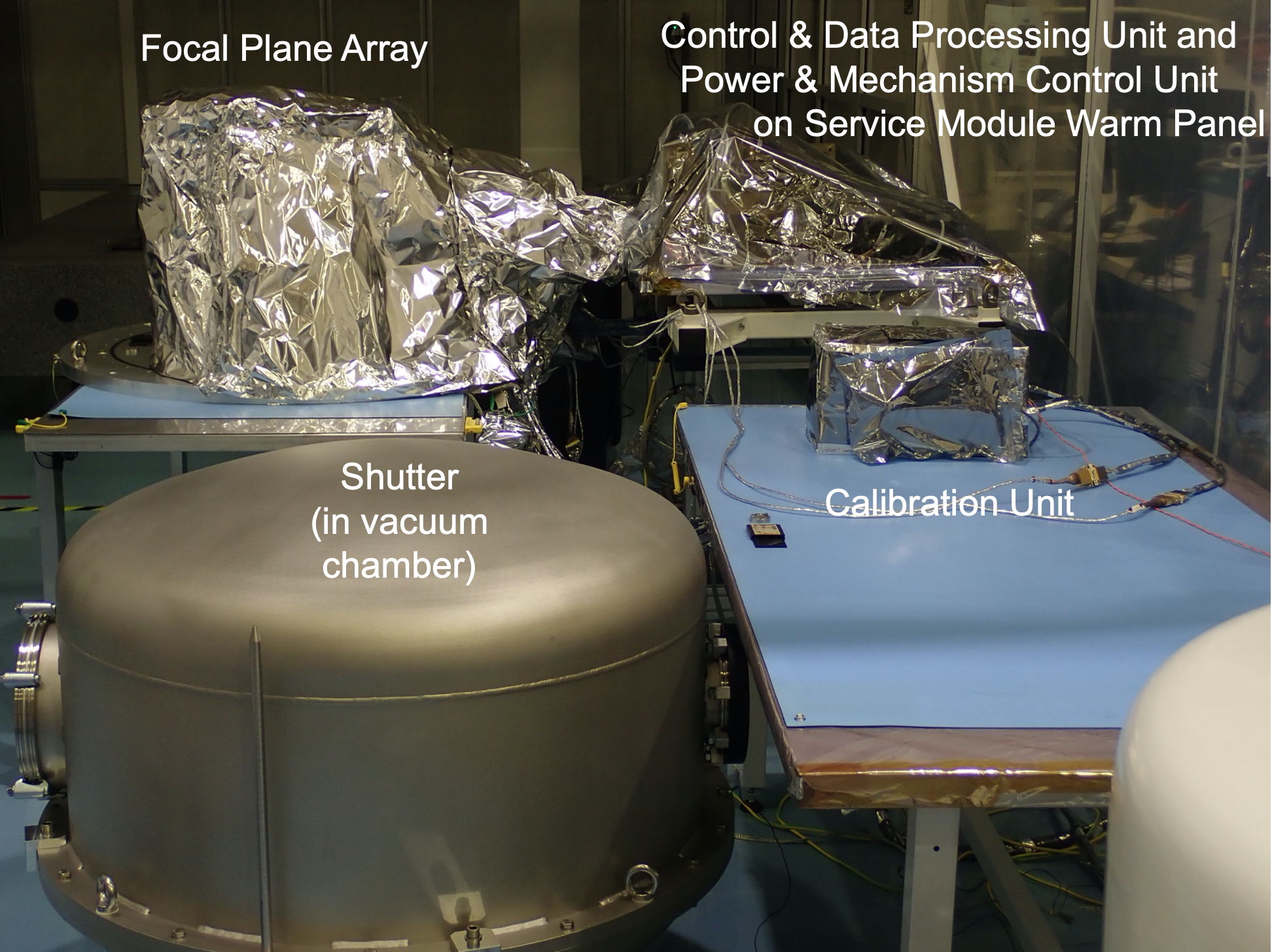}}
\caption{VIS Flight Model during final testing. This was carried out in an ISO-5 clean area in the \Euclid facilities at Airbus (Toulouse); the majority of the \Euclid Payload Module is under integration behind the large black shutter. {\it Left}: The team inspecting the progress of the Full Functional Test. VIS can be seen foil-covered behind the computer screens. {\it Right}: VIS as laid out within the ISO-5 area. The foil-covered Focal Plane Array is in the background connected via 26 harnesses to the Control and Data Processing Unit which is with the Power and Mechanism Control Unit already integrated onto the Service Module Warm Panel (Fig.~\ref{fig:SVM}). In the foreground are the foil-covered Calibration Unit and the Shutter in a vacuum chamber. The foil coverings are in place to maintain the stringent levels of cleanliness for VIS, and the vacuum chamber is required to preserve the MoS$_{2}$ lubricant during Shutter activation.}
\label{fig:VIS-FM}
\end{figure*}

Figure~\ref{fig:VIS-FM} shows the testing in progress. Under the principle above, it can be seen that the whole Focal Plane Array including the detectors is at ambient, and no optical stimulation is used to check the instrument, this having being measured at block- and Focal Plane Array-level. Instead, thermally generated electrons within the CCDs can be used to check the end-to-end performance, and the readout noise performance can be measured by reverse-clocking the CCDs. By this point, the application software in the Control and Data Processing Unit included the procedures to manage failure modes and out-of-limit parameters from the instrument.

Beyond the environmental and functional testing, the scientific performance and calibrations are described in Sect.~\ref{sec:performance}.

\section{VIS operation}
\label{sec:operation}

\begin{figure}[htbp!]
\center{\includegraphics[width=0.95\columnwidth] {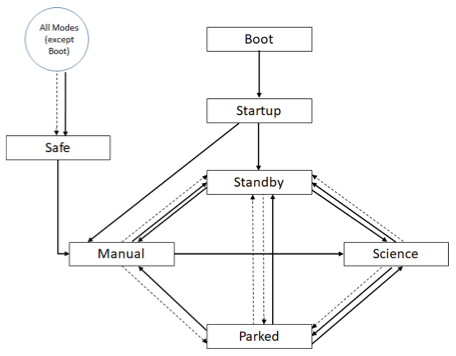}}
\caption{VIS instrument modes, showing permitted mode transitions. Solid lines denote nominal transitions while dotted ones can be triggered by a failure mode. All science and calibration exposures are taken in Science Mode.}
\label{fig:Modes}
\end{figure}

Once VIS is switched on, it boots itself and waits for a command to Start-up and proceed to Standby. From there it can be commanded to  Science Mode (Fig.~\ref{fig:Modes}) for all of the normal operations. For engineering and low-level investigations it can instead move to Manual Mode. If a failure condition that can affect the instrument safety is triggered in any mode, VIS puts itself in Safe Mode with only the Control and Data Processing Unit powered; from here it can transition back to Manual Mode on command. Because it is important to minimise thermal disturbances in the Payload Module, so that normal operations can be resumed after an anomaly as soon as possible, there is also a Parked Mode. This is a holding mode to be used for maintaining the power to the Focal Plane Array when it is safe to do so, for example when other activities in the Spacecraft require a pause in VIS operations, or when there are VIS failures such as an incorrect Shutter position. The instrument state is left unchanged on transition to this mode. In Standby Mode the Focal Plane Array is off; we reiterate that the Focal Plane Array power is controlled directly from the Spacecraft on request from the Control and Data Processing Unit. These modes, and  other aspects of the operation of VIS are detailed in \cite{IOCD} and \cite{UMan}.

The Science Mode operations allow seven types of scientific operations, including normal science exposures; bias exposures to measure the electrical signal corresponding to no illumination; dark exposures which measure the dark current in the CCDs; flat-field exposures using the Calibration Unit for illumination for several purposes but mainly to measure the pixel-to-pixel non-uniformity; linearity exposures to quantify the nonlinearity in the CCD output node and Readout Electronics; and charge injection and trap pumping exposures for calibrating the radiation damage. In addition, to mitigate the effects of radiation damage at the expense of a small loss of detector area, it is possible to inject charge into the normal science exposures: on readout of the CCD these lines of charge fill the traps caused by radiation damage, so that the measured shape of the galaxies is less distorted. 

VIS and NISP will operate together to perform the \Euclid survey, so their operation must be synchronised \citep{ROS} with the consequence of some compromises in optimal observational efficiency. Because NISP takes both spectroscopic and photometric exposures of a field, the sequence is arranged such that VIS exposes at the same time as the NISP spectroscopic exposures, nominally for 566\,s, and then during the shorter NISP photometric exposures VIS takes some shorter science exposures to increase the dynamic range of the images for PSF calibration and performs calibrations of the non-science exposure types listed above. While this latter reduces the fraction of time that VIS is carrying out science exposures to approximately 0.8, the calibrations will be essential in establishing the state of the instrument for each exposure. Further constraints on the synchronisation occur because the CCDs take 72\,s to read out, because the Control and Data Processing Unit must compress and transfer images before the next image is ready, and because there can be no VIS exposure during the operation of the NISP mechanisms owing to the disturbance in the Spacecraft pointing. 

Each of the above sequences is repeated four times with 50--100\,arcsec displacements in an S-shaped pattern to expose those parts of the field which fall in the gaps of the detectors on both NISP and VIS, whose fields cover the same extent. The S-shaped pattern allows 95\% of pixels to be exposed three or four times which minimises the spurious signal introduced into the shape measurements from the exposure mask. Thereafter, there is a larger displacement repointing of the satellite to a new field. These repointings are carefully designed in the \Euclid survey \citep{Scaramella:22} to minimise the change in the orientation of the satellite to Solar radiation, and hence thermal variability in the PSF from the telescope, but at the \Euclid level of required accuracy this still requires complex and challenging modelling, as noted at the start of Sect.~\ref{sec:design}.

While calibrations for weak lensing are mostly derived from the science data itself, in particular for the PSF modelling which uses stellar point sources within the field of view, or from the regular calibrations taken during the NISP photometric exposures, some specific calibrations are required at typical intervals of a month -- the linearity exposures are a case in point -- and these are accommodated in the survey planning.

\section{VIS performance}
\label{sec:performance}

In terms of the requirements set on VIS discussed in Sect.~\ref{sec:rationale}, these are tracked in detail and are available in \cite{VCD}. No major non-conformance is recorded. 

In this section we rather discuss in broad terms the anticipated prelaunch and immediate post-launch commissioning performance of the VIS instrument and the weak-lensing channel as a whole, insofar as it affects weak-lensing measurements. As noted in Sect.~\ref{sec:rationale}, the various aspects of NISP performance, the external photometry, and the \Euclid Science Ground Segment are also critical for the overall performance of the weak-lensing probe.

\subsection{VIS ground calibrations}
\label{subsec:GRCAL}

The primary aspects of VIS prelaunch performance were obtained through ground calibrations at block level (three CCDs, a Readout Electronics and its Power Supply Unit; Sect.~\ref{subsubsec:block}). Comprehensive reports of these activities are available in \cite{Azzollini:21} and \cite{Skottfelt:21}, with a summary here below. Limited testing was also carried out at operational temperature with all 12 blocks in the Focal Plane Array to check and confirm the results of the individual block calibrations, and these are briefly discussed in Sect.~\ref{subsec:FPACAL}, and in more detail in \cite{Azzollini:19}. Further calibrations after instrument delivery were obtained at Payload Module and satellite level with the telescope at operational temperature, and these are described in Sect.~\ref{subsec:PLM_testing}.

\subsubsection{CCD performance}
\label{subsubsec:CCD_performance}

The CCD273-84 performance is excellent, with the overall sensitivity well within specification, as noted in Fig.~\ref{fig:throughput}. The pixel-to-pixel sensitivity variation (the photon response non-uniformity) is also typically half of the maximum allowed by the specification. Deep flat-field exposures reveal at a low level the presence of `tree rings' which result from the cooling of the raw Si boule during refining. There is no specification on these and they may be sensitivity enhancements which can be addressed through the photo-response non-uniformity correction in the data processing, and/or the result of the very slight displacement of pixel boundaries.\footnote{In VIS their amplitude is much smaller than those reported in \cite{Plazas:14} and \cite{Park:17} for the CCDs of the Dark Energy Camera DECam and in the LSST Camera at the Vera Rubin Observatory, respectively. \cite{Okura:15} conclude that they will not have a significant impact on the LSST cosmological analysis.} Cosmetic defects, specifically dark and bright pixels, and bright columns, are within what was budgeted. Dark current at operating temperature is negligible. Pixel full well capacities, which set the dynamic range, are also above the minimum specification of 175\,000 e$^-$. Charge Transfer Efficiency, which measures the fraction of charge transferred from one pixel to the next during readout, is within specification in the absence of radiation damage -- appropriate to the start of the mission -- both for the transfer of rows, and, in the readout register, for the transfer to the readout node. With the correct voltage inversion of the Si substrate immediately before the exposure begins, persistence effects from bright sources are negligible.

The size and ellipticity of the PSF resulting from charge diffusion in a CCD pixel is also within specification and here the PSF size decreases with wavelength \citep{Niemi:15a}. This is because Si is more transparent to longer wavelengths, and therefore absorbs red photons closer to the electrode structure within the pixel, so they have less opportunity than blue photons to migrate laterally across pixel boundaries. This counteracts to some extent the PSF wavelength dependence of the optics.

\subsubsection{Detector chain performance}
\label{subsubsec:Detector_chain_performance}

Once connected to the Readout Electronics, the complete detector chain is compliant with the readout noise at $\leq4.5$\,e$^-$, given the gain of 3.5\,e$^-$ per least significant digital bit \citep{Szafraniec:16}. The gain is higher than the designed-for 3.3\,e$^-$ value, and results in a slightly coarser digitisation of the readout noise, but has the advantage of allowing the higher dynamic range in the flight batch of CCDs to be exploited. 

If a linear ramp of increasing voltage is applied to an analogue-to-digital converter, the digital number on the output should rise in steps in direct proportion. However, in practice, the step sizes are not uniform at the least significant bit level and all bits are affected. This is a digitisation noise sometimes called differential nonlinearity. The differential nonlinearity performance of the analogue-to-digital converter in the Readout Electronics was characterised during the on-ground testing \citep{Candini:20} for eleven\footnote{Owing to the novelty of this test, the procedure was not validated in time for the testing of the first block.} of the 12 blocks and two spares. This noise source constitutes a large fraction of the detector chain readout noise. However, there are patterns in the behaviour which may cause difficulties in background subtraction, although this is not expected given the dominance of the optical background. It is not clear yet how the characterisation that was carried out can be used to reduce the noise, but the calibration data are archived.

As the analogue-to-digital converter in the VIS Readout Electronics is seen in some instruments to display a correlated behaviour in pixel values from two conversions earlier or later \citep{Boone:18}, this was checked for VIS but found to be not present.

The electronic bias measured with the CCD connected is about 9200\,e$^-$ compared to the roughly 2000\,e$^-$ with a simulated CCD output connected. This increase was seen also in \Gaia and is understood as a characteristic of the CCD summing well, which is located immediately prior to the readout node; it proved not to be possible to avoid. The bias stability is therefore likely to be dominated by this effect, nevertheless under stable conditions it is constant to within 1 e$^{-}$ \citep{Liebing:21}.

The transient response from high to low signal levels has been found to be well-controlled and slightly underdamped on average, with overshoots or undershoots of the subsequent pixel at a fractional level of between $-3\times10^{-5}$ and $3.5\times10^{-4}$ for all 144 channels. This is important: if for example the response is slow, a tail will be evident trailing the image and this affects the shape measurement. These trails can be separated from tails arising from radiation damage trapping because the amplitude of these will not be affected by the distance of the source from the readout register.

Despite careful layout of the electronics components, electronic cross-talk within the Readout Electronics produces ghosts in adjoining channels. This is difficult to avoid, and just meets the very tight specification. The ghost is at the level of $\leq 5\times10^{-4}$ with the worst crosstalk occurring between the pairs of channels located back-to-back on opposite sides of the circuit board (Fig.~\ref{fig:ROE+RPSU}). Crosstalk from other channels is typically a factor of 3--10 lower than this. The effect on weak lensing was considered by \cite{Cross-talk} including the contribution from cosmic rays and bright stars, and, with the loss of a small fraction of discarded pixels, judged acceptable. In the ground calibrations, the repeatability at different wavelengths using optical measurements was high, as was the repeatability between measurements carried out on different blocks of Readout Electronics.

\subsubsection{Linearity}
\label{subsubsec:linearity}

 CCDs are linear in that their Si will generate photo-electrons at a rate linearly proportional to the incoming photon flux. However the degree to which the generated photo-electrons drift to adjoining pixels depends on the charge already accumulated in the pixel (the brighter-fatter effect). The PSF width therefore grows with increased charge within a pixel \citep{Niemi:15a}, as the photo-electrons generated later in an exposure experience a reducing pixel barrier caused by the accumulated charge from the electrons already in the pixel. 
 
 Although the total flux in, for example, a stellar image, is conserved in this effect, at the pixel level flux will have been transferred preferentially from brighter pixels to fainter ones, resulting in a nonlinear response for individual pixels. Subsequently, as noted in Sect.~\ref{subsubsec:detectors}, during transfer for readout, some trapping of electrons will occur as a result of lattice damage from ions, and the likelihood of this occurring depends on the size of the charge packet being transferred, with low levels of charge proportionately more affected. This is a second source of nonlinearity. 

A third source is the nonlinearity of the electronics chain, from the output node of the CCD to the analogue-to-digital converter. This was measured at Readout Electronics block-level on-ground, using calibrated shutters and light sources as less than 2.5\% for values above 25\,000 e$^-$, but rising to in excess of 10\% as the signal approached zero. There is no requirement on this nonlinearity itself, but rather on the knowledge of the nonlinearity (i.e. on the difference between the measured value and a model fit) at $\leq6\times10^{-4}$ at the end of mission \citep{PERD}, and on its stability. At low flux levels, this requirement is not appropriate, and potentially should have been couched in absolute, rather than relative terms. \cite{Azzollini:21} considers that the on-ground measurements are likely to be inaccurate, but the reason is not understood.

\subsubsection{Calibrations for charge-transfer efficiency}
\label{subsec:CTI}

The ground calibration measured the performance of the calibration methods to be used to quantify and correct for radiation damage effects, viz\@. charge injection and trap pumping. Charge-injection lines are similar over all CCDs and Readout Electronics but with different levels for the top and bottom CCD halves, because they are not mirror-imaged about the charge injection structure, as noted in Sect.~\ref{subsubsec:CCDs}. The charge levels along the CCD rows are constant within 20\% which is consistent with the specification -- some variation is desirable to constrain better the modelling characterising the radiation damage in the data analysis. The trap pumping sequences ran successfully \citep{Skottfelt:21}, demonstrating that trap species could be identified in the parallel pumping given the 4-phase electrode structure of the pixels, and in the serial pumping of the 3-phase readout register. Because too few traps are evident in the serial registers not yet affected by radiation, this was limited to the extent sufficient to prove the principle. Recommendations were made for minor improvements. 

\subsection{Performance of the integrated Focal Plane Assembly}
\label{subsec:FPACAL}

While the reference calibration for VIS was carried out at block level as described above, limited tests were also carried out with the full focal plane to confirm the earlier measurements, and specifically to check for any electromagnetic interference evident in the readout noise -- this should be negligible because of the synchronisation of all channel readouts. 

The tests were carried out during the thermal vacuum testing of the Focal Plane Array without optical stimulation. The tests included bias, charge injection and dark current at nominal operating temperature ($153\,{\rm K}$ for the CCDs and $270\,{\rm K}$ for the readout electronics), and at the extremes of the operational temperature range (145--$165\,{\rm K}$ and 245--$292\,{\rm K}$ respectively). These confirmed the operation of all 144 channels and allowed measurements of the transient response, readout noise, bias levels (and uniformity), charge injection levels, and dark noise. The results were consistent with the block level testing, with slightly better readout noise, in the range 2.17--4.06 e$^-$. 

\subsection{Payload Module and satellite-level testing}
\label{subsec:PLM_testing}

VIS was delivered to ESA in early 2020 for integration into the Payload Module (Fig.~\ref{fig:PLM}). The Assembly, Integration, and Test programme culminated in a 2-month thermal-vacuum test with optical stimulation of the instruments in mid-2021. This testing established the performance of the telescope, the relative alignment of the VIS and NISP instruments, their performance and that of the Fine Guidance Sensor, and the level of interference between these payload elements. This was the first test at operational temperatures and pressures for the full optical train up to and including VIS. Additionally, this was the first check on the alignment between the VIS Focal Plane Assembly, the Shutter, and the Calibration Unit.

A description of the telescope and optical system can be found in \cite{Venancio:14}. The predicted overall end-of-life throughput of the weak-lensing channel is slightly above requirements, owing to a better-than-expected quantum efficiency of the \Euclid CCDs offsetting a reduced optical performance consequent to a change in the third Fold Mirror coating from a multilayer dielectric to protected Ag in order to simplify the PSF modelling. The throughput (Fig.~\ref{fig:throughput}) peaks at 0.70 at $700\,{\rm nm}$, and is above the specification of 0.65 through the range 550--$750\,{\rm nm}$. Beyond this, both the predicted and required throughput fall roughly together to 0.35 at $900\,{\rm nm}$. 

A report on the telescope performance during the Payload-level Testing is in \cite{Venancio:23}. The PSF produced by the telescope optics is within specification in terms of size and ellipticity in both the static and dynamic cases. The static case refers to the fully relaxed optical system, and the dynamic case refers to the worst case performance over 11\,000\,s for a thermal perturbation caused by a defined reorientation of the spacecraft. The size is set in terms of an $R^2$ parameter defined in \cite{Massey:13} and \cite{SciRD} which weights both the core and wings of the PSF, and is essentially diffraction-limited at 800\,nm, so that its dependence is approximately proportional to wavelength.  

The rejection levels for out-of-band wavelengths (also in Fig.~\ref{fig:throughput}) are an order of magnitude or two worse than specified at certain wavelengths, again as a result of the change in the third Fold Mirror coating. This spectral leakage may cause difficulties, particularly for the stars used for the PSF modelling. This leakage also impacts the optical ghost rejection ratio, as this light at leaked wavelengths adds to the ghost more significantly than it does to the normal image. 

The scattered light from the optics depends on the optical microroughness and the optical element cleanliness, which has been the subject of substantial effort at mission level and is expected to be within specification for the weak-lensing channel. Given that the fasteners for the beams holding the CCDs in the Detector Plane Structure of the Focal Plane Array are covered with small top-hat baffles, scattered light within VIS is limited to the scattering or diffraction from these and the Shutter leaf as it opens and closes. This is expected to be negligible. Nevertheless, scattered light in optical systems at the level demanded by \Euclid is challenging to control, and is often a significant source of performance degradation.  

For the first time the optical ghosts from the dichroic were characterised. The ghost rejection is $5\times10^{-6}$ instead of $2\times10^{-6}$ owing to the Ag replacement coating on the third Fold Mirror. This means that optical ghosts will be evident in standard 566\,s exposures at the 1 e$^-$ level for stars $m_{\rm AB}\leq19$, for which there will typically be $\sim\,$100 per CCD.

\begin{figure*}
\vspace*{-7mm}
\hspace*{10mm}
\includegraphics[width=1.55\columnwidth] {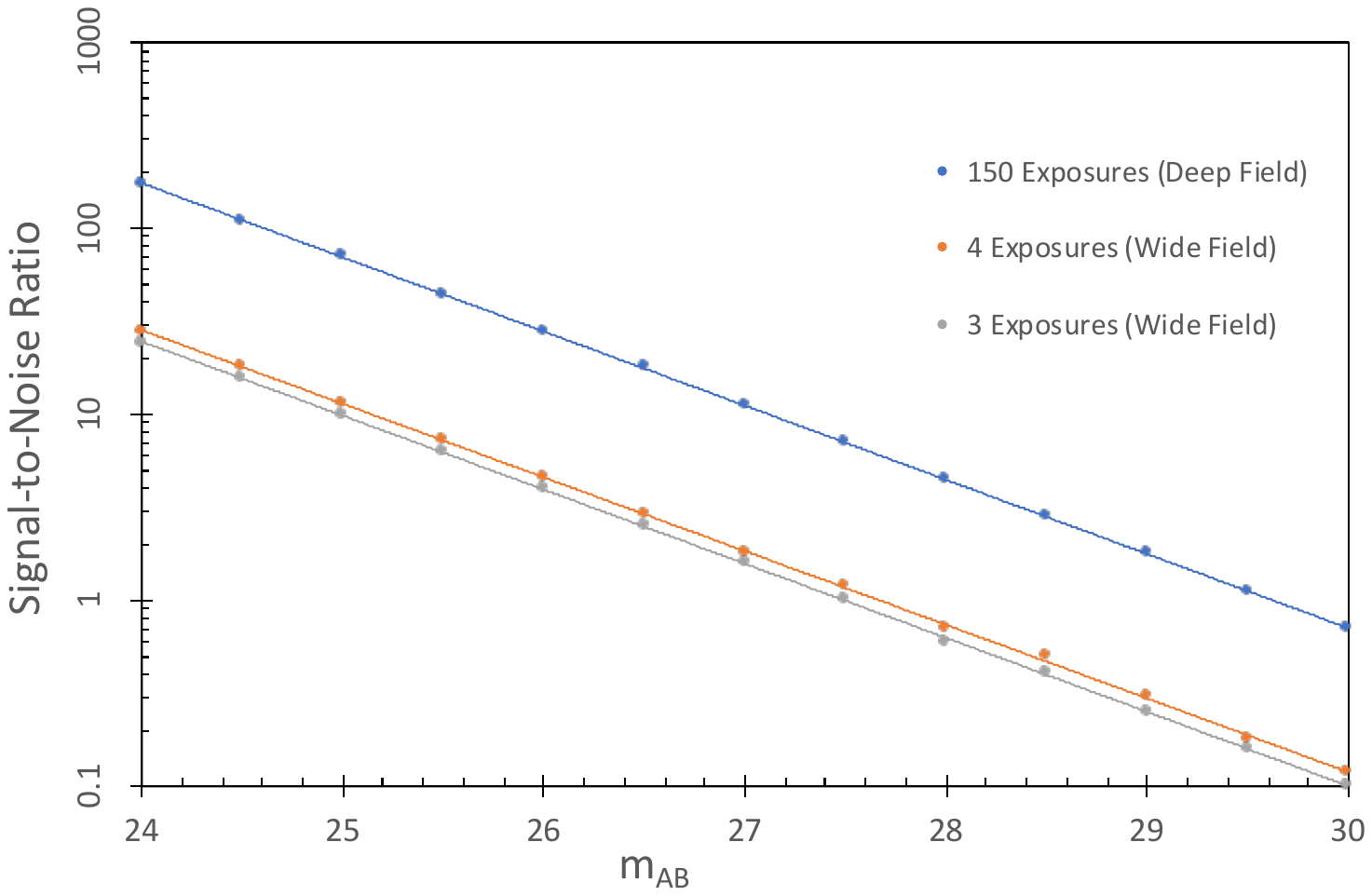}
\vspace*{-10mm}
\caption{Current best estimate of the S/N assuming average levels of zodiacal light background for the Wide (3 or 4 exposures) and Deep Survey (assuming 40 visits to the field; 150 exposures).}
\label{fig:SNratio}
\end{figure*}

\begin{figure}
\center{\includegraphics[width=0.97\columnwidth] {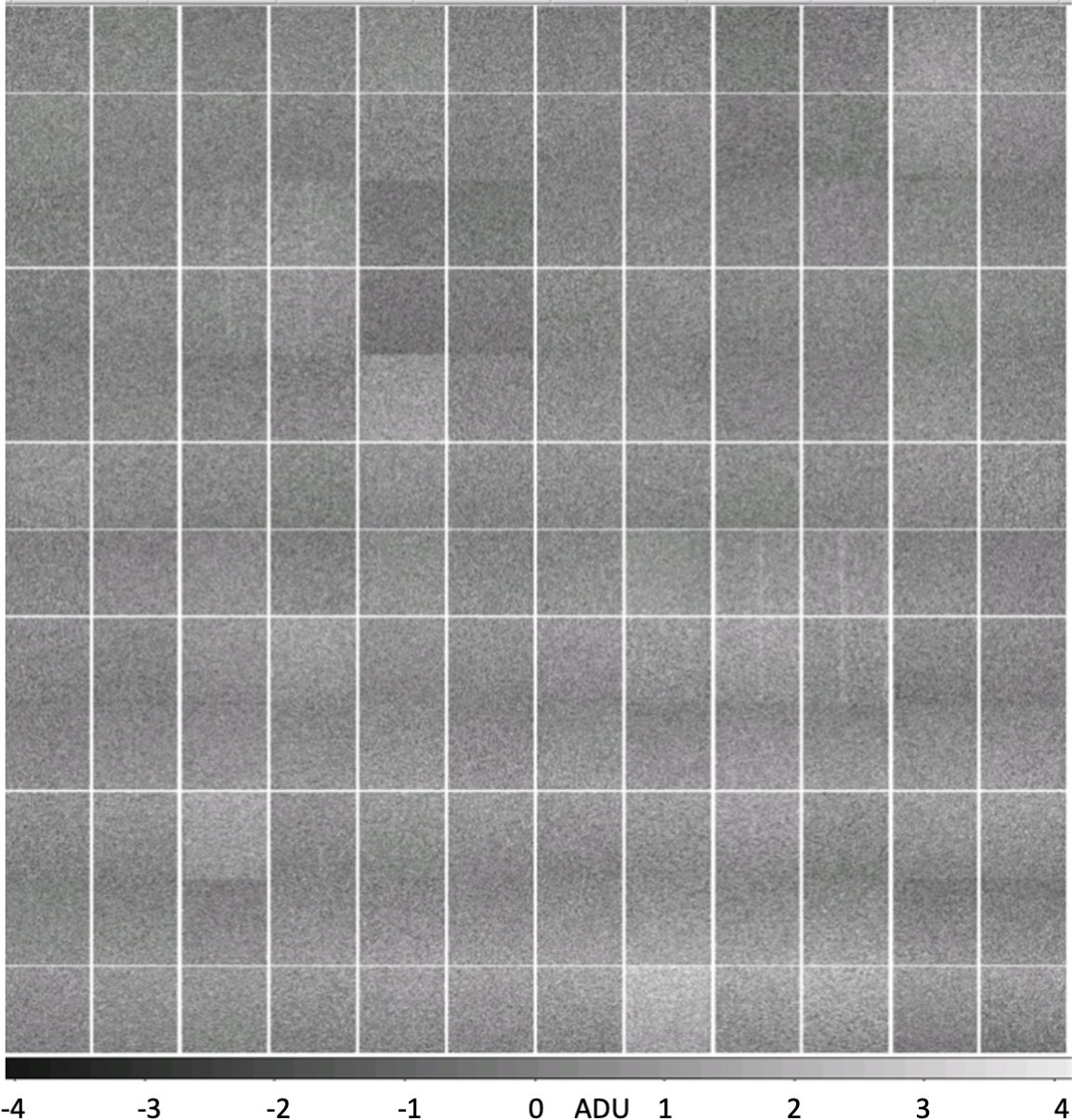}

\includegraphics[width=0.97\columnwidth] {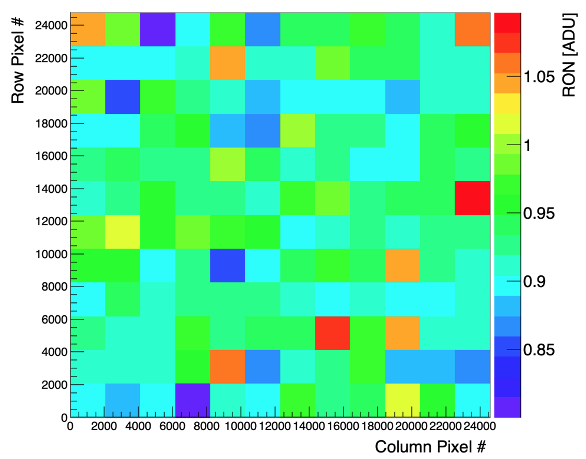}}
\caption{Payload-level testing. {\it Top}: Bias image of the full 144 quadrants in the Focal Plane Array after processing to subtract mean levels. The greyscale bar is in analogue-digital units (ADUs) where the ratio of ADUs to electrons is the gain, so $1\,{\rm ADU} = 3.5\,{\rm e}^-$. {\it Bottom}: The readout noise per quadrant in the bias image for all 144 quadrants. Again, the colour bar is in ADU, so the readout noise in the worst quadrants is $3.8\,{\rm e}^-$, within the specification of $4.5\,{\rm e}^-$.}
\label{fig:bias-noise}
\end{figure}

\begin{figure}
\centering
\includegraphics[width=0.97\columnwidth] {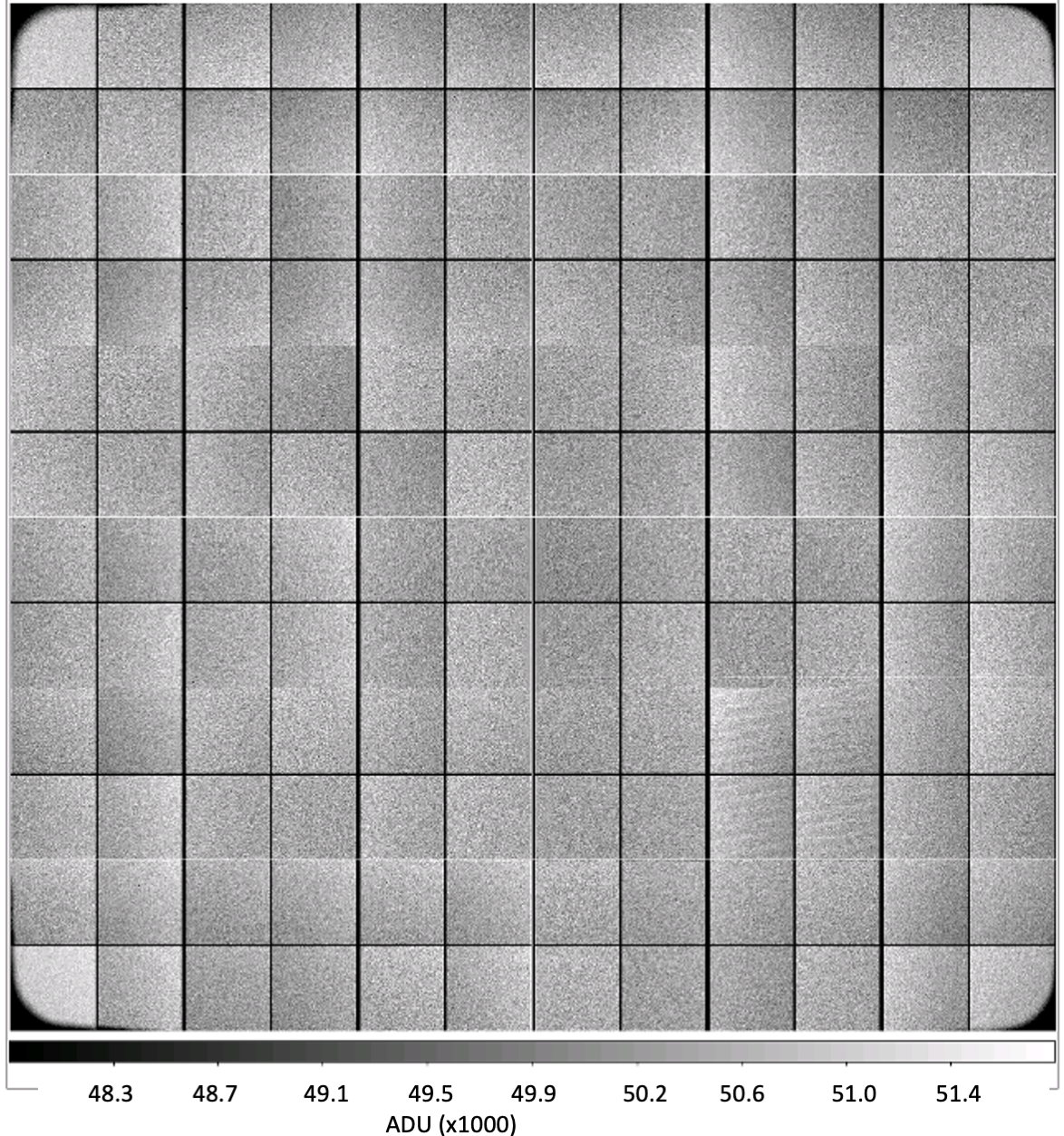}

\includegraphics[width=0.97\columnwidth] {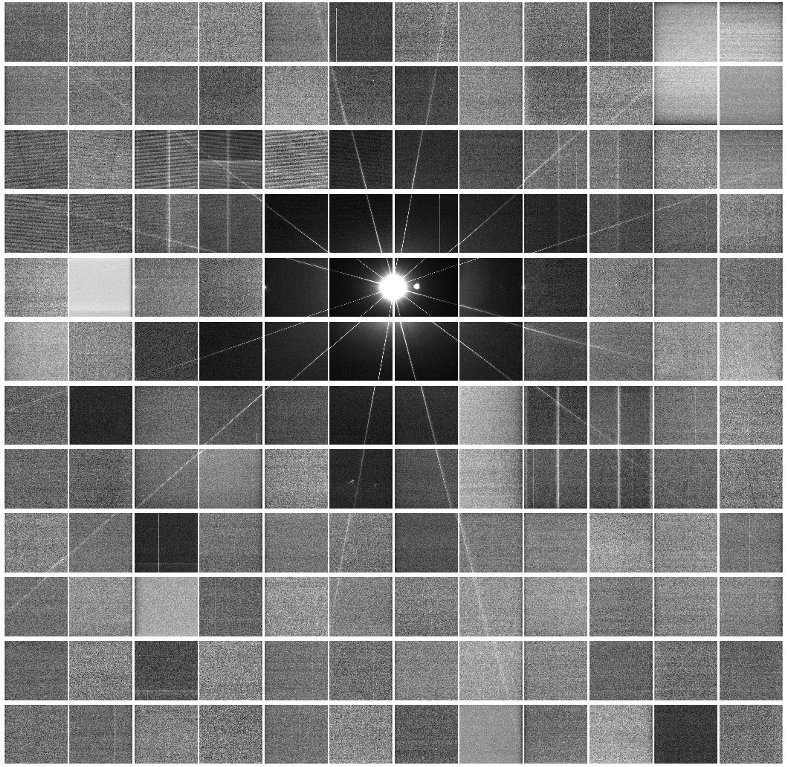}
\caption{Payload-level testing. {\it Top}: Flat-field image from the Calibration Unit at 720\,nm taken during the Payload-level testing after bias subtraction and gain correction. The reduced flux at the corners of the focal plane is as expected from the Calibration Unit design. The contrast on this image was set very high, as is evident from the greyscale bar below the image. It also indicates the correct orientation of the unit on the Payload Module baseplate. {\it Bottom}: The image of a bright source directed into the gap between CCDs near the centre of the focal plane. The 12 diffraction spikes (six each from the collimator and telescope) confirm the correct orientation and registration of the image. Also evident in this image are the faint vertical stripes on some CCDs which result from the amplifier glow in the readout register, at a level of less than 1 ADU (1 ADU $=3.5\,{\rm e}^-$) and a ghost image from the dichroic to the right of the bright source. The greyscale for each quadrant is individually calculated in this image.}
\label{fig:flat-point}
\end{figure}

Overall, the performance of VIS was consistent with or better than that from the Block-level and Focal Plane Array testing, given that their longer duration provided greater stability.  An overview is available in \cite{Cropper:21}, with more extensive analyses in \cite{Liebing:21}. Examples are shown in Figs.~\ref{fig:bias-noise} and \ref{fig:flat-point}. The long campaign allowed the stability of the instrument to be assessed with positive results. Diffraction spikes from point source images allowed the orientation and arrangement of the 144 quadrants to be confirmed. Flatfields, for the first time taken with the Calibration Unit, were as expected. Importantly, diffracted- and scattered-light levels from the shutter edge and from the Focal Plane structure supporting the detectors were negligible; also, electrical and optical interference from NISP and the Fine Guidance Sensor was undetectable. With respect to the full optical train including VIS, the characteristics of point sources from the collimator in the chamber were as specified, falling largely within a pixel; however, we note that the PSF spatial sampling becomes almost Nyquist when the spacecraft pointing error is added. VIS is expected to saturate on point sources for $m_{\rm AB}\leq18.0$ in a typical 566\,s science exposure and reach S/N $=5$ for $m_{\rm AB}=26.3$ in three exposures and $m_{\rm AB}=26.5$ in four for a \ang{;;0.6} diameter aperture with typical zodiacal light backgrounds (the dominant source of noise). For the reference Gaussian FWHM\,=\,\ang{;;0.3}, $m_{\rm AB}=24.5$ extended sources in a \ang{;;1.3} diameter aperture, VIS reaches ${\rm S/N}=15.5$ in three 566-s exposures -- the requirement is ${\rm S/N}=10$ -- and 18.0 in four. The current best estimate for the S/N of these extended sources in a \ang{;;1.3} diameter aperture as a function of $m_{\rm AB}$ is shown in Fig.~\ref{fig:SNratio}.  

At a technical level, one of the Readout Electronics, block \#7, which had earlier in the Payload-Level electromagnetic compatibility testing experienced a voltage regulation failure at high temperatures, performed nominally at operational temperatures. This is evident in Figs.~\ref{fig:bias-noise} and \ref{fig:flat-point} and confirmed the decision not to replace it. A delayed switch-on of a power-supply board in the Control and Data Processing Unit at low temperatures was corrected by replacing it with a Flight Spare board in late-2021. 

A few tests principally to make minor improvements on the operating points of the charge injection and trap pumping calibrations were carried out without optical stimulation during the satellite-level thermal vacuum testing in mid-2022.

\subsection{In-orbit commissioning}
\label{subsec:commissioning}

\begin{figure}[htbp!]
\centering
\includegraphics[width=0.97\columnwidth] {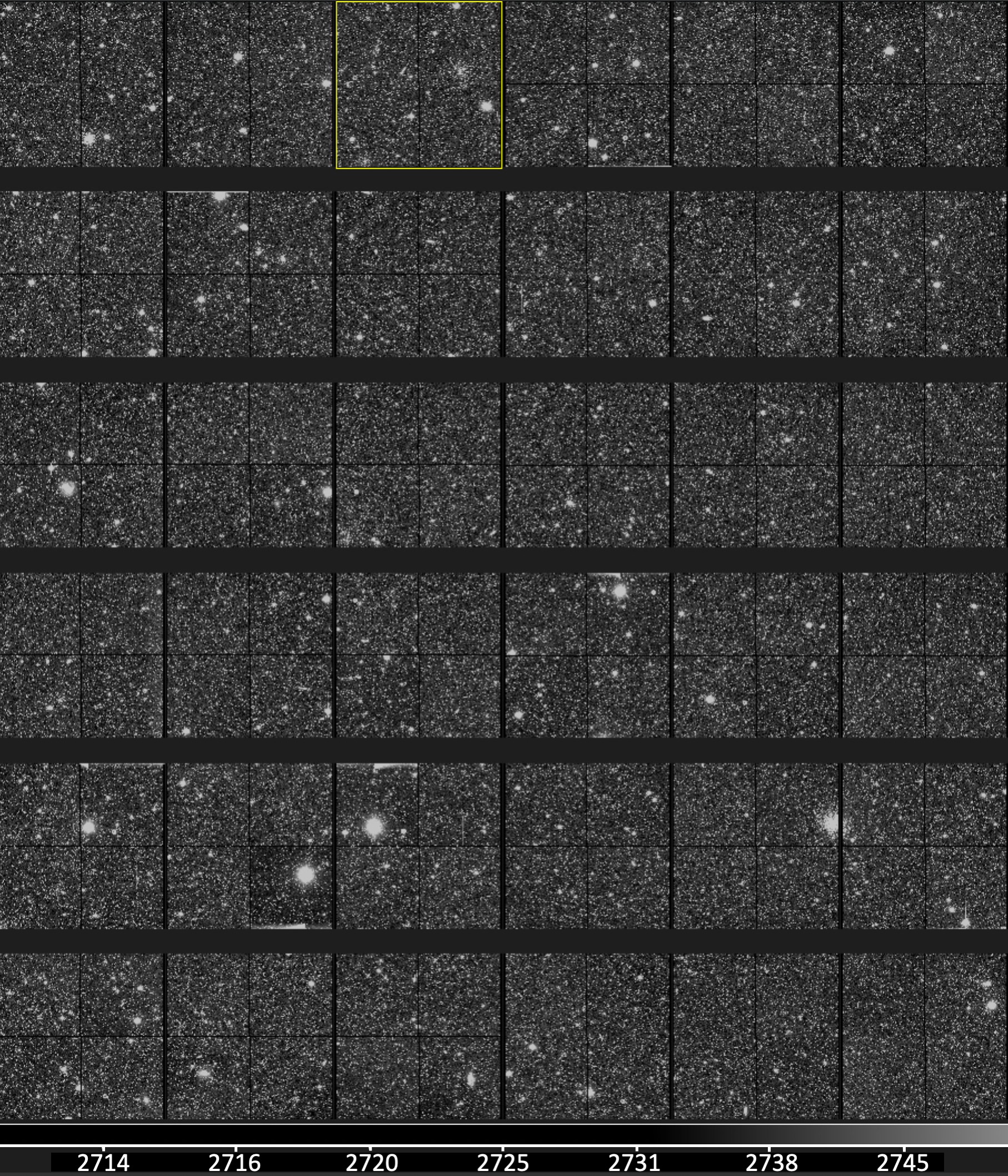}

\includegraphics[width=0.97\columnwidth] {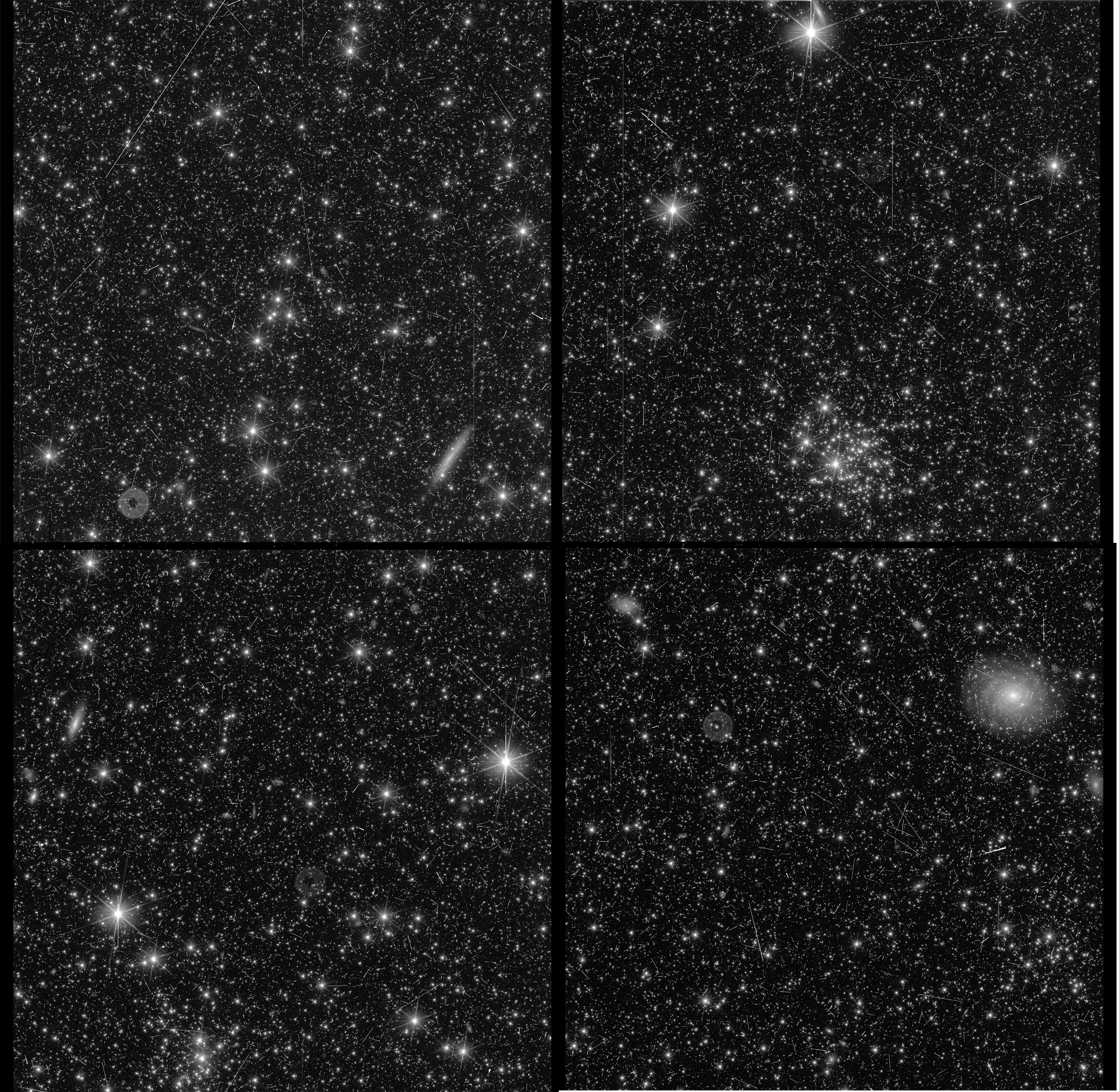}
\caption{`First Light' on-sky 566\,s exposures. {\it Top}: The full VIS mosaic of 36 detectors, after stray light avoidance measures were taken. It is shown on a log intensity scale with the intensity range shown in the grey-scale bar below the mosaic. To display it, pixels have been binned 12\texttimes12 into super-pixels to create a 2k\texttimes2k image but otherwise it is unprocessed. {\it Bottom}: The third CCD from the left-hand-side in the top row of the mosaic displayed at full resolution, again unprocessed. In addition to stars (evident from their diffraction spikes), star clusters, and galaxies, there are numerous cosmic ray events, some seen as extended streaks depending on their angle of incidence, and also optical ghosts, the most prominent of which is half way up on the left hand side.}
\label{fig:first-light}
\vspace*{-5mm}
\end{figure}

\Euclid was launched on 1 July 2023 from Cape Canaveral by a SpaceX Falcon 9 launch vehicle. The satellite and instruments were commissioned during the month-long journey towards the Second Lagrangian Point, with VIS switched on successfully on the nights of 11/12 and 12/13 July 2023. The first exposures with the shutter closed (biases and darks) were contaminated by scattered light, the origin of which was identified as originating from a thruster illuminated by the Sun at certain orientations of the spacecraft. The first open shutter observations showed fine star and galaxy images, indicating that the telescope focus was relatively good even prior to in-orbit focusing. Cosmic rays -- mostly Solar protons -- were detected at the expected rate and energies as incorporated in prelaunch simulations. An unprocessed 566\,s image taken on 23 July 2023 within a revised spacecraft pointing envelope eliminating the scattered light was released publicly and is shown in Fig.~\ref{fig:first-light}.

As the satellite commissioning progressed it became evident that as a consequence of insufficient shielding by the spacecraft, VIS detects X-rays in up to 10\% of the focal plane, depending on the level of Solar activity and the spacecraft orientation with respect to the Sun. These events are characteristic in terms of their energy and sharpness, and, as for the cosmic rays, can be removed in processing. At peak Solar X-ray flux the number of lost pixels in a single exposure in the affected region is significant; nevertheless, except in the worst cases, almost all pixels can be recovered, because all sky fields are observed with four exposures during the wide survey. The insufficient shielding however also implies greater than specified particle damage to the affected CCDs, with unavoidable reduction in charge transfer efficiency as the mission progresses. 

The \Euclid telescope was focused using VIS, and after these procedures the PSF was confirmed to be within specification. Further minor refinements have taken place during the Performance-Verification phase. It became evident that the \Euclid Fine Guidance Sensor was at times not performing to specification. This has been addressed \citep{EuclidSkyOverview}. As these effects do not pertain to VIS, these are not pursued further here. 

The VIS commissioning has demonstrated that the \Euclid VIS performance is, in all respects, as or better than specified, importantly meeting readout noise, bias stability, throughput, dynamic range, and pixel-to-pixel uniformity requirements. Using the \Euclid Science Ground Segment's colour correction to estimate VIS magnitudes from \Gaia $G$-band magnitudes, the \Gaia-based photometric zero point to produce a count rate of $1 {\rm e}^{-}\,{\rm s}^{-1}$ is $m_{\rm AB}=25.75$ for a frequency-flat spectral energy distribution. The expectation from the payload-level testing throughput measurements was $m_{\rm AB}=25.74$, so that the S/N is consistent with that represented in Fig.~\ref{fig:SNratio}. Refined characterisation and calibration is continuing with data from the Performance-Verification phase and regular calibration sequences during the survey.

\begin{figure*}
\center{\includegraphics[width=2.045\columnwidth] {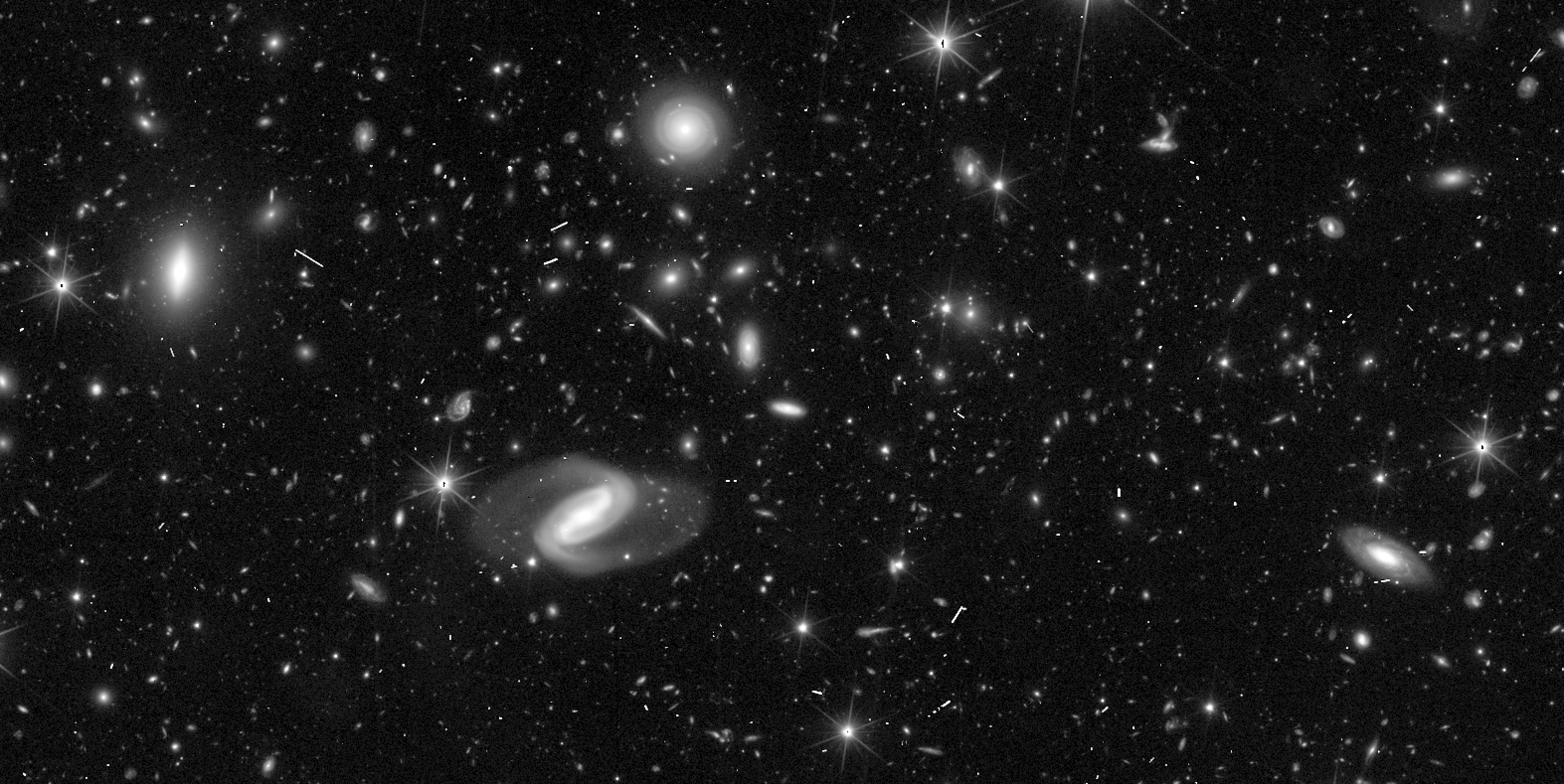}
\includegraphics[width=0.985\columnwidth] {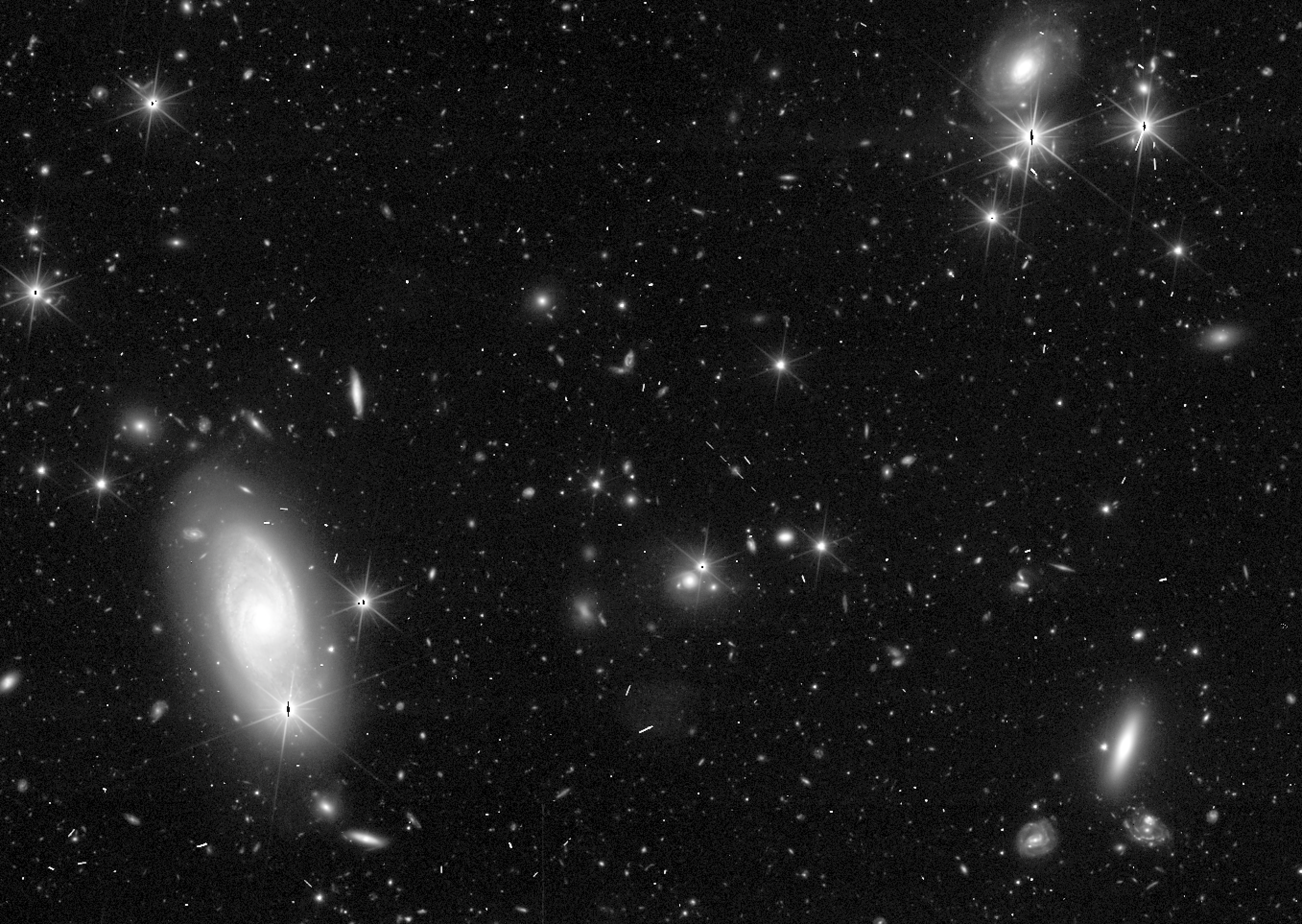}
\includegraphics[width=1.05\columnwidth] {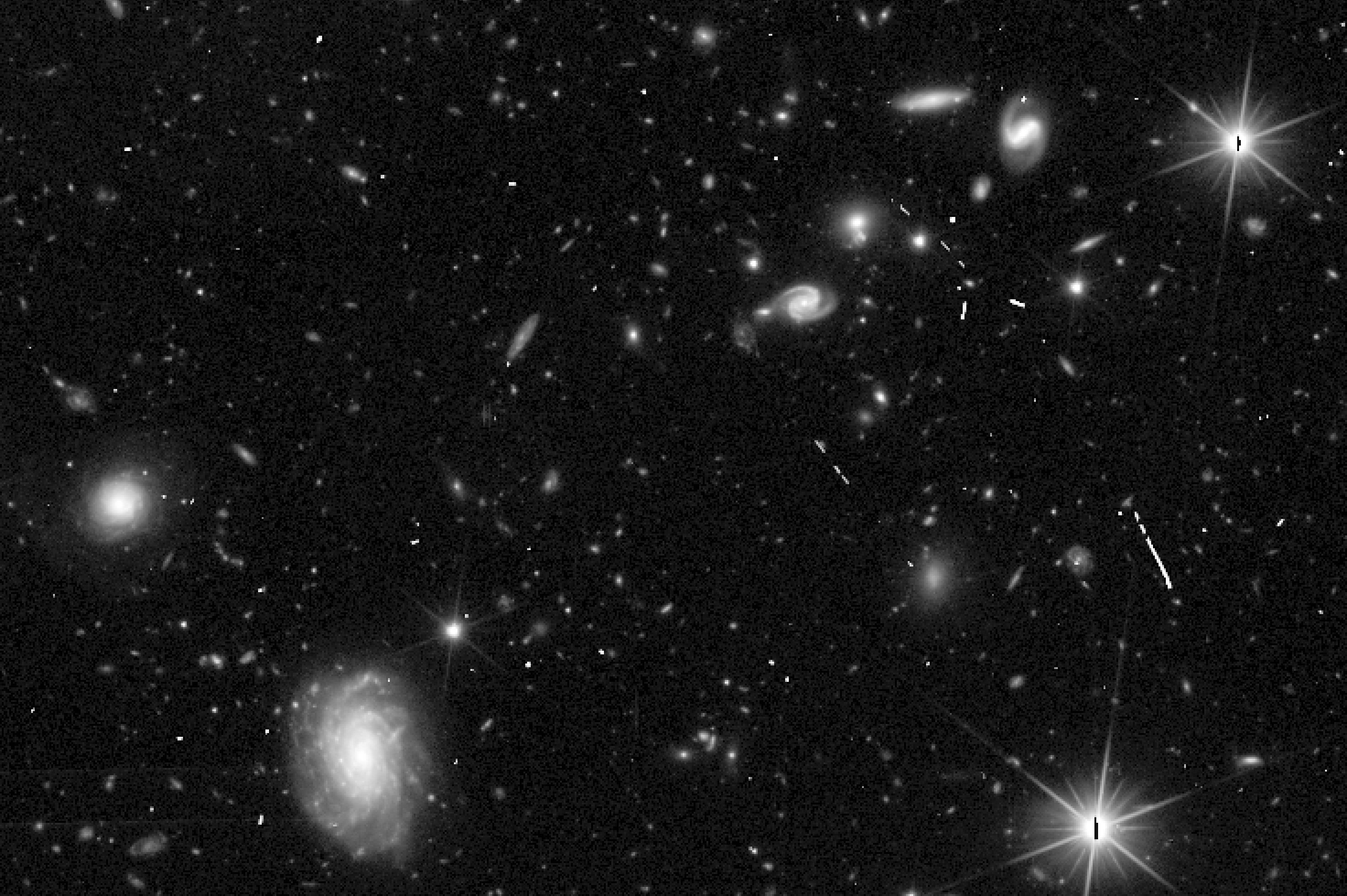}}
\caption{Three sub-images from a combined VIS image centred approximately at RA \ra{17;55;25}, Dec \ang{65;18;22} (J2000) produced from 16 nominal 566 second exposures. The prominent barred spiral galaxy at the upper left-hand side is approximately at RA \ra{17;55;15}, Dec \ang{65;04;01} (J2000). The top sub-image is $\ang{;10.5;}\times\ang{;5.3;}$ on a side while the lower two are $\ang{;9.0;}\times\ang{;6.4;}$ and $\ang{;9.8;}\times\ang{;6.5;}$ on a side. North is to the right and east is to the top. }
\label{fig:stacked}
\end{figure*}

An image centred approximately at RA \ra{17;55;25}, Dec \ang{65;18;22} (J2000) in the region of the North Ecliptic Pole was produced from 16 nominal 566 second exposures which were processed and combined using a pre-production version of the VIS processing function using on-ground calibrations. The individual exposures were combined using a weighted mean, cosmic ray flagging will be improved in future versions of the pipeline, and saturated pixels in some bright star images are shown black. Three sub-images taken from the full image are shown in Fig.~\ref{fig:stacked}. The sub-images have been re-binned from \ang{;;0.1}\,pix$^{-1}$ to \ang{;;0.2}\,pix$^{-1}$ and reach approximately $m_{\rm AB}=26$ with an S/N of $10$.

\section{\Euclid goals: open points and margins}
\label{sec:open_points}

In the years between the genesis of \Euclid and the final spacecraft testing, major advances have been made in understanding how the biases in the weak-lensing measurements can be quantified and organised, and how they may be minimised. As noted in Sect.~\ref{sec:rationale}, the work of the years to mission selection in 2011 was brought together in \cite{Cropper:13}, and the effects organised and error allowances assigned according to the understanding at that time. These were used as the basis for the design of VIS. However, some important aspects were not included in the allocations, for example the errors resulting from biases and outliers in the photometric redshifts. On the instrument itself, advancing the performance of the detectors and their associated electronics to the levels required for \Euclid identified effects not known or appreciated at the start of the project. At the same time, performance margins have arisen from new insights into how the biases and uncertainties interact. It is therefore instructive to move beyond the formal requirements here to summarise informally where VIS may meet its long-term expectations, and where there is still work to be done.

The VIS-specific effects which were not included in \cite{Cropper:13} include cosmic rays, the brighter-fatter effect, tree rings, colour-dependent pixel response non-uniformity, pathologies of analogue-to-digital converters, the transient response, the stitch-block pattern on the CCDs, and the details of the radiation effects when the detectors are irradiated cold. There are no allocations and no formal requirements for these. The brighter-fatter effect, radiation damage trapping, and the intrinsic nonlinearity of the detection chain are all interacting but separable effects. Trails from not-perfect transient responses of the Readout Electronics and the trails from the radiation-induced trapping in the CCD are also interacting but separable effects. Formally the colour dependence of the pixel response non-uniformity should require a flat-field correction dependent on the spectral energy distribution of the source at the location on the detector where it is incident. There are requirements at the lower level on the electronic cross-talk, but none at the higher levels in the Payload Elements Requirements Document \citep{PERD}. The error in the digitisation in the analogue-digital converters within the Readout Electronics was considered to be Gaussian, and as part of the detection chain readout noise; however, the differential nonlinearity measurements carried out for the VIS channels revealed patterns depending on which bits were transitioning from the previous measurement, and while the overall readout noise budget is met, there may be some residual effects resulting from its non-Gaussian nature. The stitch blocks, tree rings, and CCD metrology and optical distortion all contribute to distortion of the galaxy shape measurement, but there is no allocation for the tree rings.

Most of the knowledge of the radiation-induced trapping effects in CCDs has been gained from ambient irradiation, or in rare cases, in cold irradiation but with an ambient interval that anneal the trap sites. These results have been used in the models which correct the resulting charge trailing during readout of the detector. As noted in Sect.~\ref{subsubsec:CCDs}, irradiation and testing of the efficiency of the charge transfer during readout in radiation damaged CCDs at continuously maintained cryogenic temperatures has identified that rather than the trap time constants having discrete values, there is a broad spectrum of time constants. This requires modifications to the modelling of the charge transfer and a re-analysis of its fidelity.

Turning, on the other hand, to where performance margins have been identified, a critical point to appreciate is that the final weak-lensing performance depends on the accuracy of calibrations, be they in-frame data, dedicated observations, external observations, or simulations and models. All calibrations are imperfect at some level, and it is the difference between the accuracy of calibration and the unknown truth which results in bias -- only perfect calibrations would calibrate bias away entirely. 

Simulations and models are critical tools in reducing the biases to acceptable levels. Examples \citep{Hoekstra:17, Hoekstra:21,Cross-talk} include the effects of faint background sources, which will be important with S/N $\geq1$ at $m_{\rm AB}\geq27.5$ for the nominal survey, and the corrections for electronics ghosts. The level of some effects such as the nonlinearity of the detector electronics or the ellipticity arising the telescope optical system need be only weakly constrained if they can be well calibrated, and many residual effects can be corrected by generating simulations with and without the effect and making corrections either in the fitting to the data or in the shear catalogue. 

In some cases, calibration residuals will be evident in the detector coordinate system, perhaps only when a number of residual images are stacked. For example, inadequate correction of radiation damage trails will always result in a pattern which peaks at the pixels furthest from the readout node for each quadrant (i.e. at the centre of the CCDs) and the parameters for the correction algorithm can be adjusted to minimise these. While additive biases can generally be determined by such techniques, this does not apply to multiplicative biases; nevertheless a relationship between additive and multiplicative biases can be established through simulations, so that much of the multiplicative bias is removable \citep{Hoekstra:21}. Moreover, in \cite{Kitching:21} new ways have been identified to reduce the residual multiplicative biases. 

One of the simplifying assumptions in the error allowance organisation in \cite{Cropper:13} was that the incorrect calibrations acted independently and hence could be combined according to analytic prescriptions. \cite{EuclidCollaboration:19c} describes a methodology which allows the combination of the biases to be considered inherently according to the probability density function of their errors. All of these techniques create more margin than was considered at the earlier time. 

It is evident from the above that, at the \Euclid levels of accuracy for weak lensing, attention to the interactions of a multiplicity of effects -- several of them newly appreciated -- from the initial data capture to the cosmological analysis, is required at every level in the weak-lensing probe. While the understanding of the critical issues has evolved since their formulation, the conservatism of the mission and instrument requirements, together with the gain made in identifying margins, is expected to offset the complications and omissions arising since then. The analyses and calibrations made at satellite level to provide stability of the satellite pointing and PSF are unprecedented. The minimisation of the thermal perturbations to the telescope in the survey design has been successful. The science data processing functions can evolve along with the greater understanding of the biases in the data, and of how they can be minimised. For the VIS instrument itself, the calibration campaign has been unmatched in terms of the extent and diversity of the measurements and in the understanding of the detectors and their radiation damage susceptibility. Sufficient flexibility exists within the detector chains to optimise their operating points in the event of unexpected developments, and the on-board calibration facilities afforded by the Calibration Unit and by the charge injection and pocket pumping radiation damage monitoring set a new level of capability.

\section{Summary}
\label{sec:summary}

In this overview paper, we describe the general principles that were considered during the conception of VIS to provide a context for its final design. We then provide a description of the final build of VIS and how it was tested, and an overview of its performance prior to the formal evaluation during the \Euclid Performance-Verification phase. We emphasise the links to the other elements of the \Euclid mission -- the survey and operations, and the data processing. Towards the end of this paper we consider the impact on VIS of the development as time has passed of our understanding of weak-lensing scientific imperatives, and also the nuances of the as-built instrument performance. Many of the elements of VIS are communicated in greater detail in associated papers.

VIS provides the second-largest focal plane in space for scientific purposes, exceeded only by that of \Gaia. However, unlike in \Gaia, images from the full focal plane will be transmitted, hence its images will be larger than any others available from orbit by a significant margin. VIS is a large imager even in terms of ground-based astronomical imagers. Moreover, VIS will produce images of a  similar spatial resolution to those produced by the retired WFPC2 instrument on the \HST. Engineering such a facility for space  has been an ambitious project. 

Nevertheless, it is not the scale of the instrument that posed the most significant difficulty. The major challenges in the conception of VIS are related to the intrinsic stability needed for its operation, and the level of knowledge of its characteristics  required to understand the potential biases affecting the data it provides. Further challenges were related to understanding of how these characteristics interact with the rest of \Euclid, and in particular with the overall requirements of the weak-gravitational-lensing survey. These are particularly tightly enmeshed -- unlike in a typical observatory-class mission -- requiring a profound understanding of disciplines ranging from details of the detector physics to the statistical properties of weak-lensing power spectra. At the level of the \Euclid requirements, this has been a scientific programme in itself. 

These imperatives have translated into stringent specifications for all of the subsystems. Through its fine level of balance and momentum compensation, the shutter provides exceptionally low-disturbance actuation. The Focal Plane Array design provides a high level of stability at the focal plane by mechanically and thermally decoupling the warm Readout Electronics from the CCDs. The CCDs and Readout Electronics provide state-of-the-art performance in terms of sensitivity, noise, stability, resistance to radiation damage, and in the provision of in-flight calibration modes, all at power levels that are significantly lower than used in previous missions. The Control and Data Processing Unit performs the real-time processing from the large focal plane within the time constraints set by the efficiency requirements of the observing sequence. The sequence itself takes advantage of the parallel operation with NISP to provide a mass of calibration data for the science data processing. The VIS operation as a whole is driven by the necessity to maximise the knowledge of the instrument's state through repeatability, limited operating conditions, and intrinsic stability.

Moreover, the specifications derived for the cosmological measurements mean that \Euclid VIS can be used for a wide range of astronomy and astrophysics beyond cosmology; see for example  \citet{Signor24}, \citet{Pontinen23}, \citet{Bretonniere23}, \citet{Moriya22}, \citet{Borlaff22}, \citet{Bisigello20}, \citet{Barnett19}, and \citet{Inserra18}. Despite having only one (broad) optical band, its spatial resolution, depth, and coverage of the entire extragalactic sky will be a unique resource for galaxy morphology and, combined with the infrared photometry from NISP, for galaxy properties and evolution as a whole, producing images with finer spatial resolution at $z=0.7$ than the Sloan Digital Sky Survey at $z=0.1$. \Euclid VIS will be a powerful tool for identifying galaxies at $z>7$ through photometric dropout in the VIS band. It will be a discovery machine for strong gravitational lenses and faint structures around galaxies in the local Universe arising from mergers. Stellar populations can be resolved out to 5\,Mpc. \Euclid VIS will extend \Gaia astrometry in terms of proper motions, especially in the Deep Survey and revisited calibration fields, but also in the ${>}\,150$\,deg$^2$ of adjacent field overlaps. VIS will also be a prime resource in identifying targets for the {\it James Webb} Space Telescope, and will even be a resource for the discovery of asteroids. At ${\rm S/N}=10$, VIS is projected to reach $m_{\rm AB}>27$ over 53\,deg$^2$ in the Deep Fields, and $m_{\rm AB}\simeq25$ for the remainder of the extragalactic sky. With an improvement of approximately 10 in PSF area over the best ground-based images, VIS will provide a high-resolution optical-band context over the extragalactic sky for a wide range of astronomy and astrophysics. Perhaps VIS images will percolate into the wider public consciousness, especially when combined with lower resolution colour information from NISP or ground-based facilities, or, in the future, with the similar-resolution images from the Chinese Survey Space Telescope and, in the infrared, from the {\it Nancy Roman} Telescope.

\begin{acknowledgements}

\AckECol

\end{acknowledgements}

\bibliographystyle{aa} 
\bibliography{refs,Euclid} 

\label{LastPage}

\end{document}

%% file: authors.tex
\author{Euclid Collaboration: M.~S.~Cropper\orcid{0000-0003-4571-9468}\thanks{\email{m.cropper@ucl.ac.uk}}\inst{\ref{aff1}}
\and A.~Al-Bahlawan\inst{\ref{aff1}}
\and J.~Amiaux\inst{\ref{aff2}}
\and S.~Awan\inst{\ref{aff1}}
\and R.~Azzollini\orcid{0000-0002-0438-0886}\inst{\ref{aff1}}
\and K.~Benson\inst{\ref{aff1}}
\and M.~Berthe\inst{\ref{aff2}}
\and J.~Boucher\inst{\ref{aff1}}
\and E.~Bozzo\orcid{0000-0002-8201-1525}\inst{\ref{aff3}}
\and C.~Brockley-Blatt\inst{\ref{aff1}}
\and G.~P.~Candini\orcid{0000-0001-9481-8206}\inst{\ref{aff1}}
\and C.~Cara\inst{\ref{aff2}}
\and R.~A.~Chaudery\inst{\ref{aff1}}
\and R.~E.~Cole\orcid{0000-0002-7093-7320}\inst{\ref{aff1}}
\and P.~Danto\inst{\ref{aff4}}
\and J.~Denniston\inst{\ref{aff1}}
\and A.~M.~Di~Giorgio\orcid{0000-0002-4767-2360}\inst{\ref{aff5}}
\and B.~Dryer\orcid{0000-0001-7925-9768}\inst{\ref{aff6}}
\and J.-P.~Dubois\inst{\ref{aff7}}
\and J.~Endicott\inst{\ref{aff8}}
\and M.~Farina\orcid{0000-0002-3089-7846}\inst{\ref{aff5}}
\and E.~Galli\inst{\ref{aff5}}
\and L.~Genolet\inst{\ref{aff3}}
\and J.~P.~D.~Gow\orcid{0000-0002-5208-7924}\inst{\ref{aff9}}
\and P.~Guttridge\inst{\ref{aff1}}
\and M.~Hailey\inst{\ref{aff1}}
\and D.~Hall\inst{\ref{aff9}}
\and C.~Harper\inst{\ref{aff10}}
\and H.~Hoekstra\orcid{0000-0002-0641-3231}\inst{\ref{aff11}}
\and A.~D.~Holland\inst{\ref{aff9},\ref{aff8}}
\and B.~Horeau\inst{\ref{aff2}}
\and D.~Hu\inst{\ref{aff1}}
\and R.~E.~James\inst{\ref{aff1}}
\and A.~Khalil\inst{\ref{aff1}}
\and R.~King\inst{\ref{aff10}}
\and T.~Kitching\orcid{0000-0002-4061-4598}\inst{\ref{aff1}}
\and R.~Kohley\inst{\ref{aff12}}
\and C.~Larcheveque\inst{\ref{aff13}}
\and A.~Lawrenson\inst{\ref{aff14},\ref{aff1}}
\and P.~Liebing\inst{\ref{aff1}}
\and S.~J.~Liu\orcid{0000-0001-7680-2139}\inst{\ref{aff5}}
\and J.~Martignac\inst{\ref{aff2}}
\and R.~Massey\orcid{0000-0002-6085-3780}\inst{\ref{aff15},\ref{aff16}}
\and H.~J.~McCracken\orcid{0000-0002-9489-7765}\inst{\ref{aff17}}
\and L.~Miller\orcid{0000-0002-3376-6200}\inst{\ref{aff18}}
\and N.~Murray\inst{\ref{aff9}}
\and R.~Nakajima\inst{\ref{aff19}}
\and S.-M.~Niemi\inst{\ref{aff20}}
\and J.~W.~Nightingale\orcid{0000-0002-8987-7401}\inst{\ref{aff21},\ref{aff16}}
\and S.~Paltani\orcid{0000-0002-8108-9179}\inst{\ref{aff3}}
\and A.~Pendem\inst{\ref{aff1}}
\and A.~Philippon\inst{\ref{aff7}}
\and C.~Plana\inst{\ref{aff1}}
\and P.~Pool\inst{\ref{aff22}}
\and S.~Pottinger\inst{\ref{aff1}}
\and G.~D.~Racca\inst{\ref{aff20}}
\and J.~Rhodes\orcid{0000-0002-4485-8549}\inst{\ref{aff23}}
\and A.~Rousseau\inst{\ref{aff1}}
\and K.~Ruane\inst{\ref{aff1}}
\and M.~Salatti\inst{\ref{aff24}}
\and J.-C.~Salvignol\inst{\ref{aff20}}
\and A.~Sciortino\inst{\ref{aff25}}
\and A.~Short\inst{\ref{aff20}}
\and J.~Skottfelt\orcid{0000-0003-1310-8283}\inst{\ref{aff9}}
\and S.~J.~A.~Smit\inst{\ref{aff1}}
\and I.~Swindells\inst{\ref{aff22}}
\and M.~Szafraniec\inst{\ref{aff20}}
\and P.~D.~Thomas\inst{\ref{aff1}}
\and W.~Thomas\inst{\ref{aff1}}
\and E.~Tommasi\inst{\ref{aff24}}
\and S.~Tosti\inst{\ref{aff7}}
\and F.~Visticot\inst{\ref{aff2}}
\and D.~M.~Walton\inst{\ref{aff1}}
\and G.~Willis\inst{\ref{aff1}}
\and B.~Winter\inst{\ref{aff1}}
\and N.~Aghanim\orcid{0000-0002-6688-8992}\inst{\ref{aff7}}
\and B.~Altieri\orcid{0000-0003-3936-0284}\inst{\ref{aff12}}
\and A.~Amara\inst{\ref{aff26}}
\and S.~Andreon\orcid{0000-0002-2041-8784}\inst{\ref{aff27}}
\and N.~Auricchio\orcid{0000-0003-4444-8651}\inst{\ref{aff28}}
\and H.~Aussel\orcid{0000-0002-1371-5705}\inst{\ref{aff2}}
\and C.~Baccigalupi\orcid{0000-0002-8211-1630}\inst{\ref{aff29},\ref{aff30},\ref{aff31},\ref{aff32}}
\and M.~Baldi\orcid{0000-0003-4145-1943}\inst{\ref{aff33},\ref{aff28},\ref{aff34}}
\and A.~Balestra\orcid{0000-0002-6967-261X}\inst{\ref{aff35}}
\and S.~Bardelli\orcid{0000-0002-8900-0298}\inst{\ref{aff28}}
\and A.~Basset\inst{\ref{aff4}}
\and R.~Bender\orcid{0000-0001-7179-0626}\inst{\ref{aff36},\ref{aff37}}
\and F.~Bernardeau\inst{\ref{aff38},\ref{aff17}}
\and C.~Bodendorf\inst{\ref{aff36}}
\and T.~Boenke\inst{\ref{aff20}}
\and D.~Bonino\orcid{0000-0002-3336-9977}\inst{\ref{aff39}}
\and E.~Branchini\orcid{0000-0002-0808-6908}\inst{\ref{aff40},\ref{aff41},\ref{aff27}}
\and M.~Brescia\orcid{0000-0001-9506-5680}\inst{\ref{aff42},\ref{aff43},\ref{aff44}}
\and J.~Brinchmann\orcid{0000-0003-4359-8797}\inst{\ref{aff45}}
\and S.~Camera\orcid{0000-0003-3399-3574}\inst{\ref{aff46},\ref{aff47},\ref{aff39}}
\and V.~Capobianco\orcid{0000-0002-3309-7692}\inst{\ref{aff39}}
\and C.~Carbone\orcid{0000-0003-0125-3563}\inst{\ref{aff48}}
\and V.~F.~Cardone\inst{\ref{aff49},\ref{aff50}}
\and J.~Carretero\orcid{0000-0002-3130-0204}\inst{\ref{aff51},\ref{aff52}}
\and R.~Casas\orcid{0000-0002-8165-5601}\inst{\ref{aff53},\ref{aff54}}
\and S.~Casas\orcid{0000-0002-4751-5138}\inst{\ref{aff55}}
\and F.~J.~Castander\orcid{0000-0001-7316-4573}\inst{\ref{aff54},\ref{aff53}}
\and M.~Castellano\orcid{0000-0001-9875-8263}\inst{\ref{aff49}}
\and G.~Castignani\orcid{0000-0001-6831-0687}\inst{\ref{aff28}}
\and S.~Cavuoti\orcid{0000-0002-3787-4196}\inst{\ref{aff43},\ref{aff44}}
\and A.~Cimatti\inst{\ref{aff56}}
\and C.~Colodro-Conde\inst{\ref{aff57}}
\and G.~Congedo\orcid{0000-0003-2508-0046}\inst{\ref{aff58}}
\and C.~J.~Conselice\orcid{0000-0003-1949-7638}\inst{\ref{aff59}}
\and L.~Conversi\orcid{0000-0002-6710-8476}\inst{\ref{aff60},\ref{aff12}}
\and Y.~Copin\orcid{0000-0002-5317-7518}\inst{\ref{aff61}}
\and F.~Courbin\orcid{0000-0003-0758-6510}\inst{\ref{aff62}}
\and H.~M.~Courtois\orcid{0000-0003-0509-1776}\inst{\ref{aff63}}
\and M.~Crocce\orcid{0000-0002-9745-6228}\inst{\ref{aff54},\ref{aff64}}
\and J.-G.~Cuby\orcid{0000-0002-8767-1442}\inst{\ref{aff65},\ref{aff66}}
\and J.-C.~Cuillandre\orcid{0000-0002-3263-8645}\inst{\ref{aff2}}
\and A.~Da~Silva\orcid{0000-0002-6385-1609}\inst{\ref{aff67},\ref{aff68}}
\and H.~Degaudenzi\orcid{0000-0002-5887-6799}\inst{\ref{aff3}}
\and G.~De~Lucia\orcid{0000-0002-6220-9104}\inst{\ref{aff30}}
\and J.~Dinis\orcid{0000-0001-5075-1601}\inst{\ref{aff67},\ref{aff68}}
\and C.~Dolding\inst{\ref{aff1}}
\and M.~Douspis\orcid{0000-0003-4203-3954}\inst{\ref{aff7}}
\and C.~A.~J.~Duncan\inst{\ref{aff59}}
\and X.~Dupac\inst{\ref{aff12}}
\and S.~Dusini\orcid{0000-0002-1128-0664}\inst{\ref{aff69}}
\and A.~Ealet\orcid{0000-0003-3070-014X}\inst{\ref{aff61}}
\and M.~Fabricius\orcid{0000-0002-7025-6058}\inst{\ref{aff36},\ref{aff37}}
\and S.~Farrens\orcid{0000-0002-9594-9387}\inst{\ref{aff2}}
\and S.~Ferriol\inst{\ref{aff61}}
\and P.~Fosalba\orcid{0000-0002-1510-5214}\inst{\ref{aff53},\ref{aff64}}
\and S.~Fotopoulou\orcid{0000-0002-9686-254X}\inst{\ref{aff70}}
\and M.~Frailis\orcid{0000-0002-7400-2135}\inst{\ref{aff30}}
\and E.~Franceschi\orcid{0000-0002-0585-6591}\inst{\ref{aff28}}
\and P.~Franzetti\inst{\ref{aff48}}
\and P.-A.~Frugier\inst{\ref{aff2}}
\and M.~Fumana\orcid{0000-0001-6787-5950}\inst{\ref{aff48}}
\and S.~Galeotta\orcid{0000-0002-3748-5115}\inst{\ref{aff30}}
\and B.~Garilli\orcid{0000-0001-7455-8750}\inst{\ref{aff48}}
\and K.~George\orcid{0000-0002-1734-8455}\inst{\ref{aff37}}
\and W.~Gillard\orcid{0000-0003-4744-9748}\inst{\ref{aff71}}
\and B.~Gillis\orcid{0000-0002-4478-1270}\inst{\ref{aff58}}
\and C.~Giocoli\orcid{0000-0002-9590-7961}\inst{\ref{aff28},\ref{aff72}}
\and P.~G\'omez-Alvarez\orcid{0000-0002-8594-5358}\inst{\ref{aff73},\ref{aff12}}
\and B.~R.~Granett\orcid{0000-0003-2694-9284}\inst{\ref{aff27}}
\and A.~Grazian\orcid{0000-0002-5688-0663}\inst{\ref{aff35}}
\and F.~Grupp\inst{\ref{aff36},\ref{aff37}}
\and L.~Guzzo\orcid{0000-0001-8264-5192}\inst{\ref{aff74},\ref{aff27}}
\and S.~V.~H.~Haugan\orcid{0000-0001-9648-7260}\inst{\ref{aff75}}
\and O.~Herent\inst{\ref{aff17}}
\and J.~Hoar\inst{\ref{aff12}}
\and M.~S.~Holliman\inst{\ref{aff58}}
\and W.~Holmes\inst{\ref{aff23}}
\and I.~Hook\orcid{0000-0002-2960-978X}\inst{\ref{aff76}}
\and F.~Hormuth\inst{\ref{aff77},\ref{aff78}}
\and A.~Hornstrup\orcid{0000-0002-3363-0936}\inst{\ref{aff79},\ref{aff80}}
\and P.~Hudelot\inst{\ref{aff17}}
\and S.~Ili\'c\orcid{0000-0003-4285-9086}\inst{\ref{aff81},\ref{aff82}}
\and K.~Jahnke\orcid{0000-0003-3804-2137}\inst{\ref{aff78}}
\and M.~Jhabvala\inst{\ref{aff83}}
\and B.~Joachimi\orcid{0000-0001-7494-1303}\inst{\ref{aff84}}
\and E.~Keih\"anen\orcid{0000-0003-1804-7715}\inst{\ref{aff85}}
\and S.~Kermiche\orcid{0000-0002-0302-5735}\inst{\ref{aff71}}
\and A.~Kiessling\orcid{0000-0002-2590-1273}\inst{\ref{aff23}}
\and M.~Kilbinger\orcid{0000-0001-9513-7138}\inst{\ref{aff2}}
\and B.~Kubik\orcid{0009-0006-5823-4880}\inst{\ref{aff61}}
\and K.~Kuijken\orcid{0000-0002-3827-0175}\inst{\ref{aff11}}
\and M.~K\"ummel\orcid{0000-0003-2791-2117}\inst{\ref{aff37}}
\and M.~Kunz\orcid{0000-0002-3052-7394}\inst{\ref{aff86}}
\and H.~Kurki-Suonio\orcid{0000-0002-4618-3063}\inst{\ref{aff87},\ref{aff88}}
\and O.~Lahav\orcid{0000-0002-1134-9035}\inst{\ref{aff84}}
\and R.~Laureijs\inst{\ref{aff20}}
\and S.~Ligori\orcid{0000-0003-4172-4606}\inst{\ref{aff39}}
\and P.~B.~Lilje\orcid{0000-0003-4324-7794}\inst{\ref{aff75}}
\and V.~Lindholm\orcid{0000-0003-2317-5471}\inst{\ref{aff87},\ref{aff88}}
\and I.~Lloro\inst{\ref{aff89}}
\and J.~Lorenzo~Alvarez\orcid{0000-0002-6845-993X}\inst{\ref{aff20}}
\and G.~Mainetti\inst{\ref{aff90}}
\and D.~Maino\inst{\ref{aff74},\ref{aff48},\ref{aff91}}
\and E.~Maiorano\orcid{0000-0003-2593-4355}\inst{\ref{aff28}}
\and O.~Mansutti\orcid{0000-0001-5758-4658}\inst{\ref{aff30}}
\and S.~Marcin\inst{\ref{aff92}}
\and O.~Marggraf\orcid{0000-0001-7242-3852}\inst{\ref{aff19}}
\and K.~Markovic\orcid{0000-0001-6764-073X}\inst{\ref{aff23}}
\and M.~Martinelli\orcid{0000-0002-6943-7732}\inst{\ref{aff49},\ref{aff50}}
\and N.~Martinet\orcid{0000-0003-2786-7790}\inst{\ref{aff66}}
\and F.~Marulli\orcid{0000-0002-8850-0303}\inst{\ref{aff93},\ref{aff28},\ref{aff34}}
\and D.~C.~Masters\orcid{0000-0001-5382-6138}\inst{\ref{aff94}}
\and S.~Maurogordato\inst{\ref{aff95}}
\and E.~Medinaceli\orcid{0000-0002-4040-7783}\inst{\ref{aff28}}
\and S.~Mei\orcid{0000-0002-2849-559X}\inst{\ref{aff96}}
\and M.~Melchior\inst{\ref{aff92}}
\and Y.~Mellier\inst{\ref{aff97},\ref{aff17}}
\and M.~Meneghetti\orcid{0000-0003-1225-7084}\inst{\ref{aff28},\ref{aff34}}
\and E.~Merlin\orcid{0000-0001-6870-8900}\inst{\ref{aff49}}
\and G.~Meylan\inst{\ref{aff62}}
\and J.~J.~Mohr\orcid{0000-0002-6875-2087}\inst{\ref{aff37},\ref{aff36}}
\and M.~Moresco\orcid{0000-0002-7616-7136}\inst{\ref{aff93},\ref{aff28}}
\and L.~Moscardini\orcid{0000-0002-3473-6716}\inst{\ref{aff93},\ref{aff28},\ref{aff34}}
\and C.~Neissner\orcid{0000-0001-8524-4968}\inst{\ref{aff98},\ref{aff52}}
\and R.~C.~Nichol\orcid{0000-0003-0939-6518}\inst{\ref{aff26}}
\and T.~Nutma\inst{\ref{aff99},\ref{aff11}}
\and C.~Padilla\orcid{0000-0001-7951-0166}\inst{\ref{aff98}}
\and K.~Paech\orcid{0000-0003-0625-2367}\inst{\ref{aff36}}
\and F.~Pasian\orcid{0000-0002-4869-3227}\inst{\ref{aff30}}
\and J.~A.~Peacock\orcid{0000-0002-1168-8299}\inst{\ref{aff58}}
\and K.~Pedersen\inst{\ref{aff100}}
\and W.~J.~Percival\orcid{0000-0002-0644-5727}\inst{\ref{aff101},\ref{aff102},\ref{aff103}}
\and V.~Pettorino\inst{\ref{aff20}}
\and S.~Pires\orcid{0000-0002-0249-2104}\inst{\ref{aff2}}
\and G.~Polenta\orcid{0000-0003-4067-9196}\inst{\ref{aff104}}
\and M.~Poncet\inst{\ref{aff4}}
\and L.~A.~Popa\inst{\ref{aff105}}
\and L.~Pozzetti\orcid{0000-0001-7085-0412}\inst{\ref{aff28}}
\and F.~Raison\orcid{0000-0002-7819-6918}\inst{\ref{aff36}}
\and R.~Rebolo\inst{\ref{aff57},\ref{aff106}}
\and A.~Refregier\inst{\ref{aff107}}
\and A.~Renzi\orcid{0000-0001-9856-1970}\inst{\ref{aff108},\ref{aff69}}
\and G.~Riccio\inst{\ref{aff43}}
\and Hans-Walter~Rix\orcid{0000-0003-4996-9069}\inst{\ref{aff78}}
\and E.~Romelli\orcid{0000-0003-3069-9222}\inst{\ref{aff30}}
\and M.~Roncarelli\orcid{0000-0001-9587-7822}\inst{\ref{aff28}}
\and C.~Rosset\orcid{0000-0003-0286-2192}\inst{\ref{aff96}}
\and E.~Rossetti\orcid{0000-0003-0238-4047}\inst{\ref{aff33}}
\and H.~J.~A.~Rottgering\orcid{0000-0001-8887-2257}\inst{\ref{aff11}}
\and B.~Rusholme\orcid{0000-0001-7648-4142}\inst{\ref{aff109}}
\and R.~Saglia\orcid{0000-0003-0378-7032}\inst{\ref{aff37},\ref{aff36}}
\and Z.~Sakr\orcid{0000-0002-4823-3757}\inst{\ref{aff110},\ref{aff82},\ref{aff111}}
\and A.~G.~S\'anchez\orcid{0000-0003-1198-831X}\inst{\ref{aff36}}
\and D.~Sapone\orcid{0000-0001-7089-4503}\inst{\ref{aff112}}
\and M.~Sauvage\orcid{0000-0002-0809-2574}\inst{\ref{aff2}}
\and R.~Scaramella\orcid{0000-0003-2229-193X}\inst{\ref{aff49},\ref{aff50}}
\and J.~A.~Schewtschenko\inst{\ref{aff58}}
\and M.~Schirmer\orcid{0000-0003-2568-9994}\inst{\ref{aff78}}
\and P.~Schneider\orcid{0000-0001-8561-2679}\inst{\ref{aff19}}
\and T.~Schrabback\orcid{0000-0002-6987-7834}\inst{\ref{aff113}}
\and A.~Secroun\orcid{0000-0003-0505-3710}\inst{\ref{aff71}}
\and E.~Sefusatti\orcid{0000-0003-0473-1567}\inst{\ref{aff30},\ref{aff29},\ref{aff31}}
\and G.~Seidel\orcid{0000-0003-2907-353X}\inst{\ref{aff78}}
\and M.~Seiffert\orcid{0000-0002-7536-9393}\inst{\ref{aff23}}
\and S.~Serrano\orcid{0000-0002-0211-2861}\inst{\ref{aff53},\ref{aff114},\ref{aff54}}
\and C.~Sirignano\orcid{0000-0002-0995-7146}\inst{\ref{aff108},\ref{aff69}}
\and G.~Sirri\orcid{0000-0003-2626-2853}\inst{\ref{aff34}}
\and L.~Stanco\orcid{0000-0002-9706-5104}\inst{\ref{aff69}}
\and J.-L.~Starck\orcid{0000-0003-2177-7794}\inst{\ref{aff2}}
\and J.~Steinwagner\inst{\ref{aff36}}
\and P.~Tallada-Cresp\'{i}\orcid{0000-0002-1336-8328}\inst{\ref{aff51},\ref{aff52}}
\and D.~Tavagnacco\orcid{0000-0001-7475-9894}\inst{\ref{aff30}}
\and A.~N.~Taylor\inst{\ref{aff58}}
\and H.~I.~Teplitz\orcid{0000-0002-7064-5424}\inst{\ref{aff94}}
\and I.~Tereno\inst{\ref{aff67},\ref{aff115}}
\and R.~Toledo-Moreo\orcid{0000-0002-2997-4859}\inst{\ref{aff116}}
\and F.~Torradeflot\orcid{0000-0003-1160-1517}\inst{\ref{aff52},\ref{aff51}}
\and I.~Tutusaus\orcid{0000-0002-3199-0399}\inst{\ref{aff82}}
\and E.~A.~Valentijn\inst{\ref{aff99}}
\and L.~Valenziano\orcid{0000-0002-1170-0104}\inst{\ref{aff28},\ref{aff117}}
\and T.~Vassallo\orcid{0000-0001-6512-6358}\inst{\ref{aff37},\ref{aff30}}
\and G.~Verdoes~Kleijn\orcid{0000-0001-5803-2580}\inst{\ref{aff99}}
\and A.~Veropalumbo\orcid{0000-0003-2387-1194}\inst{\ref{aff27},\ref{aff41},\ref{aff118}}
\and S.~Wachter\inst{\ref{aff119}}
\and Y.~Wang\orcid{0000-0002-4749-2984}\inst{\ref{aff94}}
\and J.~Weller\orcid{0000-0002-8282-2010}\inst{\ref{aff37},\ref{aff36}}
\and G.~Zamorani\orcid{0000-0002-2318-301X}\inst{\ref{aff28}}
\and J.~Zoubian\inst{\ref{aff71}}
\and E.~Zucca\orcid{0000-0002-5845-8132}\inst{\ref{aff28}}
\and A.~Biviano\orcid{0000-0002-0857-0732}\inst{\ref{aff30},\ref{aff29}}
\and M.~Bolzonella\orcid{0000-0003-3278-4607}\inst{\ref{aff28}}
\and A.~Boucaud\orcid{0000-0001-7387-2633}\inst{\ref{aff96}}
\and C.~Burigana\orcid{0000-0002-3005-5796}\inst{\ref{aff120},\ref{aff117}}
\and M.~Calabrese\orcid{0000-0002-2637-2422}\inst{\ref{aff121},\ref{aff48}}
\and P.~Casenove\inst{\ref{aff4}}
\and D.~Di~Ferdinando\inst{\ref{aff34}}
\and J.~A.~Escartin~Vigo\inst{\ref{aff36}}
\and G.~Fabbian\orcid{0000-0002-3255-4695}\inst{\ref{aff122},\ref{aff123}}
\and R.~Farinelli\inst{\ref{aff28}}
\and F.~Finelli\orcid{0000-0002-6694-3269}\inst{\ref{aff28},\ref{aff117}}
\and J.~Gracia-Carpio\inst{\ref{aff36}}
\and H.~Israel\orcid{0000-0002-3045-4412}\inst{\ref{aff124}}
\and N.~Mauri\orcid{0000-0001-8196-1548}\inst{\ref{aff56},\ref{aff34}}
\and H.~N.~Nguyen-Kim\inst{\ref{aff17}}
\and A.~Pezzotta\orcid{0000-0003-0726-2268}\inst{\ref{aff36}}
\and M.~P\"ontinen\orcid{0000-0001-5442-2530}\inst{\ref{aff87}}
\and C.~Porciani\orcid{0000-0002-7797-2508}\inst{\ref{aff19}}
\and V.~Scottez\inst{\ref{aff97},\ref{aff125}}
\and M.~Tenti\orcid{0000-0002-4254-5901}\inst{\ref{aff34}}
\and M.~Viel\orcid{0000-0002-2642-5707}\inst{\ref{aff29},\ref{aff30},\ref{aff32},\ref{aff31},\ref{aff126}}
\and M.~Wiesmann\orcid{0009-0000-8199-5860}\inst{\ref{aff75}}
\and Y.~Akrami\orcid{0000-0002-2407-7956}\inst{\ref{aff127},\ref{aff128}}
\and V.~Allevato\orcid{0000-0001-7232-5152}\inst{\ref{aff43}}
\and S.~Anselmi\orcid{0000-0002-3579-9583}\inst{\ref{aff69},\ref{aff108},\ref{aff129}}
\and E.~Aubourg\orcid{0000-0002-5592-023X}\inst{\ref{aff96},\ref{aff130}}
\and M.~Ballardini\orcid{0000-0003-4481-3559}\inst{\ref{aff131},\ref{aff28},\ref{aff132}}
\and D.~Bertacca\orcid{0000-0002-2490-7139}\inst{\ref{aff108},\ref{aff35},\ref{aff69}}
\and M.~Bethermin\orcid{0000-0002-3915-2015}\inst{\ref{aff133},\ref{aff66}}
\and A.~Blanchard\orcid{0000-0001-8555-9003}\inst{\ref{aff82}}
\and L.~Blot\orcid{0000-0002-9622-7167}\inst{\ref{aff134},\ref{aff129}}
\and S.~Borgani\orcid{0000-0001-6151-6439}\inst{\ref{aff135},\ref{aff29},\ref{aff30},\ref{aff31}}
\and A.~S.~Borlaff\orcid{0000-0003-3249-4431}\inst{\ref{aff136},\ref{aff137}}
\and S.~Bruton\orcid{0000-0002-6503-5218}\inst{\ref{aff138}}
\and R.~Cabanac\orcid{0000-0001-6679-2600}\inst{\ref{aff82}}
\and A.~Calabro\orcid{0000-0003-2536-1614}\inst{\ref{aff49}}
\and G.~Calderone\orcid{0000-0002-7738-5389}\inst{\ref{aff30}}
\and G.~Canas-Herrera\orcid{0000-0003-2796-2149}\inst{\ref{aff20},\ref{aff139}}
\and A.~Cappi\inst{\ref{aff28},\ref{aff95}}
\and C.~S.~Carvalho\inst{\ref{aff115}}
\and T.~Castro\orcid{0000-0002-6292-3228}\inst{\ref{aff30},\ref{aff31},\ref{aff29},\ref{aff126}}
\and K.~C.~Chambers\orcid{0000-0001-6965-7789}\inst{\ref{aff140}}
\and R.~Chary\orcid{0000-0001-7583-0621}\inst{\ref{aff94}}
\and S.~Contarini\orcid{0000-0002-9843-723X}\inst{\ref{aff36}}
\and A.~R.~Cooray\orcid{0000-0002-3892-0190}\inst{\ref{aff141}}
\and O.~Cordes\inst{\ref{aff19}}
\and M.~Costanzi\orcid{0000-0001-8158-1449}\inst{\ref{aff135},\ref{aff30},\ref{aff29}}
\and O.~Cucciati\orcid{0000-0002-9336-7551}\inst{\ref{aff28}}
\and S.~Davini\orcid{0000-0003-3269-1718}\inst{\ref{aff41}}
\and B.~De~Caro\inst{\ref{aff48}}
\and G.~Desprez\inst{\ref{aff142}}
\and A.~D\'iaz-S\'anchez\orcid{0000-0003-0748-4768}\inst{\ref{aff143}}
\and S.~Di~Domizio\orcid{0000-0003-2863-5895}\inst{\ref{aff40},\ref{aff41}}
\and H.~Dole\orcid{0000-0002-9767-3839}\inst{\ref{aff7}}
\and S.~Escoffier\orcid{0000-0002-2847-7498}\inst{\ref{aff71}}
\and A.~G.~Ferrari\orcid{0009-0005-5266-4110}\inst{\ref{aff56},\ref{aff34}}
\and P.~G.~Ferreira\orcid{0000-0002-3021-2851}\inst{\ref{aff18}}
\and I.~Ferrero\orcid{0000-0002-1295-1132}\inst{\ref{aff75}}
\and A.~Finoguenov\orcid{0000-0002-4606-5403}\inst{\ref{aff87}}
\and A.~Fontana\orcid{0000-0003-3820-2823}\inst{\ref{aff49}}
\and F.~Fornari\orcid{0000-0003-2979-6738}\inst{\ref{aff117}}
\and L.~Gabarra\orcid{0000-0002-8486-8856}\inst{\ref{aff18}}
\and K.~Ganga\orcid{0000-0001-8159-8208}\inst{\ref{aff96}}
\and J.~Garc\'ia-Bellido\orcid{0000-0002-9370-8360}\inst{\ref{aff127}}
\and V.~Gautard\inst{\ref{aff144}}
\and E.~Gaztanaga\orcid{0000-0001-9632-0815}\inst{\ref{aff54},\ref{aff53},\ref{aff145}}
\and F.~Giacomini\orcid{0000-0002-3129-2814}\inst{\ref{aff34}}
\and F.~Gianotti\orcid{0000-0003-4666-119X}\inst{\ref{aff28}}
\and G.~Gozaliasl\orcid{0000-0002-0236-919X}\inst{\ref{aff146},\ref{aff87}}
\and A.~Gregorio\orcid{0000-0003-4028-8785}\inst{\ref{aff135},\ref{aff30},\ref{aff31}}
\and A.~Hall\orcid{0000-0002-3139-8651}\inst{\ref{aff58}}
\and W.~G.~Hartley\inst{\ref{aff3}}
\and H.~Hildebrandt\orcid{0000-0002-9814-3338}\inst{\ref{aff147}}
\and J.~Hjorth\orcid{0000-0002-4571-2306}\inst{\ref{aff148}}
\and M.~Huertas-Company\orcid{0000-0002-1416-8483}\inst{\ref{aff57},\ref{aff149},\ref{aff150},\ref{aff151}}
\and O.~Ilbert\orcid{0000-0002-7303-4397}\inst{\ref{aff66}}
\and A.~Jimenez~Mu\~noz\orcid{0009-0004-5252-185X}\inst{\ref{aff152}}
\and S.~Joudaki\orcid{0000-0001-8820-673X}\inst{\ref{aff145}}
\and J.~J.~E.~Kajava\orcid{0000-0002-3010-8333}\inst{\ref{aff153},\ref{aff154}}
\and V.~Kansal\orcid{0000-0002-4008-6078}\inst{\ref{aff155},\ref{aff156}}
\and D.~Karagiannis\orcid{0000-0002-4927-0816}\inst{\ref{aff157},\ref{aff158}}
\and C.~C.~Kirkpatrick\inst{\ref{aff85}}
\and F.~Lacasa\orcid{0000-0002-7268-3440}\inst{\ref{aff159},\ref{aff7},\ref{aff86}}
\and J.~Le~Graet\orcid{0000-0001-6523-7971}\inst{\ref{aff71}}
\and L.~Legrand\orcid{0000-0003-0610-5252}\inst{\ref{aff160}}
\and G.~Libet\inst{\ref{aff4}}
\and A.~Loureiro\orcid{0000-0002-4371-0876}\inst{\ref{aff161},\ref{aff162}}
\and J.~Macias-Perez\orcid{0000-0002-5385-2763}\inst{\ref{aff152}}
\and M.~Magliocchetti\orcid{0000-0001-9158-4838}\inst{\ref{aff5}}
\and C.~Mancini\orcid{0000-0002-4297-0561}\inst{\ref{aff48}}
\and F.~Mannucci\orcid{0000-0002-4803-2381}\inst{\ref{aff163}}
\and R.~Maoli\orcid{0000-0002-6065-3025}\inst{\ref{aff164},\ref{aff49}}
\and C.~J.~A.~P.~Martins\orcid{0000-0002-4886-9261}\inst{\ref{aff165},\ref{aff45}}
\and S.~Matthew\orcid{0000-0001-8448-1697}\inst{\ref{aff58}}
\and L.~Maurin\orcid{0000-0002-8406-0857}\inst{\ref{aff7}}
\and C.~J.~R.~McPartland\orcid{0000-0003-0639-025X}\inst{\ref{aff80},\ref{aff166}}
\and R.~B.~Metcalf\orcid{0000-0003-3167-2574}\inst{\ref{aff93},\ref{aff28}}
\and M.~Migliaccio\inst{\ref{aff167},\ref{aff168}}
\and M.~Miluzio\inst{\ref{aff12},\ref{aff169}}
\and P.~Monaco\orcid{0000-0003-2083-7564}\inst{\ref{aff135},\ref{aff30},\ref{aff31},\ref{aff29}}
\and C.~Moretti\orcid{0000-0003-3314-8936}\inst{\ref{aff32},\ref{aff126},\ref{aff30},\ref{aff29},\ref{aff31}}
\and G.~Morgante\inst{\ref{aff28}}
\and S.~Nadathur\orcid{0000-0001-9070-3102}\inst{\ref{aff145}}
\and Nicholas~A.~Walton\orcid{0000-0003-3983-8778}\inst{\ref{aff170}}
\and J.~Odier\orcid{0000-0002-1650-2246}\inst{\ref{aff152}}
\and M.~Oguri\orcid{0000-0003-3484-399X}\inst{\ref{aff171},\ref{aff172}}
\and L.~Patrizii\inst{\ref{aff34}}
\and V.~Popa\inst{\ref{aff105}}
\and D.~Potter\orcid{0000-0002-0757-5195}\inst{\ref{aff173}}
\and A.~Pourtsidou\orcid{0000-0001-9110-5550}\inst{\ref{aff58},\ref{aff174}}
\and P.~Reimberg\orcid{0000-0003-3410-0280}\inst{\ref{aff97}}
\and I.~Risso\orcid{0000-0003-2525-7761}\inst{\ref{aff118}}
\and P.-F.~Rocci\inst{\ref{aff7}}
\and R.~P.~Rollins\orcid{0000-0003-1291-1023}\inst{\ref{aff58}}
\and M.~Sahl\'en\orcid{0000-0003-0973-4804}\inst{\ref{aff175}}
\and C.~Scarlata\orcid{0000-0002-9136-8876}\inst{\ref{aff138}}
\and J.~Schaye\orcid{0000-0002-0668-5560}\inst{\ref{aff11}}
\and A.~Schneider\orcid{0000-0001-7055-8104}\inst{\ref{aff173}}
\and M.~Schultheis\inst{\ref{aff95}}
\and M.~Sereno\orcid{0000-0003-0302-0325}\inst{\ref{aff28},\ref{aff34}}
\and F.~Shankar\orcid{0000-0001-8973-5051}\inst{\ref{aff176}}
\and G.~Sikkema\inst{\ref{aff99}}
\and A.~Silvestri\orcid{0000-0001-6904-5061}\inst{\ref{aff139}}
\and P.~Simon\inst{\ref{aff19}}
\and A.~Spurio~Mancini\orcid{0000-0001-5698-0990}\inst{\ref{aff177},\ref{aff1}}
\and J.~Stadel\orcid{0000-0001-7565-8622}\inst{\ref{aff173}}
\and S.~A.~Stanford\orcid{0000-0003-0122-0841}\inst{\ref{aff178}}
\and K.~Tanidis\inst{\ref{aff18}}
\and C.~Tao\orcid{0000-0001-7961-8177}\inst{\ref{aff71}}
\and N.~Tessore\orcid{0000-0002-9696-7931}\inst{\ref{aff84}}
\and G.~Testera\inst{\ref{aff41}}
\and M.~Tewes\orcid{0000-0002-1155-8689}\inst{\ref{aff19}}
\and R.~Teyssier\orcid{0000-0001-7689-0933}\inst{\ref{aff179}}
\and S.~Toft\orcid{0000-0003-3631-7176}\inst{\ref{aff80},\ref{aff180},\ref{aff166}}
\and S.~Tosi\orcid{0000-0002-7275-9193}\inst{\ref{aff40},\ref{aff41}}
\and A.~Troja\orcid{0000-0003-0239-4595}\inst{\ref{aff108},\ref{aff69}}
\and M.~Tucci\inst{\ref{aff3}}
\and C.~Valieri\inst{\ref{aff34}}
\and J.~Valiviita\orcid{0000-0001-6225-3693}\inst{\ref{aff87},\ref{aff88}}
\and D.~Vergani\orcid{0000-0003-0898-2216}\inst{\ref{aff28}}
\and F.~Vernizzi\orcid{0000-0003-3426-2802}\inst{\ref{aff38}}
\and G.~Verza\orcid{0000-0002-1886-8348}\inst{\ref{aff181},\ref{aff123}}
\and P.~Vielzeuf\orcid{0000-0003-2035-9339}\inst{\ref{aff71}}
\and J.~R.~Weaver\orcid{0000-0003-1614-196X}\inst{\ref{aff182}}
\and L.~Zalesky\orcid{0000-0001-5680-2326}\inst{\ref{aff140}}
\and I.~A.~Zinchenko\inst{\ref{aff37}}
\and M.~Archidiacono\orcid{0000-0003-4952-9012}\inst{\ref{aff74},\ref{aff91}}
\and F.~Atrio-Barandela\orcid{0000-0002-2130-2513}\inst{\ref{aff183}}
\and T.~Bouvard\orcid{0009-0002-7959-312X}\inst{\ref{aff184}}
\and F.~Caro\inst{\ref{aff49}}
\and P.~Dimauro\orcid{0000-0001-7399-2854}\inst{\ref{aff49},\ref{aff185}}
\and P.-A.~Duc\orcid{0000-0003-3343-6284}\inst{\ref{aff133}}
\and Y.~Fang\inst{\ref{aff37}}
\and A.~M.~N.~Ferguson\inst{\ref{aff58}}
\and T.~Gasparetto\orcid{0000-0002-7913-4866}\inst{\ref{aff30}}
\and C.~M.~Gutierrez\orcid{0000-0001-7854-783X}\inst{\ref{aff186}}
\and I.~Kova{\v{c}}i{\'{c}}\orcid{0000-0001-6751-3263}\inst{\ref{aff187}}
\and S.~Kruk\orcid{0000-0001-8010-8879}\inst{\ref{aff12}}
\and A.~M.~C.~Le~Brun\orcid{0000-0002-0936-4594}\inst{\ref{aff129}}
\and T.~I.~Liaudat\orcid{0000-0002-9104-314X}\inst{\ref{aff130}}
\and A.~Montoro\orcid{0000-0003-4730-8590}\inst{\ref{aff54},\ref{aff53}}
\and A.~Mora\orcid{0000-0002-1922-8529}\inst{\ref{aff188}}
\and C.~Murray\inst{\ref{aff96}}
\and L.~Pagano\orcid{0000-0003-1820-5998}\inst{\ref{aff131},\ref{aff132}}
\and D.~Paoletti\orcid{0000-0003-4761-6147}\inst{\ref{aff28},\ref{aff117}}
\and E.~Sarpa\orcid{0000-0002-1256-655X}\inst{\ref{aff32},\ref{aff126},\ref{aff31}}
\and A.~Viitanen\orcid{0000-0001-9383-786X}\inst{\ref{aff85},\ref{aff49}}
\and J.~Lesgourgues\orcid{0000-0001-7627-353X}\inst{\ref{aff55}}
\and J.~Mart\'{i}n-Fleitas\orcid{0000-0002-8594-569X}\inst{\ref{aff188}}
\and D.~Scott\orcid{0000-0002-6878-9840}\inst{\ref{aff189}}}
										   
\institute{Mullard Space Science Laboratory, University College London, Holmbury St Mary, Dorking, Surrey RH5 6NT, UK\label{aff1}
\and
Universit\'e Paris-Saclay, Universit\'e Paris Cit\'e, CEA, CNRS, AIM, 91191, Gif-sur-Yvette, France\label{aff2}
\and
Department of Astronomy, University of Geneva, ch. d'Ecogia 16, 1290 Versoix, Switzerland\label{aff3}
\and
Centre National d'Etudes Spatiales -- Centre spatial de Toulouse, 18 avenue Edouard Belin, 31401 Toulouse Cedex 9, France\label{aff4}
\and
INAF-Istituto di Astrofisica e Planetologia Spaziali, via del Fosso del Cavaliere, 100, 00100 Roma, Italy\label{aff5}
\and
School of Physical Sciences, The Open University, Milton Keynes, MK7 6AA, UK\label{aff6}
\and
Universit\'e Paris-Saclay, CNRS, Institut d'astrophysique spatiale, 91405, Orsay, France\label{aff7}
\and
XCAM Limited, 2 Stone Circle Road, Northampton, NN3 8RF, UK\label{aff8}
\and
Centre for Electronic Imaging, Open University, Walton Hall, Milton Keynes, MK7~6AA, UK\label{aff9}
\and
UK Space Agency, Swindon, SN2 1SZ, UK\label{aff10}
\and
Leiden Observatory, Leiden University, Einsteinweg 55, 2333 CC Leiden, The Netherlands\label{aff11}
\and
ESAC/ESA, Camino Bajo del Castillo, s/n., Urb. Villafranca del Castillo, 28692 Villanueva de la Ca\~nada, Madrid, Spain\label{aff12}
\and
APCO Technologies, Chemin de Champex 10, 1860 Aigle, Switzerland\label{aff13}
\and
Surrey Satellite Technology Limited, Tycho House, 20 Stephenson Road, Surrey Research Park, Guildford, GU2 7YE, UK\label{aff14}
\and
Department of Physics, Centre for Extragalactic Astronomy, Durham University, South Road, DH1 3LE, UK\label{aff15}
\and
Department of Physics, Institute for Computational Cosmology, Durham University, South Road, DH1 3LE, UK\label{aff16}
\and
Institut d'Astrophysique de Paris, UMR 7095, CNRS, and Sorbonne Universit\'e, 98 bis boulevard Arago, 75014 Paris, France\label{aff17}
\and
Department of Physics, Oxford University, Keble Road, Oxford OX1 3RH, UK\label{aff18}
\and
Universit\"at Bonn, Argelander-Institut f\"ur Astronomie, Auf dem H\"ugel 71, 53121 Bonn, Germany\label{aff19}
\and
European Space Agency/ESTEC, Keplerlaan 1, 2201 AZ Noordwijk, The Netherlands\label{aff20}
\and
School of Mathematics, Statistics and Physics, Newcastle University, Herschel Building, Newcastle-upon-Tyne, NE1 7RU, UK\label{aff21}
\and
Teledyne E2V, 106, Waterhouse Lane, Chelmsford, CM1 2QU, UK\label{aff22}
\and
Jet Propulsion Laboratory, California Institute of Technology, 4800 Oak Grove Drive, Pasadena, CA, 91109, USA\label{aff23}
\and
Italian Space Agency, via del Politecnico snc, 00133 Roma, Italy\label{aff24}
\and
OHB Italia, Via Gallarate 150, 20151 Milano, Italy\label{aff25}
\and
School of Mathematics and Physics, University of Surrey, Guildford, Surrey, GU2 7XH, UK\label{aff26}
\and
INAF-Osservatorio Astronomico di Brera, Via Brera 28, 20122 Milano, Italy\label{aff27}
\and
INAF-Osservatorio di Astrofisica e Scienza dello Spazio di Bologna, Via Piero Gobetti 93/3, 40129 Bologna, Italy\label{aff28}
\and
IFPU, Institute for Fundamental Physics of the Universe, via Beirut 2, 34151 Trieste, Italy\label{aff29}
\and
INAF-Osservatorio Astronomico di Trieste, Via G. B. Tiepolo 11, 34143 Trieste, Italy\label{aff30}
\and
INFN, Sezione di Trieste, Via Valerio 2, 34127 Trieste TS, Italy\label{aff31}
\and
SISSA, International School for Advanced Studies, Via Bonomea 265, 34136 Trieste TS, Italy\label{aff32}
\and
Dipartimento di Fisica e Astronomia, Universit\`a di Bologna, Via Gobetti 93/2, 40129 Bologna, Italy\label{aff33}
\and
INFN-Sezione di Bologna, Viale Berti Pichat 6/2, 40127 Bologna, Italy\label{aff34}
\and
INAF-Osservatorio Astronomico di Padova, Via dell'Osservatorio 5, 35122 Padova, Italy\label{aff35}
\and
Max Planck Institute for Extraterrestrial Physics, Giessenbachstr. 1, 85748 Garching, Germany\label{aff36}
\and
Universit\"ats-Sternwarte M\"unchen, Fakult\"at f\"ur Physik, Ludwig-Maximilians-Universit\"at M\"unchen, Scheinerstrasse 1, 81679 M\"unchen, Germany\label{aff37}
\and
Institut de Physique Th\'eorique, CEA, CNRS, Universit\'e Paris-Saclay 91191 Gif-sur-Yvette Cedex, France\label{aff38}
\and
INAF-Osservatorio Astrofisico di Torino, Via Osservatorio 20, 10025 Pino Torinese (TO), Italy\label{aff39}
\and
Dipartimento di Fisica, Universit\`a di Genova, Via Dodecaneso 33, 16146, Genova, Italy\label{aff40}
\and
INFN-Sezione di Genova, Via Dodecaneso 33, 16146, Genova, Italy\label{aff41}
\and
Department of Physics "E. Pancini", University Federico II, Via Cinthia 6, 80126, Napoli, Italy\label{aff42}
\and
INAF-Osservatorio Astronomico di Capodimonte, Via Moiariello 16, 80131 Napoli, Italy\label{aff43}
\and
INFN section of Naples, Via Cinthia 6, 80126, Napoli, Italy\label{aff44}
\and
Instituto de Astrof\'isica e Ci\^encias do Espa\c{c}o, Universidade do Porto, CAUP, Rua das Estrelas, PT4150-762 Porto, Portugal\label{aff45}
\and
Dipartimento di Fisica, Universit\`a degli Studi di Torino, Via P. Giuria 1, 10125 Torino, Italy\label{aff46}
\and
INFN-Sezione di Torino, Via P. Giuria 1, 10125 Torino, Italy\label{aff47}
\and
INAF-IASF Milano, Via Alfonso Corti 12, 20133 Milano, Italy\label{aff48}
\and
INAF-Osservatorio Astronomico di Roma, Via Frascati 33, 00078 Monteporzio Catone, Italy\label{aff49}
\and
INFN-Sezione di Roma, Piazzale Aldo Moro, 2 - c/o Dipartimento di Fisica, Edificio G. Marconi, 00185 Roma, Italy\label{aff50}
\and
Centro de Investigaciones Energ\'eticas, Medioambientales y Tecnol\'ogicas (CIEMAT), Avenida Complutense 40, 28040 Madrid, Spain\label{aff51}
\and
Port d'Informaci\'{o} Cient\'{i}fica, Campus UAB, C. Albareda s/n, 08193 Bellaterra (Barcelona), Spain\label{aff52}
\and
Institut d'Estudis Espacials de Catalunya (IEEC),  Edifici RDIT, Campus UPC, 08860 Castelldefels, Barcelona, Spain\label{aff53}
\and
Institute of Space Sciences (ICE, CSIC), Campus UAB, Carrer de Can Magrans, s/n, 08193 Barcelona, Spain\label{aff54}
\and
Institute for Theoretical Particle Physics and Cosmology (TTK), RWTH Aachen University, 52056 Aachen, Germany\label{aff55}
\and
Dipartimento di Fisica e Astronomia "Augusto Righi" - Alma Mater Studiorum Universit\`a di Bologna, Viale Berti Pichat 6/2, 40127 Bologna, Italy\label{aff56}
\and
Instituto de Astrof\'isica de Canarias, Calle V\'ia L\'actea s/n, 38204, San Crist\'obal de La Laguna, Tenerife, Spain\label{aff57}
\and
Institute for Astronomy, University of Edinburgh, Royal Observatory, Blackford Hill, Edinburgh EH9 3HJ, UK\label{aff58}
\and
Jodrell Bank Centre for Astrophysics, Department of Physics and Astronomy, University of Manchester, Oxford Road, Manchester M13 9PL, UK\label{aff59}
\and
European Space Agency/ESRIN, Largo Galileo Galilei 1, 00044 Frascati, Roma, Italy\label{aff60}
\and
Universit\'e Claude Bernard Lyon 1, CNRS/IN2P3, IP2I Lyon, UMR 5822, Villeurbanne, F-69100, France\label{aff61}
\and
Institute of Physics, Laboratory of Astrophysics, Ecole Polytechnique F\'ed\'erale de Lausanne (EPFL), Observatoire de Sauverny, 1290 Versoix, Switzerland\label{aff62}
\and
UCB Lyon 1, CNRS/IN2P3, IUF, IP2I Lyon, 4 rue Enrico Fermi, 69622 Villeurbanne, France\label{aff63}
\and
Institut de Ciencies de l'Espai (IEEC-CSIC), Campus UAB, Carrer de Can Magrans, s/n Cerdanyola del Vall\'es, 08193 Barcelona, Spain\label{aff64}
\and
Canada-France-Hawaii Telescope, 65-1238 Mamalahoa Hwy, Kamuela, HI 96743, USA\label{aff65}
\and
Aix-Marseille Universit\'e, CNRS, CNES, LAM, Marseille, France\label{aff66}
\and
Departamento de F\'isica, Faculdade de Ci\^encias, Universidade de Lisboa, Edif\'icio C8, Campo Grande, PT1749-016 Lisboa, Portugal\label{aff67}
\and
Instituto de Astrof\'isica e Ci\^encias do Espa\c{c}o, Faculdade de Ci\^encias, Universidade de Lisboa, Campo Grande, 1749-016 Lisboa, Portugal\label{aff68}
\and
INFN-Padova, Via Marzolo 8, 35131 Padova, Italy\label{aff69}
\and
School of Physics, HH Wills Physics Laboratory, University of Bristol, Tyndall Avenue, Bristol, BS8 1TL, UK\label{aff70}
\and
Aix-Marseille Universit\'e, CNRS/IN2P3, CPPM, Marseille, France\label{aff71}
\and
Istituto Nazionale di Fisica Nucleare, Sezione di Bologna, Via Irnerio 46, 40126 Bologna, Italy\label{aff72}
\and
FRACTAL S.L.N.E., calle Tulip\'an 2, Portal 13 1A, 28231, Las Rozas de Madrid, Spain\label{aff73}
\and
Dipartimento di Fisica "Aldo Pontremoli", Universit\`a degli Studi di Milano, Via Celoria 16, 20133 Milano, Italy\label{aff74}
\and
Institute of Theoretical Astrophysics, University of Oslo, P.O. Box 1029 Blindern, 0315 Oslo, Norway\label{aff75}
\and
Department of Physics, Lancaster University, Lancaster, LA1 4YB, UK\label{aff76}
\and
Felix Hormuth Engineering, Goethestr. 17, 69181 Leimen, Germany\label{aff77}
\and
Max-Planck-Institut f\"ur Astronomie, K\"onigstuhl 17, 69117 Heidelberg, Germany\label{aff78}
\and
Technical University of Denmark, Elektrovej 327, 2800 Kgs. Lyngby, Denmark\label{aff79}
\and
Cosmic Dawn Center (DAWN), Denmark\label{aff80}
\and
Universit\'e Paris-Saclay, CNRS/IN2P3, IJCLab, 91405 Orsay, France\label{aff81}
\and
Institut de Recherche en Astrophysique et Plan\'etologie (IRAP), Universit\'e de Toulouse, CNRS, UPS, CNES, 14 Av. Edouard Belin, 31400 Toulouse, France\label{aff82}
\and
NASA Goddard Space Flight Center, Greenbelt, MD 20771, USA\label{aff83}
\and
Department of Physics and Astronomy, University College London, Gower Street, London WC1E 6BT, UK\label{aff84}
\and
Department of Physics and Helsinki Institute of Physics, Gustaf H\"allstr\"omin katu 2, 00014 University of Helsinki, Finland\label{aff85}
\and
Universit\'e de Gen\`eve, D\'epartement de Physique Th\'eorique and Centre for Astroparticle Physics, 24 quai Ernest-Ansermet, CH-1211 Gen\`eve 4, Switzerland\label{aff86}
\and
Department of Physics, P.O. Box 64, 00014 University of Helsinki, Finland\label{aff87}
\and
Helsinki Institute of Physics, Gustaf H{\"a}llstr{\"o}min katu 2, University of Helsinki, Helsinki, Finland\label{aff88}
\and
NOVA optical infrared instrumentation group at ASTRON, Oude Hoogeveensedijk 4, 7991PD, Dwingeloo, The Netherlands\label{aff89}
\and
Centre de Calcul de l'IN2P3/CNRS, 21 avenue Pierre de Coubertin 69627 Villeurbanne Cedex, France\label{aff90}
\and
INFN-Sezione di Milano, Via Celoria 16, 20133 Milano, Italy\label{aff91}
\and
University of Applied Sciences and Arts of Northwestern Switzerland, School of Engineering, 5210 Windisch, Switzerland\label{aff92}
\and
Dipartimento di Fisica e Astronomia "Augusto Righi" - Alma Mater Studiorum Universit\`a di Bologna, via Piero Gobetti 93/2, 40129 Bologna, Italy\label{aff93}
\and
Infrared Processing and Analysis Center, California Institute of Technology, Pasadena, CA 91125, USA\label{aff94}
\and
Universit\'e C\^{o}te d'Azur, Observatoire de la C\^{o}te d'Azur, CNRS, Laboratoire Lagrange, Bd de l'Observatoire, CS 34229, 06304 Nice cedex 4, France\label{aff95}
\and
Universit\'e Paris Cit\'e, CNRS, Astroparticule et Cosmologie, 75013 Paris, France\label{aff96}
\and
Institut d'Astrophysique de Paris, 98bis Boulevard Arago, 75014, Paris, France\label{aff97}
\and
Institut de F\'{i}sica d'Altes Energies (IFAE), The Barcelona Institute of Science and Technology, Campus UAB, 08193 Bellaterra (Barcelona), Spain\label{aff98}
\and
Kapteyn Astronomical Institute, University of Groningen, PO Box 800, 9700 AV Groningen, The Netherlands\label{aff99}
\and
Department of Physics and Astronomy, University of Aarhus, Ny Munkegade 120, DK-8000 Aarhus C, Denmark\label{aff100}
\and
Waterloo Centre for Astrophysics, University of Waterloo, Waterloo, Ontario N2L 3G1, Canada\label{aff101}
\and
Department of Physics and Astronomy, University of Waterloo, Waterloo, Ontario N2L 3G1, Canada\label{aff102}
\and
Perimeter Institute for Theoretical Physics, Waterloo, Ontario N2L 2Y5, Canada\label{aff103}
\and
Space Science Data Center, Italian Space Agency, via del Politecnico snc, 00133 Roma, Italy\label{aff104}
\and
Institute of Space Science, Str. Atomistilor, nr. 409 M\u{a}gurele, Ilfov, 077125, Romania\label{aff105}
\and
Departamento de Astrof\'isica, Universidad de La Laguna, 38206, La Laguna, Tenerife, Spain\label{aff106}
\and
Institute for Particle Physics and Astrophysics, Dept. of Physics, ETH Zurich, Wolfgang-Pauli-Strasse 27, 8093 Zurich, Switzerland\label{aff107}
\and
Dipartimento di Fisica e Astronomia "G. Galilei", Universit\`a di Padova, Via Marzolo 8, 35131 Padova, Italy\label{aff108}
\and
Caltech/IPAC, 1200 E. California Blvd., Pasadena, CA 91125, USA\label{aff109}
\and
Institut f\"ur Theoretische Physik, University of Heidelberg, Philosophenweg 16, 69120 Heidelberg, Germany\label{aff110}
\and
Universit\'e St Joseph; Faculty of Sciences, Beirut, Lebanon\label{aff111}
\and
Departamento de F\'isica, FCFM, Universidad de Chile, Blanco Encalada 2008, Santiago, Chile\label{aff112}
\and
Universit\"at Innsbruck, Institut f\"ur Astro- und Teilchenphysik, Technikerstr. 25/8, 6020 Innsbruck, Austria\label{aff113}
\and
Satlantis, University Science Park, Sede Bld 48940, Leioa-Bilbao, Spain\label{aff114}
\and
Instituto de Astrof\'isica e Ci\^encias do Espa\c{c}o, Faculdade de Ci\^encias, Universidade de Lisboa, Tapada da Ajuda, 1349-018 Lisboa, Portugal\label{aff115}
\and
Universidad Polit\'ecnica de Cartagena, Departamento de Electr\'onica y Tecnolog\'ia de Computadoras,  Plaza del Hospital 1, 30202 Cartagena, Spain\label{aff116}
\and
INFN-Bologna, Via Irnerio 46, 40126 Bologna, Italy\label{aff117}
\and
Dipartimento di Fisica, Universit\`a degli studi di Genova, and INFN-Sezione di Genova, via Dodecaneso 33, 16146, Genova, Italy\label{aff118}
\and
Carnegie Observatories, Pasadena, CA 91101, USA\label{aff119}
\and
INAF, Istituto di Radioastronomia, Via Piero Gobetti 101, 40129 Bologna, Italy\label{aff120}
\and
Astronomical Observatory of the Autonomous Region of the Aosta Valley (OAVdA), Loc. Lignan 39, I-11020, Nus (Aosta Valley), Italy\label{aff121}
\and
School of Physics and Astronomy, Cardiff University, The Parade, Cardiff, CF24 3AA, UK\label{aff122}
\and
Center for Computational Astrophysics, Flatiron Institute, 162 5th Avenue, 10010, New York, NY, USA\label{aff123}
\and
Ernst-Reuter-Str. 4e, 31224 Peine, Germany\label{aff124}
\and
Junia, EPA department, 41 Bd Vauban, 59800 Lille, France\label{aff125}
\and
ICSC - Centro Nazionale di Ricerca in High Performance Computing, Big Data e Quantum Computing, Via Magnanelli 2, Bologna, Italy\label{aff126}
\and
Instituto de F\'isica Te\'orica UAM-CSIC, Campus de Cantoblanco, 28049 Madrid, Spain\label{aff127}
\and
CERCA/ISO, Department of Physics, Case Western Reserve University, 10900 Euclid Avenue, Cleveland, OH 44106, USA\label{aff128}
\and
Laboratoire Univers et Th\'eorie, Observatoire de Paris, Universit\'e PSL, Universit\'e Paris Cit\'e, CNRS, 92190 Meudon, France\label{aff129}
\and
IRFU, CEA, Universit\'e Paris-Saclay 91191 Gif-sur-Yvette Cedex, France\label{aff130}
\and
Dipartimento di Fisica e Scienze della Terra, Universit\`a degli Studi di Ferrara, Via Giuseppe Saragat 1, 44122 Ferrara, Italy\label{aff131}
\and
Istituto Nazionale di Fisica Nucleare, Sezione di Ferrara, Via Giuseppe Saragat 1, 44122 Ferrara, Italy\label{aff132}
\and
Universit\'e de Strasbourg, CNRS, Observatoire astronomique de Strasbourg, UMR 7550, 67000 Strasbourg, France\label{aff133}
\and
Kavli Institute for the Physics and Mathematics of the Universe (WPI), University of Tokyo, Kashiwa, Chiba 277-8583, Japan\label{aff134}
\and
Dipartimento di Fisica - Sezione di Astronomia, Universit\`a di Trieste, Via Tiepolo 11, 34131 Trieste, Italy\label{aff135}
\and
NASA Ames Research Center, Moffett Field, CA 94035, USA\label{aff136}
\and
Bay Area Environmental Research Institute, Moffett Field, California 94035, USA\label{aff137}
\and
Minnesota Institute for Astrophysics, University of Minnesota, 116 Church St SE, Minneapolis, MN 55455, USA\label{aff138}
\and
Institute Lorentz, Leiden University, Niels Bohrweg 2, 2333 CA Leiden, The Netherlands\label{aff139}
\and
Institute for Astronomy, University of Hawaii, 2680 Woodlawn Drive, Honolulu, HI 96822, USA\label{aff140}
\and
Department of Physics \& Astronomy, University of California Irvine, Irvine CA 92697, USA\label{aff141}
\and
Department of Astronomy \& Physics and Institute for Computational Astrophysics, Saint Mary's University, 923 Robie Street, Halifax, Nova Scotia, B3H 3C3, Canada\label{aff142}
\and
Departamento F\'isica Aplicada, Universidad Polit\'ecnica de Cartagena, Campus Muralla del Mar, 30202 Cartagena, Murcia, Spain\label{aff143}
\and
CEA Saclay, DFR/IRFU, Service d'Astrophysique, Bat. 709, 91191 Gif-sur-Yvette, France\label{aff144}
\and
Institute of Cosmology and Gravitation, University of Portsmouth, Portsmouth PO1 3FX, UK\label{aff145}
\and
Department of Computer Science, Aalto University, PO Box 15400, Espoo, FI-00 076, Finland\label{aff146}
\and
Ruhr University Bochum, Faculty of Physics and Astronomy, Astronomical Institute (AIRUB), German Centre for Cosmological Lensing (GCCL), 44780 Bochum, Germany\label{aff147}
\and
DARK, Niels Bohr Institute, University of Copenhagen, Jagtvej 155, 2200 Copenhagen, Denmark\label{aff148}
\and
Instituto de Astrof\'isica de Canarias (IAC); Departamento de Astrof\'isica, Universidad de La Laguna (ULL), 38200, La Laguna, Tenerife, Spain\label{aff149}
\and
Universit\'e PSL, Observatoire de Paris, Sorbonne Universit\'e, CNRS, LERMA, 75014, Paris, France\label{aff150}
\and
Universit\'e Paris-Cit\'e, 5 Rue Thomas Mann, 75013, Paris, France\label{aff151}
\and
Univ. Grenoble Alpes, CNRS, Grenoble INP, LPSC-IN2P3, 53, Avenue des Martyrs, 38000, Grenoble, France\label{aff152}
\and
Department of Physics and Astronomy, Vesilinnantie 5, 20014 University of Turku, Finland\label{aff153}
\and
Serco for European Space Agency (ESA), Camino bajo del Castillo, s/n, Urbanizacion Villafranca del Castillo, Villanueva de la Ca\~nada, 28692 Madrid, Spain\label{aff154}
\and
ARC Centre of Excellence for Dark Matter Particle Physics, Melbourne, Australia\label{aff155}
\and
Centre for Astrophysics \& Supercomputing, Swinburne University of Technology,  Hawthorn, Victoria 3122, Australia\label{aff156}
\and
School of Physics and Astronomy, Queen Mary University of London, Mile End Road, London E1 4NS, UK\label{aff157}
\and
Department of Physics and Astronomy, University of the Western Cape, Bellville, Cape Town, 7535, South Africa\label{aff158}
\and
Universit\'e Libre de Bruxelles (ULB), Service de Physique Th\'eorique CP225, Boulevard du Triophe, 1050 Bruxelles, Belgium\label{aff159}
\and
ICTP South American Institute for Fundamental Research, Instituto de F\'{\i}sica Te\'orica, Universidade Estadual Paulista, S\~ao Paulo, Brazil\label{aff160}
\and
Oskar Klein Centre for Cosmoparticle Physics, Department of Physics, Stockholm University, Stockholm, SE-106 91, Sweden\label{aff161}
\and
Astrophysics Group, Blackett Laboratory, Imperial College London, London SW7 2AZ, UK\label{aff162}
\and
INAF-Osservatorio Astrofisico di Arcetri, Largo E. Fermi 5, 50125, Firenze, Italy\label{aff163}
\and
Dipartimento di Fisica, Sapienza Universit\`a di Roma, Piazzale Aldo Moro 2, 00185 Roma, Italy\label{aff164}
\and
Centro de Astrof\'{\i}sica da Universidade do Porto, Rua das Estrelas, 4150-762 Porto, Portugal\label{aff165}
\and
Niels Bohr Institute, University of Copenhagen, Jagtvej 128, 2200 Copenhagen, Denmark\label{aff166}
\and
Dipartimento di Fisica, Universit\`a di Roma Tor Vergata, Via della Ricerca Scientifica 1, Roma, Italy\label{aff167}
\and
INFN, Sezione di Roma 2, Via della Ricerca Scientifica 1, Roma, Italy\label{aff168}
\and
HE Space for European Space Agency (ESA), Camino bajo del Castillo, s/n, Urbanizacion Villafranca del Castillo, Villanueva de la Ca\~nada, 28692 Madrid, Spain\label{aff169}
\and
Institute of Astronomy, University of Cambridge, Madingley Road, Cambridge CB3 0HA, UK\label{aff170}
\and
Center for Frontier Science, Chiba University, 1-33 Yayoi-cho, Inage-ku, Chiba 263-8522, Japan\label{aff171}
\and
Department of Physics, Graduate School of Science, Chiba University, 1-33 Yayoi-Cho, Inage-Ku, Chiba 263-8522, Japan\label{aff172}
\and
Department of Astrophysics, University of Zurich, Winterthurerstrasse 190, 8057 Zurich, Switzerland\label{aff173}
\and
Higgs Centre for Theoretical Physics, School of Physics and Astronomy, The University of Edinburgh, Edinburgh EH9 3FD, UK\label{aff174}
\and
Theoretical astrophysics, Department of Physics and Astronomy, Uppsala University, Box 515, 751 20 Uppsala, Sweden\label{aff175}
\and
School of Physics \& Astronomy, University of Southampton, Highfield Campus, Southampton SO17 1BJ, UK\label{aff176}
\and
Department of Physics, Royal Holloway, University of London, TW20 0EX, UK\label{aff177}
\and
Department of Physics and Astronomy, University of California, Davis, CA 95616, USA\label{aff178}
\and
Department of Astrophysical Sciences, Peyton Hall, Princeton University, Princeton, NJ 08544, USA\label{aff179}
\and
Cosmic Dawn Center (DAWN)\label{aff180}
\and
Center for Cosmology and Particle Physics, Department of Physics, New York University, New York, NY 10003, USA\label{aff181}
\and
Department of Astronomy, University of Massachusetts, Amherst, MA 01003, USA\label{aff182}
\and
Departamento de F{\'\i}sica Fundamental. Universidad de Salamanca. Plaza de la Merced s/n. 37008 Salamanca, Spain\label{aff183}
\and
Thales~Services~S.A.S., 290 All\'ee du Lac, 31670 Lab\`ege, France\label{aff184}
\and
Observatorio Nacional, Rua General Jose Cristino, 77-Bairro Imperial de Sao Cristovao, Rio de Janeiro, 20921-400, Brazil\label{aff185}
\and
Instituto de Astrof\'\i sica de Canarias, c/ Via Lactea s/n, La Laguna E-38200, Spain. Departamento de Astrof\'\i sica de la Universidad de La Laguna, Avda. Francisco Sanchez, La Laguna, E-38200, Spain\label{aff186}
\and
Sterrenkundig Observatorium, Universiteit Gent, Krijgslaan 281 S9, 9000 Gent, Belgium\label{aff187}
\and
Aurora Technology for European Space Agency (ESA), Camino bajo del Castillo, s/n, Urbanizacion Villafranca del Castillo, Villanueva de la Ca\~nada, 28692 Madrid, Spain\label{aff188}
\and
Department of Physics and Astronomy, University of British Columbia, Vancouver, BC V6T 1Z1, Canada\label{aff189}}